\documentclass[rmp,aps,twocolumn,notitlepage,longbibliography]{revtex4-1}
\usepackage{amssymb,amsfonts,amsmath,amstext,graphicx,hyperref}
\usepackage{tikz}
\usepackage{graphicx}
\usepackage[shortlabels]{enumitem}
\usepackage[normalem]{ulem}
\hypersetup{
	colorlinks=true,
        citecolor=red,
	linkcolor=blue,
	filecolor=magenta,      
	urlcolor=cyan,
}
\usepackage{xcolor}
\usepackage{braket}
\usepackage{lipsum}
\usepackage{orcidlink}

\usepackage{makecell}

\begin{document}
\title{Quantum Synchronization}

\author{Parvinder Solanki~\orcidlink{0000-0001-7477-4453}}\thanks{equal contribution}
\affiliation{Institut f\"{u}r Theoretische Physik and Center for Integrated Quantum Science and Technology,  Universit\"{a}t T\"{u}bingen, Auf der Morgenstelle 14, 72076 T\"{u}bingen, Germany}

\author{Albert Cabot~\orcidlink{0000-0001-7440-2254}}\thanks{equal contribution}
\affiliation{Institut f\"{u}r Theoretische Physik and Center for Integrated Quantum Science and Technology,  Universit\"{a}t T\"{u}bingen, Auf der Morgenstelle 14, 72076 T\"{u}bingen, Germany}
\affiliation{Institute for Cross-Disciplinary Physics and Complex Systems (IFISC) UIB-CSIC, Campus Universitat Illes Balears, 07122, Palma de Mallorca, Spain}

\author{Fernando Iemini~\orcidlink{0000-0003-1645-9492}}
\affiliation{Instituto de F\'isica, Universidade Federal Fluminense, Av. Gal. Milton Tavares de Souza s/n, Gragoat\'a, 24210-346 Niter\'oi, Rio de Janeiro, Brazil}

\author{Federico Carollo~\orcidlink{0000-0002-6961-7143}}
\affiliation{Dipartimento di Fisica, Sapienza Università di Roma, Piazzale Aldo Moro 5, 00185 Rome, Italy}

\author{Midhun Krishna~\orcidlink{0000-0003-2643-2437}}
\affiliation{Department of Physics, Indian Institute of Technology Bombay, Maharashtra 400076, India}

\author{Yeshma Ibrahim~\orcidlink{0009-0009-0460-5430}}
\affiliation{Department of Physics, Indian Institute of Technology Bombay, Maharashtra 400076, India}

\author{Michal Hajdu\v{s}ek~\orcidlink{0000-0002-8319-9566}}%
\affiliation{Graduate School of Media Design, Keio University, Yokohama, Kanagawa 223-0061, Japan}%

\author{Igor Lesanovsky~\orcidlink{0000-0001-9660-9467}}
\affiliation{Institut f\"{u}r Theoretische Physik and Center for Integrated Quantum Science and Technology,  Universit\"{a}t T\"{u}bingen, Auf der Morgenstelle 14, 72076 T\"{u}bingen, Germany}
\affiliation{School of Physics and Astronomy and Centre for the Mathematics and Theoretical Physics of Quantum Non-Equilibrium Systems, The University of Nottingham, Nottingham, NG7 2RD, United Kingdom}

\author{Rosario Fazio~\orcidlink{0000-0002-7793-179X}}
\affiliation{The Abdus Salam International Center for Theoretical Physics, Strada Costiera 11, 34151 Trieste, Italy}
\affiliation{Dipartimento di Fisica ``E. Pancini", Universit\`a di Napoli ``Federico II'', Monte S. Angelo, I-80126 Napoli, Italy}

\author{Roberta Zambrini~\orcidlink{0000-0002-9896-3563}}
\affiliation{IFISC, Institute for Cross-Disciplinary Physics and Complex Systems (UIB-CSIC)\\
Campus Universitat Illes Balears, 07122, Palma de Mallorca, Spain}

\author{Sai Vinjanampathy~\orcidlink{0000-0002-5919-5442}}
\email{sai@phy.iitb.ac.in}
\affiliation{Department of Physics, Indian Institute of Technology Bombay, Maharashtra 400076, India}
\affiliation{Centre of Excellence in Quantum Information, Computing, Science and Technology, IIT Bombay, Maharashtra 400076, India}

\begin{abstract}
Natural and engineered classical systems are replete with examples of synchronization, understood as the adjustment of rhythms of physical systems. Such synchronization is at the heart of the stability of several classical technologies, such as mechanical bridges and electrical networks. Given the advent of quantum simulation and computation technologies, it is natural to study a quantum analogue of synchronization and explore novel applications. This review surveys synchronization in few and many-body quantum systems, measures that quantify them, and their applications to quantum technologies.
\end{abstract}

\maketitle
\tableofcontents

\section{Overview}
\label{Sec:overview}

Linear systems are the benchmark for simple and exactly solvable models leading to oscillatory behavior and linear response. However, generic classical dynamics is nonlinear. Almost all dynamics of real physical systems exhibit a regime of nonlinearity that gives rise to rich dynamical phenomena, such as chaos and synchronization \cite{strogatz2015nonlinear}.  
The natural world is replete with synchronous behavior, which underlies circadian rhythms, the beating of the heart, the flashing of fireflies, and medical conditions such as epilepsy \cite{pikovsky2001synchronization}. 
Likewise, engineered systems such as electrical networks are susceptible to collapse due to synchronization-induced instabilities \cite{machowski2020power}. 
Hence, synchronization is considered a fundamental classical phenomenon to understand natural and artificial systems.

In recent decades, advances in quantum simulation, computation, and atomic, molecular, and optical systems have enabled the realization of mechanical resonators operating with only a few phonons, and other physical systems close to their quantum ground state \cite{nakahara2008quantum}. 
Alongside these developments, the growing ability to manipulate and control quantum fluctuations has paved the way for the extension of synchronization concepts into the quantum regime. 
Motivated by this, there has been intense activity in studying quantum generalizations of nonlinear dynamical phenomena, foremost among these being the study of quantum synchronization.  
The purpose of this review is to provide an exposition of the fundamental ideas of quantum synchronization and review the progress in the field. 
In the service of pedagogical completeness, we succinctly introduce the main ideas of classical synchronization that have been studied and generalized in quantum systems.  
In few-body quantum systems, several approaches to quantum synchronization have emerged in the recent past. These include systems whose mean-field limits correspond to well-studied classical models, in contrast with synchronization of finite quantum systems.  Additionally, efforts are currently underway to understand how synchronization and time-dependent collective phenomena manifest in the realm of many-body physics. Some of these studies relate to mean-field models that have made strong connections to classical non-linear systems; examples include the Kuramoto model and the van der Pol oscillator. Some others, such as time crystal physics, represent novel many-body phase transition physics whose dynamics exhibit synchronous behavior. In several such models, we review how synchronization, when probed via measurement, interacts with the evolution of the system. 
Finally, various measures, motivated by semiclassical analysis, phase-space considerations, and quantum information-theoretic arguments, have been proposed and studied. Our review brings together these studies and presents them alongside motivating experiments in the field. Finally, we also review disparate applications of quantum synchronization to quantum technologies, from sensors to heat engines.

We begin the review with an introduction to classical synchronization in Sec.~\ref{ch:intro}, followed by a discussion of techniques necessary for studying quantum synchronization in Sec.~\ref{sec.intro.qsync}. Quantum synchronization in few-body and many-body systems is discussed in Sec.~\ref{sec:Few_Body_QSynch} and Sec.~\ref{sec:Many_Body_QSynch}, respectively. 
Section \ref{sec:thermodynamics} is devoted to a discussion on the thermodynamics of quantum synchronization and its relationship to quantum thermal machines.
The applications of quantum synchronization to various technologies are discussed in \ref{sec:applications_sync}.
Finally, we provide a summary and outlook in Sec.~\ref{sec:summary_outlook}.

\section{Introduction to Classical Synchronization}\label{ch:intro}
 
Motivated by the early work of Huygens~\cite{bennett2002huygens}, synchronization has been studied for over three centuries now. 
Within the paradigmatic framework of self‑sustained oscillators, synchronization emerges when two or more such oscillators adjust their rhythms with each other or to an external drive. 
In this section, we review those concepts from nonlinear dynamics and classical synchronization required for a self-contained study of quantum synchronization. 
We briefly review the key ideas of entrainment and mutual synchronization. 
In doing so, we discuss a few paradigmatic models that capture the essence of classical synchronization in a way that motivates the quantum discussion. 
The study of synchronization is quite mature in the classical regime, and we refer the motivated reader to several textbooks in the field \cite{strogatz2015nonlinear,pikovsky2001synchronization,balanov2009simple}. 
\subsection{Self-sustained oscillators}\label{ch:classical_sync}

A key concept in classical synchronization is that of a self-sustained oscillator~\cite{jenkins2013self}.
Rigorously studied in~\cite{andronov1966theory}, self-sustained oscillators display the following three features:
\begin{enumerate}
\item They do not damp with time.
\item Oscillations are not due to periodic driving.
\item The characteristics of self-oscillations, such as amplitude and period, are determined by the parameters of the system, not by the initial conditions.
\end{enumerate}
Examples of such oscillators are bountiful in nature; beating heart cells~\cite{vanderPol1928heartbeat}, flashing fireflies~\cite{blair1915luminous,sarfati2021self}, firing neurons~\cite{fitzhugh1961impulses,nagumo1962active,cebrian2024six} and grandfather pendulum clocks~\cite{goldsztein2021synchronization}.
A counterexample to a self-sustained oscillator is a mass suspended on a spring that oscillates due to an initial kick. 
In the absence of dissipation, the amplitude of the oscillations depends on the strength of the initial kick. 
However, any physically reasonable model of an oscillator dissipates energy. 
Oscillations will therefore eventually damp out, bringing the mass back to its initial resting position.

For any general model, we denote the rate at which the oscillator loses energy by $P_{\text{diss}}$. 
To maintain steady oscillations over time in the presence of dissipation, a self-sustained oscillator requires a source of energy. 
We denote the power supplied by this energy source as $P_{\text{gain}}$. 
The relationship between the supplied and dissipated power determines how the amplitude, denoted by $A$, changes over time. 
The oscillations are damped when $P_{\text{gain}}<P_{\text{diss}}$, while the oscillations increase in amplitude for $P_{\text{gain}}>P_{\text{diss}}$.
Finally, when the supplied power exactly compensates for the dissipated power and $P_{\text{gain}}=P_{\text{diss}}$, the oscillation amplitude remains unchanged.
We denote this steady amplitude, also known as a \emph{radial fixed point} of the system, by $A^*$.
We are interested in this last case and will discuss the conditions under which it arises.

\begin{figure}
    \centering
    \includegraphics[width=\columnwidth]{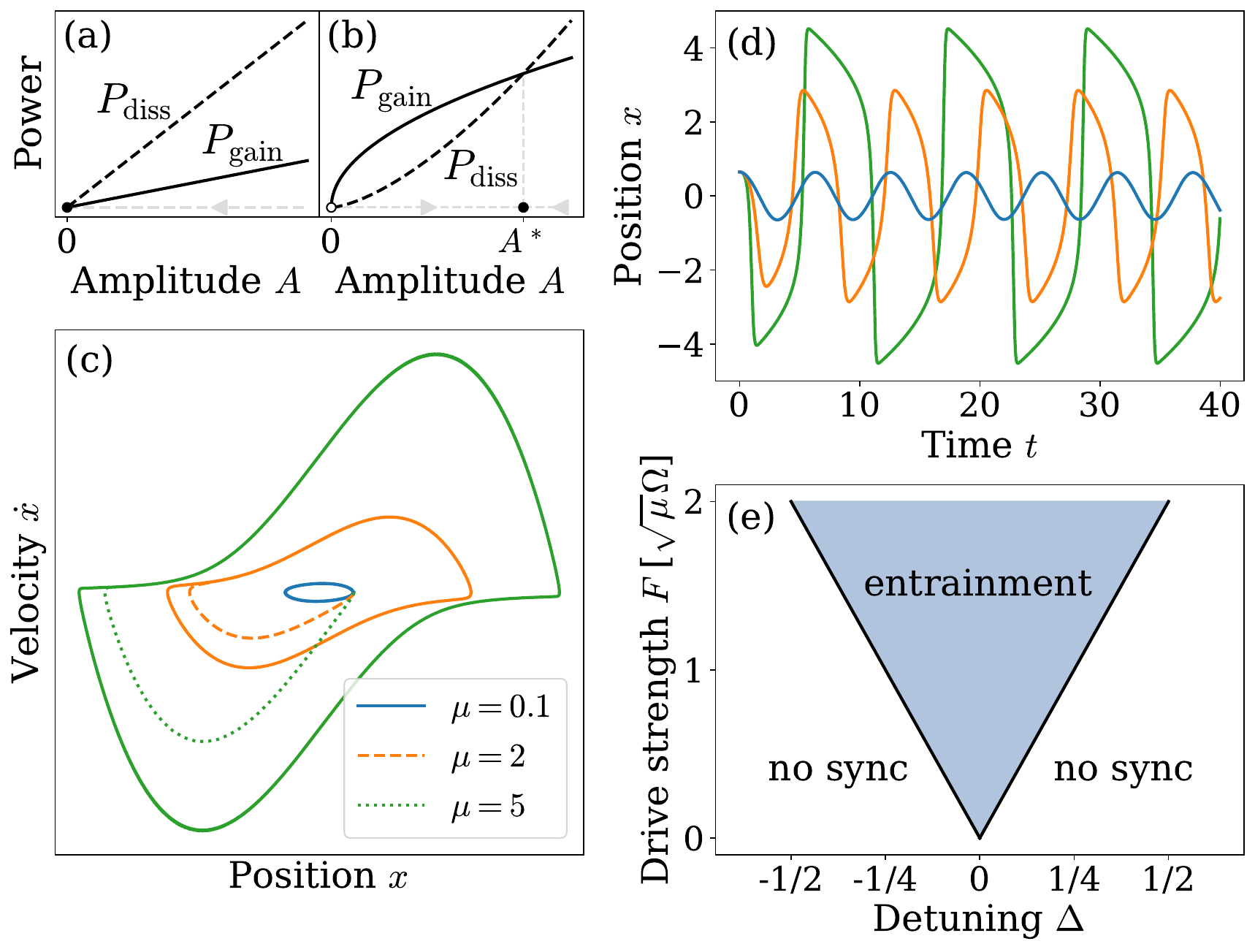}
    \caption{(a) Linear systems cannot support self-sustaining oscillations. For these systems, the only fixed point for the amplitude is $A^*=0$. (b) Non-linear systems, on the other hand, can support non-trivial stable amplitudes. (c) Increasing the non-linearity $\mu$ in Eq.~(\ref{eq:undriven_vdP_classical}) results in limit cycles that further deviate from harmonic (elliptical) trajectories. This is clearly observed from the time series for the position of the van der Pol oscillator in (d). (e) Arnold tongue marking the parameter region where a driven vdP becomes entrained.}
    \label{fig:Introduction-self_sustained}
\end{figure}

For generic nonlinear oscillators, the supplied and dissipated power are functions of the oscillator's amplitude.
This puts an immediate restriction on the type of system capable of displaying steady oscillations.
This is exemplified by considering a harmonic oscillator, where both $P_{\text{gain}}$ and $P_{\text{diss}}$ are linear functions of the amplitude $A$, as shown in Fig.~\ref{fig:Introduction-self_sustained}(a).
The steady amplitude condition of $P_{\text{gain}}=P_{\text{diss}}$ is satisfied only by the trivial case of $A^*=0$ when the oscillator is at rest.
The stability of this fixed point depends on whether there is net supplied power or net dissipated power.
Figure~\ref{fig:Introduction-self_sustained}(a) depicts the case where $P_{\text{gain}}<P_{\text{diss}}$.
The fixed point $A^*=0$ is stable, which means that any perturbations away from this fixed point will dampen and the system will return to $A^*=0$.
This is depicted by the gray arrow in Fig.~\ref{fig:Introduction-self_sustained}(a), showing that any finite amplitude of oscillations will evolve towards the stable fixed point, represented by the solid black circle.
We can imagine the situation where $P_{\text{gain}}>P_{\text{diss}}$, leading to an unstable fixed point at $A^*=0$.
Any perturbation away from the fixed point will get amplified, and the amplitude will keep increasing.
The fact that linear systems only have trivial solutions to the condition $P_{\text{gain}}=P_{\text{diss}}$ means that any self-sustained oscillator must be a non-linear system.

For non-linear systems, the supplied and dissipated powers are no longer restricted to being linear functions of the amplitude.
The steady amplitude condition $P_{\text{gain}}=P_{\text{diss}}$ can now support a non-trivial fixed point $A^*$, as shown in Fig.~\ref{fig:Introduction-self_sustained}(b).
In fact, the depicted system has two fixed points; the original trivial fixed point when the system is at rest, and a new one at a finite amplitude $A^*$.
The stability of these two fixed points depends on the relationship between the supplied and dissipated power.
In order for the non-trivial fixed point to be stable, we require that $P_{\text{gain}}>P_{\text{diss}}$ for amplitudes $0<A<A^*$, and that $P_{\text{gain}}<P_{\text{diss}}$ for amplitudes $A^*<A$.
Dynamical systems with qualitatively similar non-linear relationships between the supplied and dissipated power can support self-sustaining oscillations.

A canonical example of a self-sustained oscillator is the van der Pol (vdP) oscillator~\cite{vanderPol1926relaxation,ginoux2012van},
\begin{equation}
    \ddot{x} - (\mu-x^2)\dot{x}+\omega_0^2x=0,
    \label{eq:undriven_vdP_classical}
\end{equation}
where $x$ is the position, dots represent time derivatives, $\omega_0$ is the natural frequency, and $\mu$ is a scalar quantifying the degree of non-linearity of the oscillator.
For $\mu<0$, no self-sustained oscillations exist and the only stable fixed point is at the origin.
At $\mu=0$, the system undergoes a Hopf bifurcation. Therefore, the fixed point at the origin becomes unstable and a stable attractive trajectory is born, known as a \emph{limit cycle}~\cite{strogatz2015nonlinear}. 
We can observe that the sign of the damping term in Eq.~(\ref{eq:undriven_vdP_classical}) depends on position $x$.
When $x^2<\mu$, the negative damping term acts as an energy supply, resulting in an amplitude increase.
When $x^2>\mu$, positive damping acts as dissipation, leading to a decrease in amplitude.
This demonstrates the interplay of energy supply and dissipation, resulting in limit cycles with stable self-sustained oscillations.

For small non-linearities, where $0<\mu\ll1$, the limit cycle can be well approximated by harmonic oscillations.
Self-sustained oscillators in this regime are therefore known as \emph{quasi-harmonic}.
Figure~\ref{fig:Introduction-self_sustained}(c) shows the limit cycle trajectories in the phase space of a vdP oscillator.
For a weak non-linearity, shown in blue, the quasi-harmonic nature of the oscillations is evident from the elliptic shape of the limit cycle, as well as from the near sinusoidal variation of the position $x$ with time, depicted in Fig.~\ref{fig:Introduction-self_sustained}(d).
For larger non-linearity $\mu$, the oscillations are no longer quasi-harmonic as can be observed in Fig.~\ref{fig:Introduction-self_sustained}(c).
The shape of the limit cycles, shown in orange and green,  deviates from a simple sinusoidal behavior of a simple harmonic motion and exhibits other frequency contributions due to the non-linearity in the equations of motion, as also seen in the corresponding position evolution in Fig.~\ref{fig:Introduction-self_sustained}(d).

A crucial characteristic of self-sustained oscillators is how they respond to perturbations.
We have seen that the amplitude of oscillations is stable, meaning that small perturbations away from the limit cycle trajectory in phase space decay quickly.
However, the phase of the oscillations behaves in a distinctly different way.
Perturbations of the phase do not decay, as is the case for amplitude perturbations; therefore, the phase is not stable.
At the same time, they do not diverge, as with unstable dynamic variables. 
Phase perturbations remain constant, and therefore the phase is said to be \emph{neutrally stable}~\cite{pikovsky2001synchronization}. 
It is this characteristic of self-sustained oscillators that is behind their ability to synchronize either to an external drive or to each other, as we will see below.

\subsection{Entrainment and mutual synchronization in classical systems}\label{sec:classical_entrain_mutual}

We consider here the synchronization of a self-sustained oscillator when coupled to an external drive or to other self-sustained oscillators. 
We discuss two important phenomena associated with classical synchronization, namely, entrainment and mutual synchronization.
The case of a single self-sustained oscillator adjusting its frequency due to an external drive is usually called \emph{frequency entrainment}.
On the other hand,  \emph{mutual synchronization} refers to the case when two or more self-sustained oscillators adjust their frequencies due to coupling between them.
We also review here mutual synchronization in the limit of a thermodynamically large number of such oscillators. Here, synchronization exhibits itself as a second-order continuous phase transition, as seen in the paradigmatic \emph{Kuramoto model}.

\subsubsection{Entrainment}

Here, we discuss how a quasi-harmonic oscillator becomes entrained to an external drive by looking at a phase-space picture that rotates at the drive frequency. 
In general, the frequency of the oscillator $\omega$ will be different from the frequency of the drive $\Omega$.
For weak driving strength, the motion in the phase space traced by the limit cycle trajectory will do so at a frequency well approximated by the detuning $\Delta=\omega-\Omega$. 
Increasing the strength of the drive will have a nontrivial effect on the dynamics of the point in phase space due to the neutral stability of the oscillator's phase mentioned above.
During some portions of its cycle, the point in phase space will be slowed down as it tries to move against the action of the drive.
In other parts of the trajectory, the point in phase space will be accelerated by the drive.
This effect will become more pronounced as the strength of the drive increases.
Upon reaching a critical driving strength, the point in phase space will cease to move in the rotating frame. 
Here, the phase difference between the oscillator ($\phi(t)$) and the drive ($\phi_{\text{drive}}(t)$) reaches a constant value,
\begin{equation}
    \phi(t)-\phi_{\text{drive}}(t) = \text{constant},
\end{equation}
meaning the oscillator oscillates at the same frequency as the drive. In other words, the oscillator is synchronized to the drive.
This effect is referred to as \emph{frequency locking} or \emph{phase locking}, and we say that the oscillator is \emph{entrained} to the external drive~\cite{pikovsky2001synchronization}. Entrainment has also been studied extensively in quantum mechanical systems, and will be reviewed in Sec.~\ref{sec:Few_Body_QSynch}.

We now go beyond the qualitative discussion and consider a specific example of entrainment. To this end, we add a periodic force of the form $F\cos(\Omega t)$ to the vdP oscillator in Eq.~(\ref{eq:undriven_vdP_classical}), where $F$ denotes the strength of the drive, and $\Omega$ its frequency. 
Solutions that are synchronized to the driving frequency will have the form $x(t)=A(t)\cos(\Omega t + \phi(t))$, assuming weak non-linearity $\mu$. Provided that the amplitude $A(t)$ and phase $\phi(t)$ vary slowly compared to the driving term $\cos(\Omega t)$, averaging their dynamics over one full cycle removes the effect of fast-oscillating terms yielding the so-called \emph{truncated equations}~\cite{balanov2009simple},
\begin{align}
    \dot{A}(t) & = \frac{\mu}{2}A(t) - \frac{1}{8}A^3(t) - \frac{F}{2\Omega}\sin\phi(t), \label{eq:truncated_vdP_amplitude} \\
    \dot{\phi}(t) & = \Delta - \frac{F}{2A(t)\Omega}\cos\phi(t). \label{eq:truncated_vdP_phase}
\end{align}
In the absence of the drive ($F=0$), the amplitude of oscillations is $A^* = 2\sqrt{\mu}$.
For a weak drive, the perturbed amplitude can be assumed to be very close to the unperturbed amplitude $A^*$.
Substituting $A^*$ into Eq.~(\ref{eq:truncated_vdP_phase}) leads to the \emph{Adler's equation}~\cite{Adler1946study},
\begin{equation}
    \dot{\phi}(t) = \Delta-\frac{F}{4\sqrt{\mu}\Omega}\cos\phi(t).
    \label{eq:vdP_Adler}
\end{equation}
The vdP oscillator becomes entrained by the external drive when $\phi(t)$ attains a stable fixed value, i.e., when solutions to $\dot{\phi}(t)=0$ exist.
This immediately leads to the synchronization condition,
\begin{equation}
    4\sqrt{\mu}\Omega|\Delta| \leq F.\label{eq:entrainment_arnold_tongue}
\end{equation}
For parameters satisfying this inequality, phase and frequency locking are possible. Otherwise, the phase drifts freely without locking. Correspondingly, the stable solution of the system bifurcates from a fixed point (entrained regime) to a limit cycle (lack of entrainment). The above condition reflects the general behavior that larger differences in frequency can be compensated for by stronger forcings. 
This mechanism underlies the well-known \textit{Arnold tongue} structure of frequency locking in nonlinear oscillators \cite{pikovsky2001synchronization, balanov2009simple}. The \textit{tongue} refers to the fact that as drive strength is increased, the system entrains to an external drive (or synchronizes with another system as discussed below) for larger values of detuning. Consequently, a generic diagram representing detuning on the x-axis and drive strength on the y-axis would produce a tongue-like structure, as depicted in Fig.~\ref{fig:Introduction-self_sustained}(e).

\subsubsection{Mutual synchronization}

As a concrete example for mutual synchronization ~\cite{chakraborty1988transition,hoppensteadt1982phase,kawato1980two}, we consider two coupled vdP oscillators,
\begin{eqnarray} \label{eq:coupled_vdP_classical}
    \ddot{x}_1 - (\mu_1-x_1^2)\dot{x}_1+\omega_1^2x_1 + g (\dot{x}_1-\dot{x}_2) & = 0, \nonumber \\
    \ddot{x}_2 - (\mu_2-x_2^2)\dot{x}_2+\omega_2^2x_2 + g (\dot{x}_2-\dot{x}_1) & = 0,
\end{eqnarray}
where the oscillators are coupled via the difference of their velocities at coupling strength $g$.
This particular type of coupling is known as \emph{dissipative coupling}~\cite{pikovsky2001synchronization}.
Other types of coupling are possible as well~\cite{aronson1987analytical}, in particular \emph{reactive coupling}, where the oscillators interact via the difference in their positions. 
Unlike in the case of entrainment,  the coupling here is bidirectional, meaning the oscillators influence each other and these have been of great interest in quantum systems (see Sec.~\ref{sec:Few_Body_QSynch}-\ref{sec:Many_Body_QSynch}).
Synchronized solutions to the above set of equations take the form $x_{1,2}(t)=A_{1,2}(t)\cos(\omega t + \phi_{1,2}(t))$, where $A_{1,2}(t)$ are the amplitudes of oscillations and $\phi_{1,2}(t)$ their phases.
Considering the simplest case, where the oscillators differ only in their natural frequencies $\omega_1\neq\omega_2$ while $\mu_1=\mu_2=\mu$ and $A_1=A_2=A$ leads to the following equations~\cite{balanov2009simple},
\begin{align}
    \dot{A}(t) & = \frac{A}{2} \left( \mu - g -\frac{1}{4}A^2 \right) + \frac{g}{2}A\cos\theta(t), \\
    \dot{\theta}(t) & = \Delta - g\sin\theta(t), \label{eq:couple_vdP_classical_phase}
\end{align}
where $\theta=\phi_2-\phi_1$ is the phase difference between the two oscillators, and $\Delta=\omega_2-\omega_1$ is the detuning.
The two oscillators become synchronized when the phase difference attains a constant value.
From Eq.~(\ref{eq:couple_vdP_classical_phase}), we see that this fixed point exists when $|\Delta|\leq g$.
\begin{figure}
    \centering
    \includegraphics[width=\columnwidth]{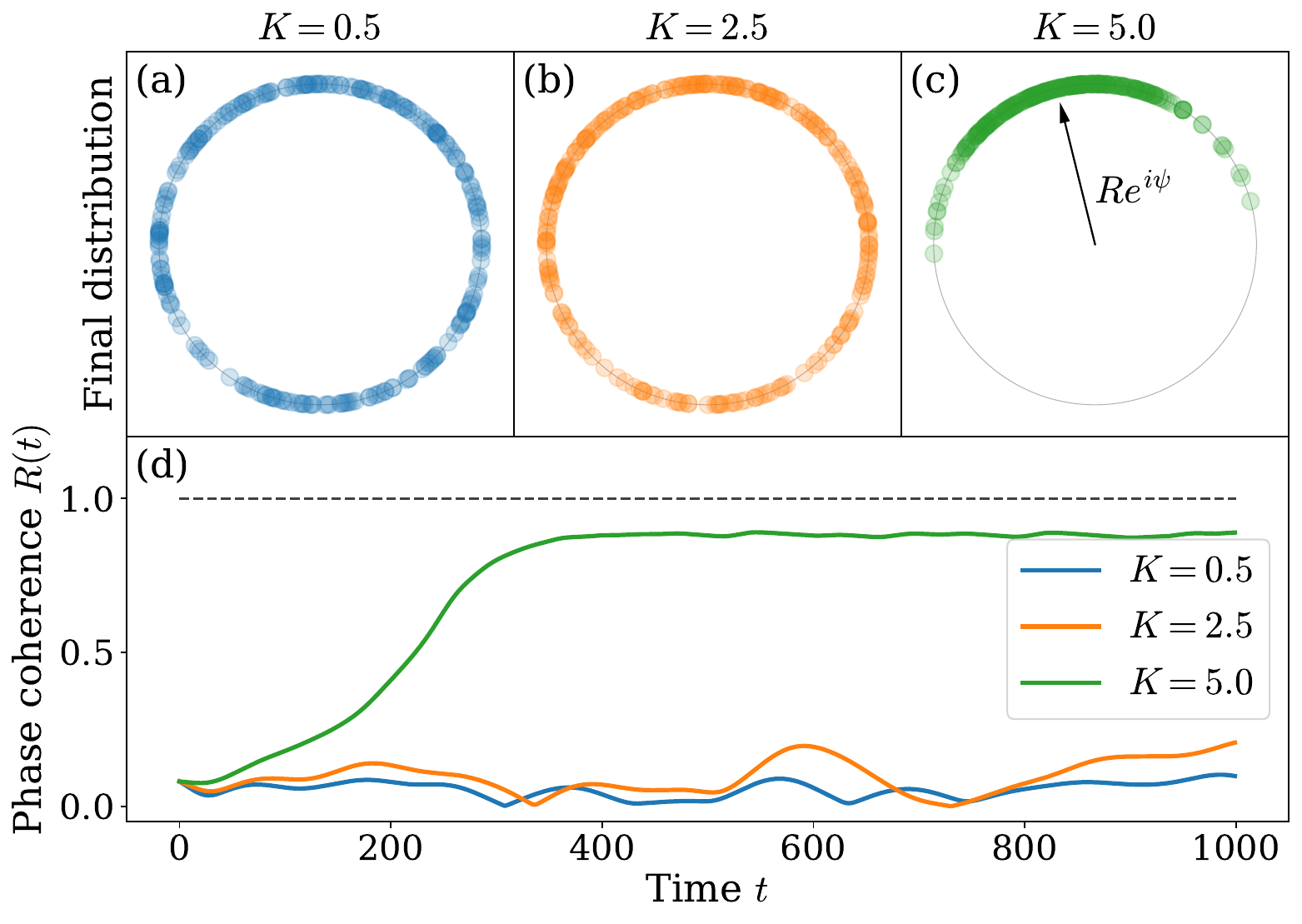}
    \caption{Time evolution of the Kuramoto model for $N=250$ phase oscillators. Initial phases are picked uniformly at random, $\phi_i(0)\in[0,2\pi)$. Natural frequencies $\omega_i$ are also chosen uniformly at random, $\omega_i\in[0,2]$. The top three panels show the phase distributions of the oscillators in the long time limit. For coupling strength below the critical value, namely $K=0.5$ in (a) and $K=2.5$ in (b), the oscillators remain unsynchronized. For $K=5.0$, which is above the critical coupling strength, the oscillators develop strong mutual synchronization as shown in (c). The black arrow shows the complex order parameter $Re^{i\psi}$ from Eq.~(\ref{eq:Kuramoto_order_parameter}). (d) Time evolution of the magnitude of the order parameter, also known as the phase coherence.}
    \label{fig:Introduction-Kuramoto}
\end{figure}
Just like Eq.~(\ref{eq:entrainment_arnold_tongue}), this condition describes the region where the system exhibits synchronization, forming an Arnold tongue similar to Fig.~\ref{fig:Introduction-self_sustained}(e).
The mutual synchronization also follows an Arnold tongue structure as a function of detuning between the oscillator frequencies and the coupling strength.

\subsubsection{Kuramoto model}

Mutual synchronization is highly relevant in many-particle systems, where it can lead to emergent collective behavior with qualitatively different signatures as compared to few-body scenarios. A paradigmatic example for mutual synchronization is the \emph{Kuramoto model}~\cite{kuramoto1975self,kuramoto2003chemical,gupta2014kuramoto,acebron2005}.  
Consider a large number $N$ of limit-cycle oscillators, each with its own natural frequency $\omega_i$,
\begin{equation}
    \dot{\phi}_i(t) = \omega_i + \frac{K}{N} \sum_{j=1}^N \sin(\phi_j(t)-\phi_i(t)), \label{eq:_classical_kuramoto}
\end{equation}
where $\phi_i$ denotes the phase of each oscillator.
This model assumes a uniform all-to-all coupling at strength $K$, and a distribution of natural frequencies given by a probability density. 
The dynamics of the phases in Eq.~(\ref{eq:_classical_kuramoto}) can be described by a set of points on a unit circle in the complex plane.
If phase $\phi_i$ is lagging behind phase $\phi_j$, it feels a pull from phase $\phi_j$ and its evolution is sped up.
On the other hand, when $\phi_i$ is ahead of $\phi_j$, its evolution is slowed down. 
Summing this effect over all phase oscillators leads to two types of behavior in the thermodynamic limit when $N\rightarrow\infty$.
For weak coupling, the phases evolve incoherently and virtually independently of one another.
Beyond a critical coupling $K>K_c$, the system undergoes a phase transition and becomes coherent.
This transition can be witnessed by the Kuramoto order parameter 
\begin{equation}
    R(t)e^{i\psi(t)} = \frac{1}{N}\sum_{j=1}^N e^{i\phi_j(t)},
    \label{eq:Kuramoto_order_parameter}
\end{equation}
where $R(t)$ measures the phase coherence and $\psi(t)$ represents the average phase.

Figure~\ref{fig:Introduction-Kuramoto} depicts the dynamics of $N=250$ phase oscillators evolving according to Eq.~(\ref{eq:_classical_kuramoto}).
The initial phases of all oscillators are uniformly distributed, as are their natural frequencies $\omega_i$.
In Figs.~\ref{fig:Introduction-Kuramoto}(a,b), where the coupling strength is below the critical coupling $K_c$, we observe that the oscillators remain unsynchronized and the distribution of phases in the long time limit is nearly uniform on the unit circle.
Figure~\ref{fig:Introduction-Kuramoto}(c) depicts the case of a coupling strength that is larger than the critical coupling, leading to mutual synchronization.
The time evolution of the phase coherence $R(t)$ is shown in Fig.~\ref{fig:Introduction-Kuramoto}(d), further highlighting the qualitatively different behavior for sub- and supercritical coupling strengths.

The assumption of uniform all-to-all coupling can be lifted, as demonstrated in~\cite{crawford1999synchronization}.
For a historical perspective on the Kuramoto model, we encourage the reader to follow the exposition in~\cite{strogatz2000kuramoto}.
Despite Kuramoto model's simplicity, it has proven to be well suited to studying mutual synchronization of large numbers of non-linear oscillators, including in quantum many-body systems (see Sec.~\ref{sec:Many_Body_QSynch} for details). 
It has been observed in a plethora of varied physical and biological systems, including yeast cells~\cite{bier2000yeast}, cardiac pacemaker cells~\cite{winfree1980geometry}, current-biased arrays of Josephson junctions~\cite{wiesenfeld1998frequency}, decentralized power grids~\cite{rohden2012self}, rhythmic applause~\cite{neda2000physics}, and animal flocking~\cite{ha2010emergent}.

\subsection{Varieties of classical synchronization and their measures} \label{sec:classical_meas} 

Coupled oscillator systems, such as the Kuromoto model, have long served as a test bed for studies on complex collective behaviors, including synchronization. 
However, these systems exhibit diverse, sometimes counterintuitive, exotic forms of synchronization that go beyond the paradigm of complete synchronization in frequency and phase. 
Natural measures of these behaviors rely on phase-space analyses, information theory, and other approaches that quantify correlations in the dynamics of different oscillators \cite{kreuz2013synchronization, VossReconstruction1997}. In the following, we discuss several types of classical synchronization phenomena and the measures that quantify them. 

\textit{Phase synchronization} refers to the situation discussed in Sec.~\ref{sec:classical_entrain_mutual} where the phases of the oscillators lock, even if the amplitudes undergo independent evolution. 
This kind of synchronization is inferred directly from the angular variable of the dynamics or from phase estimates obtained using methods such as the Hilbert transform \cite{CHOI1975285,aydore2013note}. 
For a pair of oscillators with phases $\theta_{1}(t)$ and $\theta_{2}(t)$, the phase synchronization can be quantified using the phase locking value, defined as $|\overline{ e^{-i (\theta_{1}(t) - \theta_{2}(t))}  }|$, where the overline represents the average taken over time. 
This quantity equals one when the phases of the two oscillators maintain a fixed phase difference at all times, and decreases as the phases drift from one another, reaching zero when no phase locking is present \cite{MORMANN2000358}. 
In scenarios with multiple oscillators, such as in the Kuramoto model, the global phase coherence can be quantified using the Kuramoto order parameter given by  Eq.~\eqref{eq:Kuramoto_order_parameter}. 
Phase coherence, the radial part of the Kuramoto order parameter, takes the value one when all phases are identical, implying perfect phase synchronization, and the value zero when the phases are uniformly distributed around the unit circle, reflecting the absence of any coherent alignment. 
Intermediate values between zero and one could indicate partial synchrony as shown in Fig.~\ref{fig:Introduction-Kuramoto}(d).

As discussed earlier, \textit{mutual frequency synchronization} and \textit{entrainment} to an external drive occur when the effective frequency of the system converges to that of another system or the drive, respectively \cite{pikovsky2001synchronization}.
The effective frequencies can be identified from the Fourier transforms of the variables describing the system's evolution.
Synchronization can be quantified in these scenarios using the frequency difference between the coupled oscillators for mutual synchronization and the deviation of the system frequency from the drive frequency for entrainment. 
These varieties of synchronizations exhibit Arnold tongue-like behavior in the parameter space of coupling strength and the frequency detuning. 
A generalization of frequency locking leads to \textit{higher-order synchronization}, where the frequencies of two oscillators satisfy a rational ratio $m:n$, i.e., $\omega_i/\omega_j = m/n$ \cite{pikovsky2001synchronization}. In this case, synchronization corresponds to a phase relation of the form $n\phi_i - m\phi_j = \text{const}$, rather than simple one-to-one locking \cite{coombes1999mode,schilder2007computing,valkering2000dynamics,ypey1980mutual,montoya2013construction}. 
As system parameters such as driving or coupling strength are varied, these distinct locking regimes organize into a characteristic fractal structure known as the \textit{devil’s staircase}, where plateaus corresponding to stable $m:n$ synchronized states emerge \cite{aubry1983exact,bak1986devil,bak1982commensurate}.
Such behavior naturally arises in nonlinear systems and reflects the existence of more complex, commensurate synchronized states beyond the standard $1:1$ regime.

Synchronization may constrain the dynamics of coupled chaotic systems to a lower-dimensional submanifold of the entire phase space, often called the synchronization manifold. 
Such a manifold is said to be stable if it is attractive in directions transversal to it, such that any perturbation in this direction dies down exponentially. 
This has been of considerable interest given the need to control numerous complex systems \cite{PhysRevLett.74.5028, eroglu2017synchronisation}. 
A common method used to accomplish chaos synchronization in a system of coupled identical chaotic oscillators is the replacement method, where a component of a \textit{drive} oscillator is used to replace the same component in the other \textit{response} oscillators \cite{PhysRevLett.64.821}. This is known as \textit{identical} or \textit{complete synchronization} and results in a set of synchronous, chaotic systems \cite{PhysRevLett.64.821, parlitz1999handbook}. 
In systems exhibiting \textit{complete synchronization}, all state variables of the two oscillators show convergence in their dynamics. 
If $\{\vec{x}_{1}(t),\vec{x}_{2}(t)\}$ are the state variables, complete synchronization in these systems can be quantified by the synchronization error defined as $\epsilon(t) = \| \vec{x}_{1}(t)- \vec{x}_{2}(t) \|$, where $\| \cdot \|$ is an appropriate norm on the state space of the oscillators.
The convergence of $\epsilon(t)$ to zero indicates complete synchronization.  

These quantities can also be extended to diagnose the so-called \textit{lag synchronization}, in which the trajectories of two oscillators are similar but shifted in time by a constant lag. 
Lag synchronization can be detected by introducing a time shift to one of the trajectories and checking that the synchronization error converges to zero for some fixed time shift. 
In this context, \textit{generalized synchronization} extends these definitions of chaotic synchronization to coupled nonidentical chaotic systems \cite{rulkov1995generalized, PhysRevLett.76.1816, KITTEL1998459}. 
Unlike identical synchronization, in which the trajectories of all oscillators match exactly, generalized synchronization occurs when there exists a function relating the trajectory of the drive oscillator to the response oscillators. 
The properties of this function further make for \textit{strong} and \textit{weak} general synchronization corresponding to \textit{smooth} and \textit{nonsmooth} functions, respectively \cite{KITTEL1998459}.
Generalized synchronization implies that, under a functional relation between the systems, neighboring points in the phase space of one system must map to neighboring points in the other.
This idea led to the development of several methods, including mutual false-nearest-neighbor methods and mean conditional dispersion methods to capture generalized synchronization by quantifying the degree to which this neighborhood correspondence fails~\cite{rulkov1995generalized, boccaletti2002synchronization}.

\textit{Chimeras} are another interesting phenomenon, where identical coupled oscillators spontaneously exhibit coexisting regions of dynamically disparate behaviours \cite{shima2002coexistence, abrams2004chimera}. 
They constitute regions of synchronous as well as incoherent, asynchronous, or chaotic behaviours and result from the coexistence of competing dynamical attractors in the system. 
Such a multistable system can be realized in a multitude of ways, including using non-local couplings \cite{shima2002coexistence}, purely local but strong \cite{PhysRevLett.112.144101}, nonlinear \cite{PhysRevE.97.022201}, or globally couplings with delayed feedback \cite{PhysRevLett.112.144103}. 
Earlier works used local order parameters based on spatial correlation functions to characterize the presence of chimeras \cite{abrams2004chimera, PhysRevE.89.052914, PhysRevLett.106.234102}. Among these, the strength of incoherence has emerged as a standard measure for characterizing them \cite{PhysRevE.89.052914}.
To define this for a coupled network of oscillators, we partition them into $M$ bins, each containing an equal number of oscillators. 
A standard deviation is defined for the $m$th bin, $\sigma(m)$, such that 
the strength of incoherence is $\tilde{S} = 1 - \sum_{m=1}^{M} s_m/M$, where $
s_m = \Theta\!\left(\delta - \sigma(m)\right)$ \cite{PhysRevE.89.052914}. 
Here, $\Theta\!\left(\cdot\right)$ is the Heaviside step function, and $\delta$ is a threshold that is predefined and reasonably small. The value of $\tilde{S}$ quantifies the fraction of bins that are unsynchronized, such that $\tilde{S} = 0$ and $\tilde{S} = 1$ indicate a coherent and incoherent system, respectively, and $(0 < \tilde{S} < 1)$ corresponds to a chimera state. 
Later, statistical measures, such as Pearson correlation coefficients, were applied to the time series of networks of quantum oscillators to characterize the presence of chimeras \cite{solanki2024exotic}. 

\textit{Cluster synchronization}, on the other hand, occurs when a network of coupled oscillators spontaneously partitions into two or more distinct clusters \cite{KANEKO1990137, KANEKO1992368, PhysRevE.57.276, PhysRevE.63.036216}. Oscillators within each cluster remain synchronized with one another, but a cluster as a whole remains unsynchronized with any other cluster. This is generally a consequence of the structural symmetries in the underlying network topology of the oscillators \cite{pecora2014cluster}. 
To identify the clusters formed in a network, similarity relations between pairs of oscillators can be defined using various synchronization measures such as synchronization error, phase-locking values, or correlation-based metrics \cite{LuCluster2010, PhysRevE.97.042217, della2020symmetries}.
These similarity relations can then be used to group oscillators into equivalence classes corresponding to distinct synchronization clusters.

In the list of other curious synchronization phenomena is \textit{remote synchronization}, in which two uncoupled entities synchronize through an unsynchronized intermediate entity \cite{PhysRevE.85.026208}.
Such a phenomenon has been shown to exist in systems where the intermediate entities are coherent in the sense of generalized synchronization \cite{PhysRevLett.97.123902}, as well as in the case when they are completely incoherent \cite{PhysRevLett.118.174102}. 
Such behaviors can be detected by evaluating standard synchronization measures across all pairs of oscillators and identifying pairs that show synchrony despite not being directly coupled \cite{GambuzzaAnalysis2013}.

Furthermore, with an increase in coupling strength, a collection of oscillators can undergo a sudden, discontinuous, first-order phase transition from an asynchronous to a synchronized state \cite{PhysRevE.72.046211}. This phenomenon is called an \textit{explosive synchronization} \cite{PhysRevLett.106.128701, PhysRevLett.108.168702, PhysRevE.86.056108}.

\section{Introduction to Quantum Synchronization}\label{sec.intro.qsync}

In classical mechanical systems discussed so far, the state is described as points in the phase space.
In contrast, states of quantum mechanical systems are fundamentally described as state vectors in a Hilbert space or, more generally, as density matrices.  
Over the last decade, the concept of synchronization has been extended from classical to quantum systems, prompting investigation into whether and how genuinely quantum degrees of freedom can exhibit synchronized behavior — either under external drives or through mutual interactions. The earliest efforts in this direction extended the classical understanding of synchronization to the quantum regime by mapping the density matrix for continuous-variable quantum systems to a quasi-probability distribution on phase space. 
While certain hallmarks of classical synchronization, such as frequency and phase locking, persist in the quantum regime, their manifestation is fundamentally altered by quantum fluctuations. 

In the forthcoming sections, we discuss synchronization in quantum systems. 
In classical systems, non-linearities and sources of both pumping and damping were required to set up a stable limit cycle that is then synchronized. Though sources of driving can be introduced in quantum systems systematically by adding terms to the Hamiltonian, the question of dissipation is more subtle. To preserve canonical commutation relations \cite{carmichael2013statistical}, quantum systems interacting with an environment (we interpret this as sources of pumping and damping) are modeled as \textit{open quantum systems}. 
Such systems are consistent with a state vector description of the system and environment combined (sometimes called the \textit{universe}), but produce a computationally minimal description of the system dynamics alone. 
Since synchronization typically emerges in out-of-equilibrium systems subject to drive and dissipation, open quantum systems are a natural framework to investigate quantum synchronization, as they inherently incorporate both coherent evolution and dissipative processes. 
In this section, we briefly review the fundamentals of open quantum systems weakly interacting with memoryless baths.

\subsection{Review of open quantum systems} 
We review and summarize three techniques in preparation to study synchronization of quantum systems. The first of these, addressed in Sec.~\ref{sec:master_equation}, is a description of the average dynamics of the subsystem, which in the Markovian regime gives rise to the so-called GKSL (Gorini, Kossakowski, Sudarshan and Lindblad) master equations, also known as Lindblad master equations \cite{gorini1976completely,lindblad1976generators}.
The second is a trajectory approach to open system dynamics, which models the continuous monitoring of the quantum system of interest by environmental degrees of freedom via stochastic differential equations that update the quantum state based on a measurement record, briefly reviewed in Sec.~\ref{sec.IntroQS.trajectory}. 
The master equation approach can, in fact, be thought of as an average over several of these trajectories. 
Finally, we also introduce methods that map quantum dynamics to a few classical variables in Sec.~\ref{Subsec:meanfield}. 
These include mean-field approaches, which describe the evolution of observable expectation values, and phase-reduction methods that map quantum states onto a limit cycle via a phase parameter. 
We review each of these techniques below briefly.

\subsubsection{Quantum master equation approach to dynamics}\label{sec:master_equation} 

The theory of open quantum systems addresses the dynamics of systems subject to the influence of environmental degrees of freedom, which induce irreversible dynamics in the system. When the system is weakly coupled to a macroscopic environment and there is a separation of timescales between the fast relaxation process of the environment and the dynamics of the system, one can apply the Born \cite{wiseman_nm} and Markov approximations to derive a reduced description for the open system dynamics \cite{breuer2002theory,gardiner2004quantum,carmichael2013statistical}. This approach provides a microscopic derivation of the GKSL master equation, which is a time-local equation that describes the Markovian time evolution. 
The GKSL master equation finds widespread application across a variety of scenarios such as atomic, molecular, and quantum optical systems \cite{breuer2002theory,gardiner2004quantum,carmichael2013statistical}, optomechanics \cite{aspelmeyer2014cavity}, and superconducting circuits \cite{blais2021circuit}. 
It reads  ($\hbar=1$), 
\begin{equation}
\dot{\hat{\rho}} =-i[\hat{H}, \hat{\rho}] + \sum_{j}\mathcal{D}[\hat{L}_j]\hat{\rho},
\end{equation}
where the system state given by $\hat{\rho}$ is a density matrix, a construct that generalizes the notion of state vectors on the Hilbert space to include incoherent mixtures of these vectors.
The above equation contains two types of terms: the Hamiltonian evolution due to $\hat{H}$, which describes coherent processes, and the dissipators 
\begin{equation}
\mathcal{D}[\hat{L}_j]\hat{\rho}=\hat{L}_j\hat{\rho} \hat{L}^\dagger_j-\frac{1}{2}\{\hat{L}_j^\dagger \hat{L}_j,\hat{\rho} \},
\end{equation}
with the jump operators  $\hat{L}_{j}$ which describe various incoherent processes, such as spontaneous emission, dephasing, or incoherent pumping.

Since Lindblad equations are linear in the elements of the density matrix, a linear map such as $\vert i\rangle \langle j \vert\rightarrow \vert i\rangle  \otimes \vert j\rangle$ may be used to write an equation for a vectorized form of the density matrix $\hat{\rho}$. 
Such a procedure is related to the so-called Choi-Jamio{\l}kowski isomorphism, which relates CPTP maps to quantum states in a higher dimension \cite{watrous2018theory}.
In general, for a linear operator $\hat{X} \in \mathcal{M}(\mathcal{H}_0,\mathcal{H}_1)$, where $\mathcal{M}(\mathcal{H}_0,\mathcal{H}_1)$ is the vector space of all linear operators that takes vectors from $\mathcal{H}_0$ to $\mathcal{H}_1$, the procedure $\vert X\rangle\in \mathcal{H}_1\otimes \mathcal{H}_0$ is termed \textit{row stacking} (our example above) whereas $\vert X\rangle\in \mathcal{H}_0\otimes \mathcal{H}_1$ is called the \textit{column-stacking} scheme \cite{PhysRevA.80.022339}. 
Using either procedure 
(computer programs often use column stacking, whereas row stacking is preferred by authors who wish to avoid an extra complex conjugation), we can recast the GKSL equation to the vectorized form $\vert \dot{\rho}\rangle=\mathcal{L}\vert \rho\rangle$. Here, $\mathcal{L}$ is called the Liouville \textit{superoperator}, indicating that it is an operator that acts on other operators (the density matrix in this case). A formal solution can be immediately written in the form $\vert\rho(t)\rangle=\exp(\mathcal{L}t)\vert\rho(0)\rangle$. 
For time-dependent systems, the propagator takes the form $\mathcal{P}(t)=\mathcal{T}\exp\big(\int_0^t~ds\mathcal{L}(s)\big)$. 
However, we note that the effective Liouville superoperator corresponding to time-periodic driving need not be in the GKSL form \cite{PhysRevB.101.100301}. 
In this review, we mostly deal with cases in which the time dependence of the Hamiltonian, e.g., due to periodic driving, can be removed by transforming to a co-rotating frame, resulting in a time-independent master equation. 
Since the transients and steady states of open system evolution in this case can be determined by the eigenspectrum of $\mathcal{L}$, we will now briefly review its spectral properties.

\begin{figure}
    \centering
    \includegraphics[width=0.8\columnwidth]{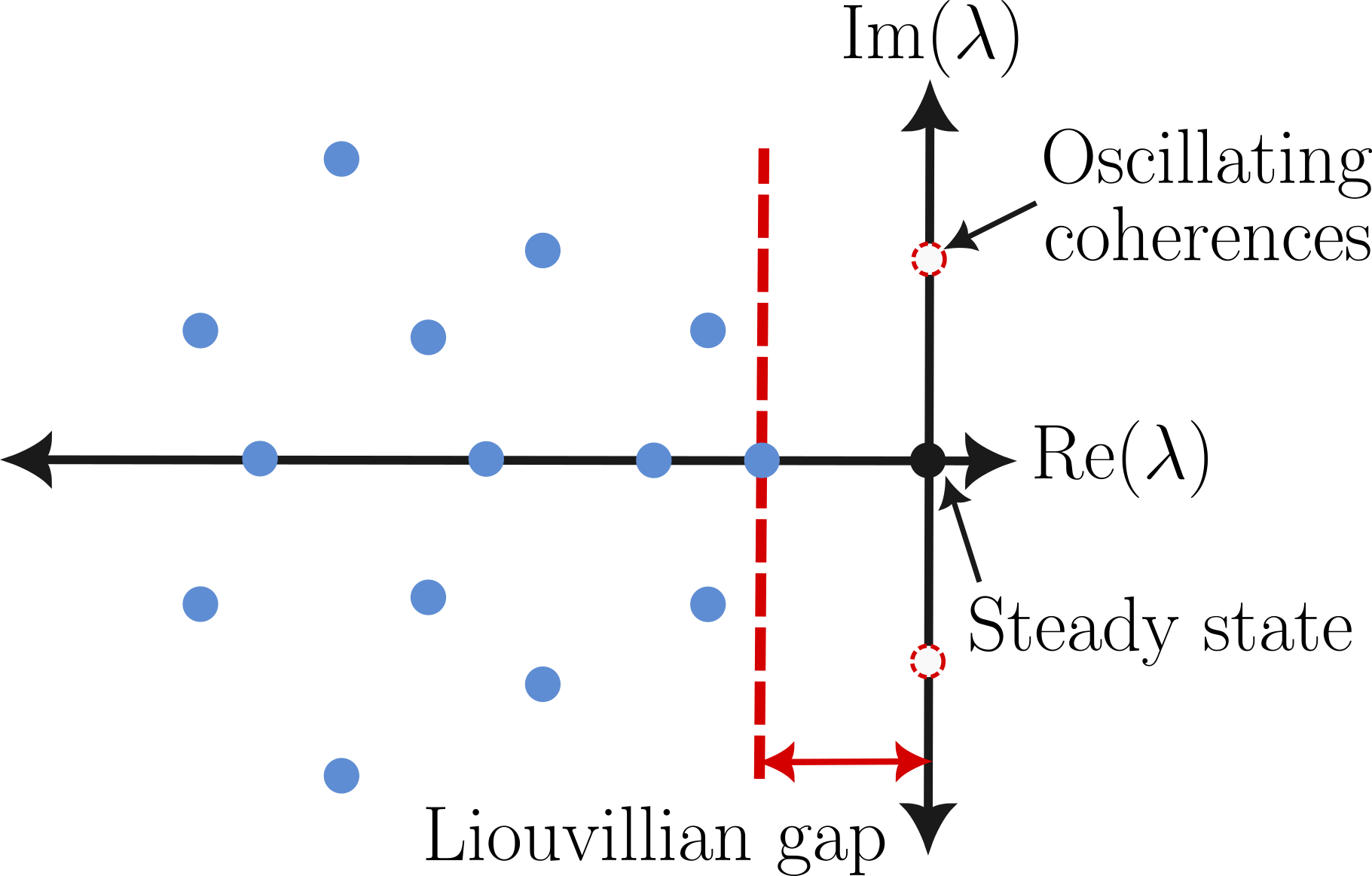}
    \caption{Illustration of the eigenvalue spectrum of a generic Liouvillian superoperator. }
    \label{fig:liouvillian_eigenspectra}
\end{figure}

Given that $\mathcal{L}$ is not Hermitian, it generally has complex eigenvalues, $\lambda_k = \alpha_k + i \beta_k$ as shown in Fig.~\ref{fig:liouvillian_eigenspectra}, with the corresponding right and left eigenvectors defined through the eigenvalue equations $\mathcal{L}\vert r_k\rangle = \lambda_k\vert r_k\rangle$ and  $\mathcal{L}^{\dagger}\vert l_k\rangle = \lambda^{*}_k\vert l_k\rangle$, respectively, where $\mathcal{L}^{\dagger}$ is the Hermitian conjugate of $\mathcal{L}$. 
Here, $k$ indexes the eigenspectrum and $\alpha_k,\beta_k\in\mathbb{R}$ with $\alpha_k\leq 0$ for consistency with Born's interpretation. 
Non-zero values of $\alpha_k$ govern the transient dynamics while $\beta_k$ dictates the corresponding oscillatory dynamics. 
Steady states are given by the part of the spectrum corresponding to $\alpha_k=0$ (solid black dot in Fig.~\ref{fig:liouvillian_eigenspectra}).
When the associated $\beta_k$ is nonzero, this can lead to sustained oscillations in the steady state, also referred to as oscillating coherences (see open red dots in Fig.~\ref{fig:liouvillian_eigenspectra}). The smallest non-zero value of $|\alpha_k|$ is often called the spectral gap of the Liouville superoperator (indicated by a red arrow in Fig.~\ref{fig:liouvillian_eigenspectra}) and dictates the longest transient timescale in the problem. The existence of steady states (at least one $\alpha_k=0$) is guaranteed by Evans' theorem \cite{evans1977irreducible} and the corresponding eigenvector represents the traceful part of the steady state. 
When the GKSL equation has a unique steady state, it is reached in the long-time limit independently of the initial conditions. 
However, the steady state properties are modified by the presence of symmetries in open quantum systems and are broadly classified as strong or weak \cite{Albert_2014, buvca2012note}. 
\textit{Strong} symmetry is realized when a unitary symmetry operator $Q$ commutes with the Hamiltonian and all the Lindblad operators. 
The existence of strong symmetry is a sufficient criterion for the existence of multiple steady states. 
In contrast, a \textit{weak} symmetry arises when the corresponding symmetry superoperator $\mathcal{Q}$, whose action on any operator $A$ is defined as $\mathcal{Q}(A) =  QAQ^{\dagger}$,  commutes with the total Liouvillian $\mathcal{L}$. We note that in this case, $Q$ need not commute individually with the Hamiltonian and the Lindblad operators.

The GKSL master equation discussed above describes the evolution of an open quantum system at the level of its ensemble-averaged state \cite{breuer2002theory}. 
In other words, it describes the evolution of the system when averaged over several measurement records, where each measurement record corresponds to a single trajectory. 
In the next section, we will review the dynamics of an open system along a single trajectory.

\subsubsection{Trajectory approach to open system dynamics}
\label{sec.IntroQS.trajectory}

In Sec.~\ref{ch:classical_sync}, we discussed the synchronization of classical systems along phase space trajectories. However, contrary to the case of classical systems,  a time-evolving quantum system cannot be directly observed without disturbing it. For instance, reconstructing the system density matrix requires quantum state tomography, which entails repeating the same experiment many times and performing measurements in a complete set of bases at different stages of the dynamics. If one is instead interested in the time evolution of a specific observable, measurements in a single basis suffice, but the experiment must still be repeated multiple times to collect statistics at each time. 
 
Continuous monitoring of a quantum system, instead, either observes mechanical motion directly, follows atomic coherences through transmitted light, or measures the electromagnetic field emitted by the system into its environment using a detector \cite{murch2013observing,PhysRevA.47.642,breuer2002theory,gardiner2004quantum}. 
The system is only weakly perturbed, and information can be continuously acquired without reinitializing or repeating the experiment. 
Depending on the detection scheme, this information may consist of a time series of photon emission events (in photodetection) or of a continuous signal associated with homodyne or heterodyne currents. 
These measurement records can be used to infer dynamical properties \cite{garrahanThermodynamicsOfQuantum2010,landi2024current} and to construct witnesses and measures of quantum synchronization (see Sec. \ref{Sect:meas}). 
The information obtained in this way is inherently stochastic and conditioned on the measurement outcomes. 
This naturally leads to a description of the system dynamics in terms of quantum trajectories, which represent single realizations of the evolution in the presence of monitoring.

We consider different classes of quantum trajectories arising from distinct measurement schemes applied to the emitted electromagnetic field  \cite{plenio1998thequantum,wiseman2009quantum}. 
First, we focus on continuous measurements of the homodyne current, in which the signal is mixed with a reference optical field using a beamsplitter before performing a quadrature measurement \cite{PhysRevA.47.642}. 
In this case, the system dynamics is described by the diffusive stochastic Schrödinger equation 
\begin{equation}
\begin{split}
d\ket{\psi}=&{dt} \left[-i\hat{H}_{\rm eff}+\sum_k \frac{\langle \hat{X}_k\rangle}{2}\left(\hat{L}_k-\frac{\langle \hat{X}_k\rangle}{4}\right)\right]\ket{\psi}\\
&+\sum_k dW_k \left(\hat{L}_k-\frac{\langle \hat{X}_k\rangle}{2}\right)\ket{\psi}\, .
\end{split}
\label{eq:homo-traj}
\end{equation}
Here, $\ket{\psi}$ is the pure state of the system in a single experimental realization, $d\ket{\psi}$ is its infinitesimal increment over the time interval $t\to t+dt$ and $\hat{H}_{\rm eff}=\hat{H}-i\sum_k \hat{L}^\dagger_k \hat{L}_k /2$. Furthermore, $\hat{X}_k=\hat{L}_k+\hat{L}_k^\dagger$ and $dW_k$ are independent real Wiener stochastic increments satisfying  $\mathbb{E} [dW_k]=0$ and $ dW_k^2 =dt$, where $\mathbb{E}[\cdot]$ denotes the average over the stochastic increments. 
This equation describes the evolution of the system conditioned on the homodyne current 
\begin{equation}
    J_k=\langle \hat{X}_k\rangle +\frac{dW_k}{dt}\, ,
    \label{eq:homod_current}
\end{equation}
which provides a noisy, time-resolved record of the emitted field. 
Similar types of diffusive trajectories arise in heterodyne detection, where complex Wiener increments are associated with complex measured currents  \cite{wiseman2009quantum}. 
It is important to note that trajectories and their properties are determined by the observables one chooses to measure.

A different type of stochastic dynamics, the so-called quantum-jump trajectories, emerges when monitoring individual photon emissions \cite{plenio1998thequantum}. In this case, the detection of a photon associated with the jump operator $\hat{L}_k$ at time $t$ induces a discontinuous state update 
$
\ket{\psi'}\propto \hat{L}_k\ket{\psi}, 
$
occurring with probability $p_k=dt \braket{\psi|\hat{L}_k^\dagger \hat{L}_k|\psi}$. In the absence of detection, with  probability $1-\sum_k p_k$, the state evolves according to  
$
\ket{\psi'}\propto e^{-i\hat{H}_{\rm eff }dt}\ket{\psi}\, .
$
The corresponding measurement record is given by the sequence of detection times and channels. 
Finally, notice that when only a subset of jump operators is monitored, or when detection is inefficient, the evolution is no longer described by pure-state trajectories. Instead, these cases require stochastic master equations for the conditional density matrix \cite{wiseman2009quantum}.

\subsubsection{Mean-field approach and phase reduction}
\label{Subsec:meanfield}
Solving open quantum dynamics is challenging even for systems with only a few degrees of freedom or for single nonlinear oscillators. For this reason, effective approaches have been developed and widely used in the study of quantum synchronization. A prominent example is the mean-field approach, which is especially accurate for oscillators in regimes of large excitations. Here, the bosonic operator $a$ is replaced by a complex variable $a\approx \alpha$, reducing the infinite hierarchy of Heisenberg equations to a nonlinear differential equation for $\alpha$. When applied to quantum trajectories, this equation becomes a stochastic Langevin equation. Such a mean-field description can become exact in certain many-body spin models in the thermodynamic limit (see Sec. \ref{sec.CTCs.spinmodels}) \cite{alickiNonlinear1983,PhysRevLett.126.230601,benattiQuantumSpinChain2018}. 

Phase reduction approaches provide a complementary framework, reducing limit cycles to the dynamics of a single variable. 
Consider a generic system that exhibits a stable limit-cycle solution with period $T$ and frequency $\omega=2\pi/T$. One can introduce a phase variable $\theta$ associated with states in the limit-cycle orbit and then extend its definition throughout their basin of attraction. The phase is defined by a phase function that assigns a phase value to each state of the system. Firstly, it is defined for states within the limit cycle, and it can always be chosen such that it evolves uniformly along the cycle, i.e., $\dot{\theta}=\omega$. This can be done, for instance, by simply identifying the phase as the time within the limit cycle multiplied by the frequency \cite{Nakao2016}. 
The definition can then be extended to all states by assigning each state the phase of the point on the limit cycle to which it asymptotically converges. 
In classical dynamical systems, this extension is achieved through the notion of isochrons, namely, sets of states that converge stroboscopically to the same point in the limit cycle \cite{Nakao2016}. 
The ability to define the phase outside of the limit-cycle orbit allows one to investigate the impact of noise and perturbations on synchronization behavior, by projecting their effects onto the dynamics of the phase, i.e.,
\begin{equation}
\dot{\theta}=\omega+\mbox{noise + perturbations}\, .
\end{equation}
Extending this idea to quantum systems is nontrivial, as there is no direct concept of phase-space trajectories. A first approach circumvents this difficulty using a semiclassical description  \cite{kato2019semiclassical}, where the quantum master equation is mapped to a Fokker-Planck equation. The system then behaves as a noisy classical oscillator, allowing standard phase-reduction techniques to be applied with quantum effects entering as noise. 
More recent work develops a fully quantum phase reduction directly at the level of the wave function \cite{setoyamaLie2024,setoyamaLie2025}. This approach considers quantum trajectories that exhibit limit-cycle behavior in the absence of noise. 
Similar to what was discussed earlier for classical systems, a phase is associated with each quantum state in the limit cycle in such a way that the phase changes at constant frequency.
For states outside the limit cycle, the phase is defined using the concept of isochrons, mentioned earlier.
This framework allows one to effectively study the effects of noise and perturbations on synchronization in the deep quantum regime, such as in single qubits or finite-level systems, where semiclassical approximations break down.

\subsection{Measures of quantum synchronization}
\label{Sect:meas}
Having established the framework of open quantum systems and their dynamics, we are naturally led to investigate the emergence of synchronization in this setting. To do so in a meaningful and systematic way, it is essential to first define appropriate measures that can quantify synchronization in the quantum regime. We therefore begin by introducing the key metrics used to characterize quantum synchronization before turning to specific examples. Defining a measure can help clarify and refine concepts, and in the last two decades, the development of various quantitative measures has been crucial for understanding the nature and significance of quantum synchronization. Compared to classical synchronization measures, moving into the quantum regime introduces the challenge of accounting for both temporal and quantum features. Seminal works on quantum entrainment \cite{Goychuk2006quantum,lee2013quantum,walter2015quantum} and mutual synchronization \cite{zhirov2008synchronization,orth2010dynamics,giorgi2012quantum,Mari2013Measures,Giorgi2013Spontaneous} displayed a variety of approaches, ranging from spectral features of quantum trajectories and phase coherence, to temporal correlations of expectation values of observables and 
quantum fluctuation reduction. 
The state of the art was reviewed almost ten years ago \cite{Galve2017Quantum}, identifying some desirable features for insightful synchronization measures such as the ability to identify truly temporal phenomena or enabling quantitative comparison in different regimes/systems. In this section, we present a concise summary of the various measures of quantum synchronization, providing a convenient reference for the reader.

\subsubsection{Temporal correlations}\label{subsec:temoral correlation}
A natural approach to assess synchronization between interacting systems is comparing the temporal correlation of local quantum observables $f_i(t)=\langle\hat O_i(t)\rangle$. Pioneering works considered the dynamics of positions and momenta of two coupled oscillators (including higher order moments or variances) \cite{Giorgi2013Spontaneous} or Pauli operators for spin systems \cite{giorgi2012quantum}. The mutual synchronization between the corresponding (local) dynamics was quantified with the Pearson factor, a normalized temporal correlation 
\begin{equation}
C_{f_i,f_j}(t,\tau)=\frac{\overline{ \delta f_i \delta f_j}}{\sqrt{\overline{  \delta f_i^2}
~\overline{\delta f_j^2} }},
\label{eq:meas_temp_correl}
\end{equation}
  where the bar stands for a time average $\overline{f}=\int_{t}^{t+\tau}dt'f(t')$ over a time window $\tau$  spanning a significant dynamics, for instance several oscillations, and $\delta f=f-\overline{ f}$.  Maximum (anti-)synchronization is established if 
$C\sim 1 (-1)$, while the value tends to vanish when the units do not display any temporal coherence.  
This is a bounded measure that captures temporal (anti-)correlation persistence in observables that can be defined for any system in finite \cite{giorgi2012quantum,cabotquantum2019,bucaDissipationInducedNonstationarity2019} and infinite \cite{Giorgi2013Spontaneous,manzano2013synchronization} dimensions.
It has been used as a standard measure of spontaneous synchronization both in theoretical \cite{manzano2013synchronization,bucaalgebraic2022,schmolkenoise2022} and experimental \cite{tao2025noise} works, useful for any form of temporal synchronization  (limit cycles, chaotic or relaxing dynamics). When considering an external driver signal, the entrainment of any observable can be characterized through the Pearson correlation with the driving signal. 
This indicator characterizes synchronization as a temporal coherence that can be compared with the occurrence of quantum correlations  \cite{Galve2017Quantum}. In \cite{giorgi2012quantum}, it was shown that transient synchronization occurs without entanglement but in the presence of mutual information and quantum discord; in dissipative and dephasing environments \cite{Giorgi2013Spontaneous}, the quantum effect can actually be more robust than synchronization.
Pearson correlations can also be analyzed at the level of quantum trajectories and compared with other correlations, as in QvdP in \cite{eshaqi-sani2020synchronization} (see also Sec~\ref{subsec:fewbody_traj}).

\subsubsection{Quantum correlations}

The quantumness of the generated quantum synchronization has been a key question. In general, the correspondence between synchronization and quantum effects does actually depend on the specific features of the dynamical system under consideration. 
A synchronization measure related to the presence of squeezed fluctuations was introduced in \cite{Mari2013Measures} for mutual synchronization of two coupled optomechanical oscillators
attaining limit cycles,
\begin{equation}
\mathcal{S}_c(t)=\left< \hat{x}_-(t)^2+\hat{p}_-(t)^2\right>^{-1},\label{syncerr}
\end{equation}
where $\hat{x}_-=(\hat{x}_1-\hat{x}_2)/\sqrt{2}$ is the difference in position, and similarly $\hat{p}_-=(\hat{p}_1-\hat{p}_2)/\sqrt{2}$. 
At variance with the classical case where the average distance can go to zero, in the quantum domain this measure is upper-bounded due to the uncertainty principle ($\mathcal{S}_c(t)\le 1$).
 
Other forms of observables between subsystems have been considered  to characterize mutual synchronization as spin-spin correlations $ \langle\hat{\sigma}_1^\alpha\hat{\sigma}_2^\alpha\rangle-\langle\hat{\sigma}_1^\alpha\rangle\langle\hat{\sigma}_2^\alpha\rangle$
 ($\alpha=x,y,z$) for 
trapped ions  \cite{hush2015trapped} or  
$\langle\hat{\sigma}_1^+\hat{\sigma}_2^-+\hat{\sigma}_2^+\hat{\sigma}_1^-\rangle$ in ensembles of dipoles \cite{zhu2015synchronization}. 
Looking at the synchronization of observables, a degree of quantumness has also been proposed in terms of the number of synchronized linearly independent and non-commuting observables \cite{Eneriz2019Degree}.

Beyond the dynamics of observables, entropic measures have also been proposed to quantify synchronization through quantum correlations. An enhancement of mutual information can occur during the synchronized dynamics of quantum systems \cite{giorgi2012quantum} and this was suggested as a control parameter for synchronization \cite{Ameri2015Mutual}. However, neither classical nor quantum correlations are in general distinctive signatures of synchronization or dynamical locking \cite{Galve2017Quantum,Eneriz2019Degree,cabotquantum2019}.

Mutual synchronization between different QvdP oscillators in quantum trajectories (see Sec.~\ref{subsec:fewbody_traj}) can be quantified also using the so-called normalized complex correlator \cite{eshaqi-sani2020synchronization}
\begin{equation}
C_\psi(t)=\frac{\langle \hat{a}_1^\dagger \hat{a}_2\rangle_{\psi(t)}}{\sqrt{\langle \hat{a}^\dagger_1 \hat{a}_1\rangle_{\psi(t)}\langle \hat{a}^\dagger_2 \hat{a}_2\rangle_{\psi(t)}}}\, ,
\label{eq:meas_quant_correl}
\end{equation}
where $\hat{a}_{k} = (\hat{x}_{k} + i\hat{p}_{k})/\sqrt{2}$ are the bosonic creation operators. The phase of $C_\psi$ encodes the relative phase between two oscillators. When $|C_\psi|$ is close to unity, the phase difference is well defined, whereas for $|C_\psi|\to0$ the oscillators are effectively uncorrelated and the phase carries no meaningful information. 

\subsubsection{Phase \& frequency locking \label{sec:phase_locking_measures}}

Synchronization can also be characterized as dynamical phase locking through the spectral content of two-time correlations of observables, defined by the power spectral density,
\begin{equation} \label{eq:psd}
    S_{\hat O}(\omega)=\int_{-\infty}^{\infty} d\omega \, e^{i\omega \tau} \langle \hat{O}^\dagger (\tau)\hat{O}(0)\rangle_\mathrm{ss}.
\end{equation}
Spectral analysis offers a more insightful characterization of synchronization than simple correlations between observables \cite{Galve2017Quantum,cabotquantum2019}, as it captures the temporal coherence of the underlying dynamics. Temporally coherent motion manifests as a dominant spectral peak at a characteristic frequency, identifying the principal oscillatory mode of the system. In the case of mutual synchronization, this frequency represents an emergent oscillation arising from interactions among coupled units, as analyzed, for instance, in coupled van der Pol oscillators \cite{walter2015quantum}, atom ensembles  \cite{xu2014synchronization}, or spin arrays \cite{cabotquantum2019}. In the presence of an external driving, the appearance of a dominant spectral peak at the driving frequency indicates perfect entrainment, as for instance reported in  VdP oscillators \cite{walter2014quantum}. Another important feature for characterizing synchronization is the sharpness of the spectral peak, often quantified by the linewidth. A narrower linewidth reflects stronger temporal coherence and more stable phase relationships, while a broader linewidth indicates weaker synchronization or greater phase fluctuations.

Moving from observable dynamics to the full quantum steady state, phase-space quasi-probability has been considered to characterize phase-locking in limit cycles. 
Quasi-probability distributions, such as the Wigner or Husimi-$Q$, are phase-space representations of quantum states, generalizing classical probability distributions by allowing negative values, constructed by decomposing the density matrix over coherent states, either quantum optics coherent states in the bosonic case or spin coherent states in the finite-dimensional case \cite{gardiner2004quantum, gerry2023introductory}.

Quantum synchronization in QvdP oscillators was assessed with the Wigner distribution in the stationary state $W(\alpha,\alpha^*)=W(|\alpha|,\phi)$ \cite{lee2013quantum,walter2014quantum}. For any state $\hat{\rho}$, the Wigner distribution is defined as \cite{gerry2023introductory, walls2008quantum},
\begin{equation}
    W(\alpha,\alpha^*)= \frac{1}{\pi^2}\int e^{\left(\eta^* \alpha - \eta \alpha^*\right)}\text{Tr}\left(\hat{\rho} e^{\eta \hat{a}^{\dagger} - \eta^* \hat{a}}\right) d^2 \eta.
\end{equation} 
In the absence of synchronization, the stationary Wigner distribution in the complex plane displays a ring-like shape, indicating that it is phase independent. The emergence of phase-locking corresponds to a breaking of this phase symmetry, where the marginal of the Wigner distribution for the phase 
\begin{equation}
    W(\phi)=\int d |\alpha| W(|\alpha|,\phi)
\end{equation}
displays a peaked function that was used as a measure of either entrainment of one QvdP oscillator \cite{lee2013quantum,walter2014quantum} and mutual synchronization when considering the phase difference of two identical coupled QvdP \cite{lee2013quantum}. 
This approach has been generalized to spin systems in \cite{PhysRevLett.121.053601} considering the Husimi representation $Q(\theta,\phi)\propto \bra{\theta,\phi}\hat{\rho}\ket{\theta,\phi}$ of the rotated extremal states $ \ket{\theta,\phi}=\exp(-i \phi \hat{S}_z)\exp(-i\theta \hat{S}_y)\ket{S,S}$. In the presence of a limit cycle, phase locking can be assessed if the marginal $
Q(\phi)=\int  d\theta\, \sin\theta\, Q(\theta,\phi)-\frac{1}{2\pi} $
is peaked around a value. These measures have been extended to both coupled oscillators \cite{lee2013quantum} and spins \cite{PhysRevLett.121.053601}, where distributions for the relative phase only can be similarly obtained from the full multimode ones, showing a peaked structure when mutual synchronization occurs (see also Sec.~\ref{sec:QvdP}-\ref{sec: finite-level}). In many-body settings, the Wigner distribution for each oscillator's reduced state has also proven useful, displaying a similar ring-to-peak transition when synchronization emerges \cite{ludwig2013quantum}.

Information measures have been proposed to assess phase locking in limit cycles \cite{Jaseem2020Generalized} as distance (such as relative entropy or trace distance) with respect to the nearest unsynchronized limit cycle states $\hat{\rho}_{\mathrm{lim}}$. 
The approach is inspired by known geometric measures that quantify certain qualities of a state (e.g., entanglement), as the distance to the nearest state lacking this quality (e.g., the set of separable states). 
Once suitable families of unsynchronized states $\hat{\rho}_{\mathrm{lim}}$ are identified, synchronization can be assessed as a minimal distance, vanishing in the absence of synchronization in spontaneous and driven settings.
This approach is discussed for various classes of limit cycle states, including diagonal limit-cycle states \cite{walter2014quantum, PhysRevLett.121.053601, roulet2018quantum, weiss2016noise}, marginal limit cycle states \cite{Ameri2015Mutual} and partially-coherent limit cycle states. 
Diagonal limit cycle states have no coherence between energy eigenstates. We denote the set of all such states by $\delta$. 
This set is convex, meaning that any mixture of diagonal states is itself a diagonal state, so the closest element of $\delta$ to any state $\hat{\rho}$ is simply its own diagonal part. 
The minimization then gives the synchronization measure which has a simple closed form, $\Omega_R(\hat{\rho}) = \min_{\hat\rho_{\mathrm{lim}}\in\delta} S(\hat\rho\|\hat\rho_{\mathrm{lim}})= S_{\mathrm{coh}}(\hat\rho)$, and is exactly given by the coherence of $\hat\rho$ in the energy eigenbasis. This can be useful in understanding entrainment \cite{walter2014quantum, roulet2018quantum} and mutual synchronization \cite{PhysRevLett.121.053601, weiss2016noise} in several systems. 
Marginal limit cycle states relax this condition such that coherences within each subsystem are now allowed, and only correlations between the two subsystems are restricted, $\hat\rho_{\mathrm{lim}} = \hat\sigma^A\otimes\hat\sigma^B$ \cite{Ameri2015Mutual}. 
In this case, the closest reference state is just given by the marginals of $\hat\rho$, so the measure reduces to $\Omega_R(\hat\rho) = S(\hat\rho^A) + S(\hat\rho^B) - S(\hat\rho)$, the quantum mutual information between the two subsystems. 
Partially-coherent limit cycle states sit between these two cases, in that they retain a chosen subset of coherences, such as those induced by a local drive, while treating any additional correlation between the subsystems as synchronization. 
This lets the measure separate genuine mutual synchronization from entrainment caused by local driving.

\subsubsection{Liouvillian spectral features} \label{Sect:Liouv}

As discussed in Sec.~\ref{sec:master_equation}, the Lindblad master equation can be recast in terms of the Liouvillian superoperator, whose spectrum encodes relaxation, oscillatory modes, metastability, and steady states. This spectral structure provides an alternative framework for characterizing the temporal dynamics in nonequilibrium phenomena such as quantum synchronization and time crystals.

An important spectral feature is the emergence of spectral gaps, where a subset of eigenvalues has real parts whose magnitudes are much smaller (in absolute value) than the rest. While such gaps are a characteristic trait of metastability in open quantum systems \cite{kasha2016towards}, they have also been shown to give rise to quantum synchronization in several settings. In particular, in the phenomenon of transient (metastable) mutual synchronization discussed below (Sec.~\ref{Sec:transient synch}), slowly decaying eigenmodes govern the temporary emergence of synchronized dynamics before reaching the steady state \cite{giorgi2012quantum,Giorgi2013Spontaneous,tindallQuantum2020,schmolkenoise2022}. This can be related to the opening of spectral gaps in the Liouvillian, where normal modes decay at widely different rates, and the long-time dynamics is dominated by the slowest decaying mode. If these slow modes have imaginary components, they can give rise to a non-equilibrium preasymptotic regime characterized by an ordered, spatially delocalized, and monochromatic oscillation, a manifestation of transient synchronization \cite{giorgi2019transient,tindallQuantum2020,schmolkenoise2022}. Accordingly, measures signaling the opening of such spectral gaps have also proven useful for characterizing the emergence of synchronization across various settings \cite{giorgi2019transient,tindallQuantum2020,schmolkenoise2022}. We note that the eigenvalues of the synchronizing eigenmodes can even have vanishing real parts, in which case the synchronous oscillation does not decay.
This regime is known as steady or stable synchronization (see Sec.~\ref{Sec:transient synch}, not to be confused with steady‑state synchronization, which concerns properties of the stationary state) and is linked to the presence of decoherence‑free subspaces \cite{giorgi2012quantum,manzano2013synchronization,cabot2018unveiling,manzano2013avoiding} and dynamical symmetries \cite{tindallQuantum2020,bucaalgebraic2022}. 
Beyond mutual synchronization, spectral‑gap openings and metastability have also been found to accompany the emergence of subharmonic entrainment \cite{cabot2021metastable}. 

Another intriguing spectral feature connected to synchronization is the presence of exceptional points, identified both in the Liouvillian \cite{cabot2021synchronization} and in the matrices governing two‑time correlations \cite{nadolny_nonreciprocal_2025}. These spectral singularities can separate regimes with multiple oscillation frequencies from those with a single frequency, thereby providing a signature of the onset of synchronization \cite{xu2015conditional,cabot2021synchronization,nadolny_nonreciprocal_2025} and entrainment \cite{cabot2021metastable} in various settings.

Finally, as discussed in Sec.~\ref{sec:sync_and_TCs}, spectral features also play a prominent role in identifying emerging collective behavior such as time‑crystalline phases. In these settings, signatures of the emerging phase can be inferred from the real parts of the dominant complex eigenmodes, which vanish in the thermodynamic limit, while their imaginary parts approach a well‑defined separation \cite{iemini_boundary_2018}. As further noted in Sec.~\ref{sec:sync_and_TCs}, the identification of dynamical symmetries can likewise provide signatures of these phenomena \cite{bucaNonstationaryCoherentQuantum2019}.
From a spectral viewpoint, synchronization of such oscillatory phases manifests through the adjustment of the purely imaginary Liouvillian eigenvalues of coupled subsystems toward a common value, thereby locking their oscillation frequencies \cite{PhysRevA.105.L020401,hajdusekSeedingCrystallizationTime2022}.

\subsubsection{Measures based on time-resolved emissions}
\label{sec:emission_measures}

The measures of quantum synchronization introduced above are mostly based on expectation values of system observables (see e.g.~Eq.~\eqref{eq:meas_temp_correl}) or on the quantum regression theorem (see~Eq.~\eqref{eq:psd}). 
In some settings, accessing time‑resolved photocurrents is more practical than measuring expectation values, and these signals can likewise reveal the onset of synchronization.
Analogous quantities (see Eq.~\eqref{eq:meas_temp_correl} and Eq.~\eqref{eq:meas_quant_correl}) can also be defined at the level of quantum trajectories and are useful for theoretical investigations \cite{eshaqi-sani2020synchronization}.  However, in these settings, they remain experimentally inaccessible as their computation would require postselection over identical trajectories. That is, since the probability of observing the same trajectory multiple times is essentially zero, these quantities cannot be obtained in experiments.  To overcome this limitation, alternative measures based on the measurement record of the emitted field have been proposed \cite{nadolnyQuantum2026}. For instance, complex homodyne currents, $J_A$ and $J_B$, can be used to infer in real time the phase relation between two quantum systems via $\phi_{AB}=\arg[J_A/J_B]$. Moreover, these currents allow one to access spectral properties through 
\begin{equation}
S(\omega)=\lim_{t\to\infty}\int_{-\infty}^{\infty}d\omega e^{-i\omega \tau}\mathbb{E}[J^*_A(t+\tau)J_A(t)]\propto S_{\hat{O}}(\omega)\, ,
    \label{eq:meas_spec_traj}
\end{equation}
where $\hat{O}$ is some operator related to the heterodyne current (see~Eq.~\eqref{eq:psd}) and $\mathbb{E}[\cdot]$ denotes the average over the stochastic measurement record. 
Such quantities provide a direct and experimentally viable route to characterize quantum synchronization in real time \cite{coxPhase2014,weiner2017phase,nataleSynchronization2026,baurBand2026} without requiring quantum state tomography or repeated projective measurements at different times.

\section{Quantum Synchronization in Few-Body Systems}\label{sec:Few_Body_QSynch}

Having discussed the measures of quantum synchronization, we can now explore the models that demonstrate it. 
Here, we discuss quantum synchronization, starting with examples closer to classical models, such as continuous-variable quantum systems, and then moving towards finite-level quantum systems with no classical counterparts. We will also review both transient (metastable) and steady (stable) synchronization, along with the trajectory approach to quantum synchronization.

\subsection{Quantum synchronization in systems that generalize classical synchronization}\label{sec:QvdP}

The quantum counterparts of the classical self-sustained oscillators discussed in Sec.~\ref{ch:classical_sync} constitute a valuable platform for understanding the transition from the classical to the quantum synchronization regime.
In this section, paradigmatic models such as the {\it quantum van der Pol oscillator} (QvdP) are introduced, and their contributions to extending synchronization to the quantum realm are reviewed. 
We discuss the entrainment and mutual synchronization features in these paradigmatic models and, furthermore, provide a current overview of other complex synchronization behaviors recently investigated in these systems.

\subsubsection{Quantum self-sustained oscillators}

Early studies of quantum synchronization were naturally motivated by extending classical models of self-sustained oscillators into the quantum regime. 
Classical paradigms such as the Rayleigh oscillator, the van der Pol oscillator, Adler equation, and the Kuramoto model have inspired investigations of their quantum counterparts across a variety of physical platforms, including trapped ions \cite{lee2013quantum,hush2015trapped}, optomechanical oscillators \cite{ludwig2013quantum,heinrich2011collective,weiss2016noise,li2020noise}, nano-mechanical oscillators \cite{bagheri2013photonic}, atomic ensembles \cite{xu2014synchronization}, cavity QED systems, and superconducting circuits \cite{hriscu2013superconducting}. 

A widely studied example of a self-sustained oscillator is the vdP oscillator, which we briefly reviewed in Sec.~\ref{ch:classical_sync}.
It has long served as a bridge between classical nonlinear dynamics and the quantum domain. 
In open quantum systems, its natural extension is a harmonic oscillator subject to linear gain and nonlinear loss \cite{lee2013quantum,walter2014quantum}, described within the Lindblad formalism as
\begin{equation}\label{eq:QvdP}
    \dot{\hat{\rho}}=-i[\omega_0\hat{a}^\dagger \hat{a},\hat{\rho}]+\gamma_1 \mathcal{D}[\hat{a}^\dagger]\hat{\rho}+\gamma_2 \mathcal{D}[\hat{a}^2]\hat{\rho},
\end{equation}
where $\hat{a}$ is the annihilation operator.
The equation of motion corresponding to $\alpha=\langle \hat{a} \rangle$ in the mean-field limit, neglecting the correlations and factorizing higher order terms such as $\langle \hat{a}^2 \rangle\approx\langle \hat{a}\rangle
^2$, is expressed as
\begin{equation}
    \dot{\alpha}=i\omega_0 \alpha+\frac{\gamma_1}{2}\alpha-\gamma_2\vert \alpha \vert^2\alpha.
\end{equation}
Writing $\alpha = r e^{i\phi}$ and separating the real and imaginary parts of $\dot{\alpha}$ yields coupled equations for the amplitude and phase that correspond to the classical undriven Stuart–Landau oscillator in Eq.~(\ref{eq:truncated_vdP_amplitude}), describing the weakly nonlinear regime of the vdP oscillator.
Recently, the quantum equivalent of the beyond weakly nonlinear vdP oscillator was also examined \cite{chia2020qvdp,arosh2021quantum}.

\begin{figure}
\includegraphics[width=0.48\textwidth]{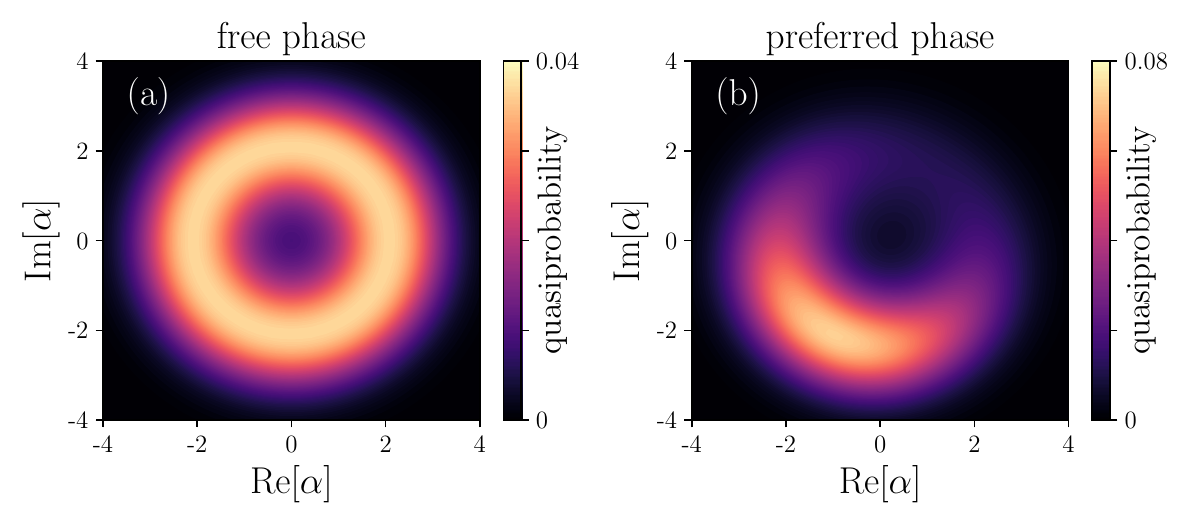}
    \caption{
    Signatures of entrainment in a driven QvdP oscillator. Wigner distribution of the stationary state for $\Delta/\gamma_1=0.1$, $\gamma_2/\gamma_1=0.25$, (a) $F/\gamma_1=0$ and (b) $F/\gamma_1=0.25$. 
    }
    \label{fig:entrainment_QvdP}
\end{figure}

Quasi-probability distributions such as the Husimi-$Q$, $P$, and Wigner functions provide a useful phase-space representation of such dynamics and have been widely employed to characterize limit cycles and their synchronization properties \cite{lee2013quantum,walter2014quantum}. 
For the quantum van der Pol (QvdP) oscillator, the steady-state distribution acquires a ring-like structure in phase space described by the quadratures of $a$, mirroring the classical limit cycle with a fixed amplitude and a free phase. 
In the quantum regime, this ring structure persists but is broadened and deformed by quantum fluctuations, as seen in Fig.~\ref{fig:entrainment_QvdP}(a). 
In the upcoming sections, we will investigate the synchronization and entrainment properties of quantum limit cycles through an external drive or by coupling with other quantum oscillators.

\subsubsection{Entrainment}

The QvdP oscillator provides a testbed for analyzing the crossover from classical to quantum entrainment \cite{lee2013quantum,walter2014quantum}. To describe this scenario, one needs to add a harmonic drive, $F (\hat{a} e^{i\omega t}+\hat{a}^\dagger e^{-i\omega t})$, to the Hamiltonian part of Eq. (\ref{eq:QvdP}). This explicit time dependence can be removed by moving to a frame co-rotating with the drive frequency,  in which the Hamiltonian reads as
\begin{equation}
\hat{H}=\Delta \hat{a}^\dagger \hat{a}+F(\hat{a}+\hat{a}^\dagger),
\label{eq:qvdP_dr}
\end{equation}
where $\Delta = \omega_0 - \omega$ is the detuning between the oscillator frequency $\omega_0$ and the drive frequency $\omega$, and the dissipative terms in Eq.~(\ref{eq:QvdP}) are unchanged. 
In the classical limit ($\gamma_1/\gamma_2\gg 1$) and for small forcing strengths, the synchronization transition is captured by the Adler equation discussed in Sec.~\ref{sec:classical_entrain_mutual} \cite{lee2013quantum,walter2014quantum,kato2019semiclassical}. In this regime, the transition from a synchronized regime to an unsynchronized regime is characterized by a saddle-node bifurcation. Increasing $F$, however, eventually changes the nature of this bifurcation to a Hopf bifurcation \cite{weiss2017underdamped}. In the quantum regime, where $\gamma_1$ is comparable or smaller than $\gamma_2$, the synchronization transition becomes a smooth crossover, due to the presence of quantum fluctuations \cite{lee2013quantum,walter2014quantum}. 
Signatures of phase and frequency locking have been identified for the QvdP, both in its steady state and in its dynamical response.

Following the synchronization measures introduced in Sec.~\ref{sec:phase_locking_measures}, steady-state phase-space representations, such as the Wigner distribution, can be used to assess the emergence of synchronization in the quantum regime \cite{lee2013quantum,walter2014quantum}. The onset of entrainment is signaled by the quasiprobability distribution gradually shifting from a ring-like shape, indicating diffusion of a free phase as shown in Fig.~\ref{fig:entrainment_QvdP}(a), to a localized lobe as in Fig.~\ref{fig:entrainment_QvdP}(b), indicating the emergence of phase-locking. 
Integrating out its radial part, the marginal probability distribution for the phase is obtained, which ensues a gradual shift from a uniform distribution to a singly-peaked one as the system becomes entrained \cite{lee2013quantum,walter2014quantum}.

Since the power spectrum $S_{\hat{O}}(\omega)$ defined in Eq.~(\ref{eq:psd}) is a key quantity that provides information about the frequency response of a system to, e.g., a weak external signal or to thermal fluctuations \cite{clerk2010introduction}, it can serve as a signature of synchronization for $\hat{O} = \hat{a}$. This experimentally accessible quantity is of use both in classical and quantum systems  \cite{zhang2012synchronization,bagheri2013photonic,weiner2017phase}. 
Theoretically, the stationary two-time correlation $\langle \hat{a}^\dagger (\tau)\hat{a}(0)\rangle_\mathrm{ss}$ can be computed using the quantum regression theorem \cite{carmichael2013statistical}. The frequency at which $S_{\hat{a}}(\omega)$ reaches its maximum (removing Dirac-delta contributions) is the so-called observed frequency, $\omega_\mathrm{obs}$. A signature of entrainment in the quantum regime is $\omega_\mathrm{obs}$ gradually shifting from $\omega_0$ to $\omega$ as the driving strength is increased \cite{walter2014quantum}. 
Generally, when the QvdP operates deeper in the quantum regime,  it is necessary to drive it more strongly to induce frequency-entrainment \cite{walter2014quantum}. The entrained response in the QvdP has been further analyzed using different approaches, such as deriving a linearized dynamics for the fluctuations \cite{weiss2017underdamped,navarrete-benlloch2017general}, or semiclassical phase reduction \cite{kato2019semiclassical}, revealing regimes in which the entrained dynamics generates long-lived quantum coherence and squeezing \cite{weiss2017underdamped}. 

Proposals based on the use of a squeezed drive \cite{sonar2018squeezing} or continuous measurement with feedback control \cite{kato2021enhancement} have been shown to enhance synchronization in the quantum regime. The squeezed drive is also capable of inducing subharmonic entrainment, i.e., phase- and frequency-locking to a fraction of the driving frequency \cite{balanov2009simple}. Both steady-state quasiprobability distributions and power spectrum measures are still capable of signaling the emergence of this other form of entrainment, although the particular features vary with respect to the driven case \cite{sonar2018squeezing,kato2019semiclassical,kato2021quantum,cabot2021metastable,cabot2024nonequilibrium, hermani_synchboost}. 

Beyond the QvdP, the quantum to classical crossover of entrainment phenomena has also been explored in other nonlinear quantum oscillators \cite{sudler2024driven}, including optomechanical systems \cite{heinrich2011collective,amitai2017synchronization}, and superconducting devices \cite{danner2021injection,hohe2025quantum}, in which similar signatures in the steady-state Wigner distribution and power spectrum have been reported.

\subsubsection{Mutual synchronization} \label{subsec:fewbody_mutualsynch}

We now consider cases of mutual synchronization between quantum self-sustained oscillators. Early studies on quantum mutual synchronization focused on minimal models such as coupled QvdP oscillators~\cite{lee2013quantum,walter2015quantum}.  
Mutual synchronization manifests through the locking of the relative phases of the oscillators. 
Various measures have been proposed to quantify this phase locking in quantum systems (see Sec.~\ref{Sect:meas}). 
For instance, the two-mode Wigner function $W(\alpha_1,\alpha_1^*,\alpha_2,\alpha_2^*)$ was introduced in \cite{lee2013quantum}, where $ \alpha_i=\langle \hat{a}_i \rangle= \vert \alpha_i \vert e^{i\phi_i} $.
A relative phase measure $W(\phi_1-\phi_2)$ can be obtained from  $W(\alpha_1,\alpha_1^*,\alpha_2,\alpha_2^*)$ by integrating out the irrelevant degree of freedoms such as $\vert \alpha_1 \vert, \vert \alpha_2 \vert$ and $\phi_1+\phi_2$.
Alternatively, \cite{walter2015quantum} employed the joint probability distribution measure $P(x_1,x_2)=\langle x_1,x_2 \vert \hat{\rho} \vert x_1,x_2 \rangle$ to investigate the mutual synchronization between two QvdPs.
The observed frequency difference, $\Delta_{\text{obs}} = \vert \omega^1_{\text{obs}} - \omega^2_{\text{obs}} \vert$, was also explored as a measure of mutual synchronization in \cite{walter2015quantum}. 
Here, the observed frequency $\omega^i_{\text{obs}}$ corresponds to the frequency for which power spectrum $S_{\hat{O}_i}(\omega)$ exhibits a maxima.
We refer to Sec.~\ref{Sect:meas} for a detailed discussion on synchronization measures, which has been a subject of intense research.

The emergence of synchronization in QvdPs depends crucially on the nature of their coupling.
For example, reactive (coherent) coupling, represented by the Hamiltonian $\hat{H}_c = g (\hat{a}_1^\dagger \hat{a}_2 + \hat{a}_2^\dagger \hat{a}_1)$, produces bistable phase locking (in-phase $|\phi_{1} - \phi_{2}|=0$ and antiphase $|\phi_{1} - \phi_{2}|=\pi$) for identical oscillators~\cite{lee2013quantum}. 
In contrast, dissipative coupling, described by $g\mathcal{D}[\hat{a}_1 \pm \hat{a}_2]$, results in a unique phase-locked state for identical QvdPs (with `+' indicating in-phase and `-' indicating antiphase synchronization)~\cite{walter2015quantum}. 
Nonlinear coupling between these oscillators has also been investigated~\cite{thomas2022nonlinearcoupling}. Beyond these canonical models, mutual synchronization has also been investigated in nanomechanical resonators \cite{li2020noise,bemani2017synchronization}.

\subsubsection{Other exotic synchronization phenomena}

In addition to the conventional synchronization scenarios discussed above, more complex synchronization phenomena have been explored in various dynamical systems.
One such phenomenon is \emph{$m:n$ phase locking}, which establishes higher-order phase relationships between oscillators instead of the typical $1:1$ phase locking \cite{balanov2009simple}. 
For instance, synchronization has been observed between two nonlinearly coupled QvdP oscillators with the coupling Hamiltonian expressed as $ \hat{H}_c = g\,(\hat{a}_1^{\dagger 2} \hat{a}_2 + \hat{a}_1^{2} \hat{a}_2^{\dagger})$~\cite{thomas2022nonlinearcoupling}.
This results in higher-order phase locking of the form $\phi_1 - 2\phi_2$. 
Furthermore, in the presence of Kerr nonlinearity, these nonlinearly coupled QvdP oscillators exhibit multiple resonant phase-locking behaviors, leading to the formation of a series of Arnold tongues.
In addition to this, the squeezing drive discussed in previous sections also leads to 2:1 phase locking \cite{sonar2018squeezing} (see also Sec.~\ref{subsec:fewbody_mutualsynch}).

Cluster synchronization and chimera states are characterized by the coexistence of synchronized and unsynchronized dynamics (see Sec.~\ref{sec:classical_meas}).  
Such phenomena have been explored in quantum settings in networks of coupled quantum QvdP oscillators \cite{bastidas2015chimera}, where synchronized QvdP oscillators have steady state squeezing along the same axis, and the unsynchronized ones have random squeezing directions.

Another interesting example is the Liénard system, which supports multiple limit cycles separated by basins of attraction. 
A quantum analogue was proposed using the QvdP with higher-order dissipators $\mathcal{D}[\hat{a}^{\dagger 3}]$ and $\mathcal{D}[\hat{a}^{4}]$, enabling the coexistence of two stable limit cycles \cite{kehrer2025twin}. 
Quantum fluctuations induce switching between different limit cycles, while Kerr nonlinearity, through its radius-dependent frequency, leads to distinct phase-locking behaviors under identical driving. 
Mutual synchronization of such twin–limit-cycle oscillators also reveals rich synchronization features.  

Synchronization of self-sustained oscillators with strong nonlinearities can extend beyond periodic dynamics to quasi-periodic and even chaotic regimes (see Sec.~\ref{sec:classical_meas}). 
Quantum signatures of such chaotic synchronization have been explored in coupled BEC-optomechanical systems~\cite{li2017quantum}, and atomic clocks \cite{patra2019chaotic}, 
highlighting how synchronization persists even in highly nontrivial dynamical phases.

Recently, non-reciprocal coupling mechanisms have also been explored in coupled vdPs \cite{lai2025nonreciprocal, 9y9w-lw92}. 
These mechanisms emerge from the interplay between coherent coupling $g (e^{i\theta}\hat{a}_1^\dagger \hat{a}_2 + e^{-i\theta}\hat{a}_2^\dagger \hat{a}_1)$ and dissipative coupling $\mathcal{D}[\hat{a}_1 + \hat{a}_2]$. 
The parameter $\theta$ allows for a transition from reciprocal coupling (where $\hat{a}_1 \leftrightarrow \hat{a}_2$) to fully unidirectional coupling (either $\hat{a}_1 \to \hat{a}_2$ or $\hat{a}_1 \leftarrow \hat{a}_2$).
This nonreciprocity leads to synchronization blockades between the undriven oscillator and interesting synchronization effects when the system is additionally influenced by an external drive \cite{9y9w-lw92}.

\subsection{Synchronization in finite level systems} \label{sec: finite-level}

As previously noted, the synchronization of quantum systems has been examined through examples inspired by classical analogues, such as harmonic and anharmonic systems. 
These are mainly continuous-variable models, where well-defined classical limit cycles emerge and underpin the notion of synchronization. 
This framework, however, does not straightforwardly extend to finite-level (discrete) quantum systems, where the concept of a limit cycle is not always well-defined.
This has led to a broader perspective on quantum synchronization in such systems, which we review here and in Sec.~\ref{subsec:blockade}-\ref{Sec:transient synch}.

One of the first works addressing the emergence of driven synchronization in finite-level systems and beyond classical analogues was \cite{zhirov2008synchronization}, in which they considered different superconducting qubits getting synchronized via a driven resonator. This system was also found to exhibit two possible metastable synchronized configurations and stochastic switching between them. 
Spontaneous synchronization between detuned qubits in the absence of driving was reported a few years later \cite{Giorgi2013Spontaneous}. Two spins coupled to a common thermal bath were shown to synchronize when dissipation dominates over dephasing, owing to a time-scale separation in the decay rates of the eigenmodes. Generalizations to localized dissipation in one qubit \cite{giorgi2016probing}, non -Markovianity \cite{karpat2021synchronization}, and spin chains \cite{cabotquantum2019,tindallQuantum2020,schmolkenoise2022}, have also been reported. This viewpoint led to the notions of transient (metastable) and steady (stable) synchronization, reviewed in  Sec. \ref{Sec:transient synch}. Here, synchronization is characterized by the correlated dynamics of observables, even in the absence of underlying limit cycles, with, for instance, qubit expectation values being correlated in the sense of the Pearson or similar measures (see Sec.~\ref{Sect:meas}).

Generalization of limit cycles to finite‑level quantum systems that do not necessarily admit a classical counterpart was also addressed. In this context, one of the starting points is the seminal work  \cite{PhysRevLett.121.053601,roulet2018quantum}, which argued that the underlying undriven limit-cycle of a state should be observable. This work noted that since the free phase for qubits amounts to an unobservable gauge degree of freedom, the limit cycle is rendered unobservable. Consequently, \cite{PhysRevLett.121.053601} identifies the three-level system as the minimal system capable of supporting a genuine quantum limit cycle, making it a natural platform for studying quantum synchronization. 
In \cite{PhysRevLett.121.053601}, the limit cycle is a steady state which is a quantum superposition of spin coherent states with a free phase in the corresponding Husimi-$Q$ distribution. 
There is, however, an alternative view of quantum synchronization in qubit systems based on statistical mixtures of spin coherent states in the resolution of their steady states~\cite {Bergli2020synchronization}. 
Such qubit systems have been investigated both theoretically \cite{PhysRevA.107.022221,entrain-2qubits-NM} and experimentally \cite{PhysRevResearch.5.033209} in several studies. 
In the phase-space representation (usually given by the Husimi-$Q$ distribution), both the qubit and qutrit cases can hence possess a free phase; and the mathematical analysis remains the same as is presented below for the qutrit case.

\begin{figure}
    \centering
    \includegraphics[width=\columnwidth]{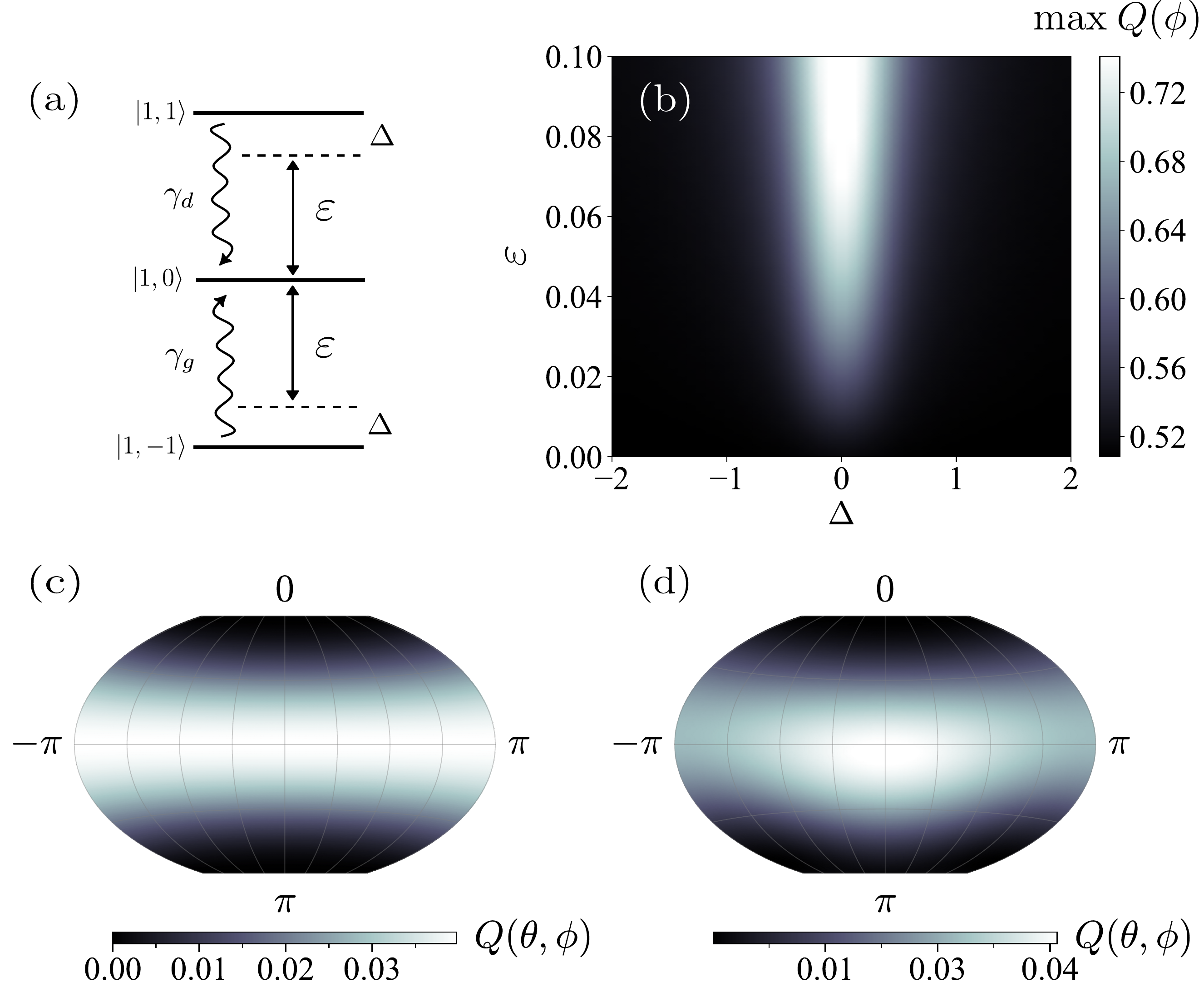}
    \caption{(a) A schematic representation of a driven qutrit. $\varepsilon$ is the strength of the external drive, $\Delta$ is the detuning between the natural frequency of the oscillator and the external drive, and $\gamma_g, \gamma_d$ denote the gain and dissipation rates respectively. (b) The synchronization measure given by the Husimi-$Q$ function shows an Arnold tongue-like behavior, indicating that the system synchronizes for a wider range of detuning for higher coupling strengths. The Husimi-$Q$ function is plotted in the steady state using a Winkel tripel earth projection for (c) $ \gamma_g/\gamma_d = 0.1, \varepsilon = 0$, (d) $\gamma_g/\gamma_d = 0.1, \varepsilon = 0.1 \gamma_g$.}
    \label{fig:qutrit_synch}
\end{figure}

In case of a qutrit, the limit cycle behavior can be understood by considering an equally spaced three-level system whose spin-coherent state is given by $\vert\theta,\phi\rangle=\exp(-i\phi \hat{S}_z)\exp(-i\theta \hat{S}_y)\vert S,S\rangle$, where the spin operators are generators of the $\mathrm{SO}(3)$ group given by the commutation relation $[\hat{S}_i,\hat{S}_j]=i\varepsilon_{ijk}\hat{S}_k$ with $i,j,k \in\{x,y,z\}$. 
The time evolution of the given spin-coherent state under the free Hamiltonian $\hat{H}_0=\omega_0\hat{S}_z$ is given by $e^{-i\hat{H}_0t}\vert\theta,\phi\rangle=\vert\theta,\phi +\omega_0 t\rangle$, thereby defining $\phi$ as a free phase that can be synchronized.
A proper limit cycle requires the system to relax to a steady state with no preferred phase, independent of initial conditions.
This was investigated in \cite{PhysRevLett.121.053601} by considering the dissipators $\gamma_d D[\hat{S}_-\hat{S}_z]\hat{\rho}$ and $\gamma_g D[\hat{S}_+\hat{S}_z]\hat{\rho}$ (as shown in Fig.~\ref{fig:qutrit_synch}(a), that stabilizes a phase-neutral state, as depicted in the Husimi $Q$-function in Fig.~\ref{fig:qutrit_synch}(c).
Upon applying an external drive, the system exhibits phase locking in the steady state as shown in Fig.~\ref{fig:qutrit_synch}(d), signaling synchronization, which also exhibits an Arnold tongue as seen in Fig.~\ref{fig:qutrit_synch}(b).
We note that there is nothing special about the Husimi distribution function, and instead the Wigner function \cite{wigner_nemoto} could have been chosen as is customary in the continuous variable case.
Synchronization was also studied in three-level systems with unequal energy level spacings \cite{PhysRevE.101.020201}. 
Here, the synchronization measure was generalized using the $\mathrm{SU}(3)$ coherent states to define the limit cycle. 
Since unequally-spaced three-level maser efficiency is a function of the energy differences, this model allows one to study the relationship between synchronization and maser output power, as reviewed in Sec.~\ref{sec:thermodynamics}. 

Building on synchronization in minimal quantum systems, mutual synchronization between two spin-1 systems has also been explored \cite{roulet2018quantum}. For a coherent interaction of the form $i\varepsilon(\hat{S}_+^A\hat{S}_-^B-h.c.)$, the individual phases $\phi_{A,B}$ remain uniformly distributed, while the relative phase becomes localized at finite coupling strength $\varepsilon$ signaling the onset of synchronization. Notably, the emergence of entanglement in this regime follows an Arnold tongue–like structure \cite{roulet2018quantum}, closely mirroring phase-synchronization measures, which suggests that synchronization can act as a resource for entanglement generation in finite-level systems. 
Mutual synchronization in this system has also been studied in \cite{bucaalgebraic2022}, where the focus was on the dynamics of expectation values. 
Interestingly, the same work also discusses a different approach to generalize limit cycles to quantum systems without classical limit, which builds on dynamical symmetries (see also Sec.~\ref{sec:sync_and_TCs}). 
Mutual synchronization of two three-level quantum heat engines \cite{PhysRevA.105.L020401}, where coherent interactions can lead to synchronization between the two thermal machines has been investigated. 
The interplay between mutual synchronization and entrainment to an external drive in degenerate thermal machines has been further investigated in \cite{PhysRevLett.131.030401}, revealing a richer dynamical structure (see Sec.~\ref{sec:thermodynamics}).
Beyond these, there are other interesting examples for finite-level systems. 
For instance, \cite{Zhang2020nonMarkovian} studied the relationship between dissipation and non-Markovianity for specific models. 

While many examples of quantum synchronization in finite-level systems exist, it is possible to generalize this argument of phase synchronization to the SU($N$) group and to a general $N$-level system. 
To understand the phase synchronization of such complex systems, the concept of the phase synchronization measure was extended in \cite{PhysRevA.108.022216} for generic $N$-level systems in terms of the density matrix formalism.
Writing the complex components of the density matrix in the polar representation as $\rho_{jk}=R_{jk}\exp(-i\chi_{jk})$, we can write a measure of synchronization for $N$-level systems \cite{tan2022half, PhysRevA.108.022216} as 
\begin{align}
S(\vec{\phi})=2\mathcal{N}\sum_{j<k}^{N}R_{jk}z_{jk}\cos\xi_{jk}. \label{eq:generalized_sync_measure}
\end{align}
Here $\mathcal{N}$ is the normalization attached to a general $SU(N)$ coherent state such that $\mathcal{N}\int d^2\alpha\vert\alpha\rangle\langle\alpha\vert=\mathbb{I}$ and $\alpha_j=r_j\exp(-i\phi_j)$ defines the polar decomposition of the coherent amplitude. 
Finally, we note that $\xi_{jk}=\phi_j-\phi_k-\chi_{jk}$ and $ z_{jk}=\int d\Omega_\theta r_j r_k$ is the integration over the population parameters with $d\Omega_\theta$ being the Haar measure of the population degree of freedom. This measure reduces to the $l_1$-norm based measure of coherence for externally driven unipartite systems and can be used to study synchronization in generic $N$-level atoms.
This measure has also been successful in understanding the interferometric cancellation of coherences leading to synchronization blockade in complex $N$-level systems \cite{PhysRevA.108.022216} as discussed below.

\subsection{Synchronization blockade}\label{subsec:blockade}
 
We now focus on a key difference between classical and quantum synchronization, namely the synchronization blockade. 
Synchronization blockade is defined as the suppression of phase-locking in a system that is otherwise capable of synchronizing at a given coupling (or drive) strength. This suppression is induced not by adjusting the coupling/drive strength itself, but by varying an auxiliary parameter that renders the system effectively transparent to the synchronizing field or to a mutually coupled oscillator. Such a suppression can be induced, for instance, by interferometric cancellation or by an interaction-induced frequency shift.

Quantum synchronization blockade was first studied in two interacting Kerr oscillators \cite{lorch2017blockade}, where non-linear jump operators were employed to stabilize steady states with population concentrated in only a few Fock states for each oscillator. 
In the extreme non-linear regime, only a single Fock state can be populated in the steady state. 
This results in uncertain phases due to phase-number uncertainty, leading to a quantum limit cycle behavior.
In the deep quantum limit, the interacting Kerr oscillators cease to synchronize under near-resonant conditions due to interaction-induced frequency shifts, resulting in a synchronization blockade.
However, by introducing a detuning, the oscillators become resonant once more, thereby restoring synchronization and overcoming the blockade.
This blockade effect can be understood in terms of the energy exchanged between the oscillators: synchronization is blocked if the extractable energy from one oscillator does not perfectly match the amount that the second oscillator can absorb and interact with. 
See \cite{PhysRevA.97.013811} for the Josephson junction circuit realization of such a system.

Similar counterintuitive behavior was also studied for the finite-level spin-1 quantum system \cite{PhysRevLett.121.053601}, discussed in the previous section.
For an externally driven spin-1 system, the synchronization is suppressed for any drive strength when the dissipation rates are equal, $\gamma_g=\gamma_d$.
In this case, the synchronization blockade results from the interferometric cancellation of coherence rather than from a hindrance to energy exchange.
See \cite{PhysRevA.99.043804} for a detailed discussion on synchronization blockade in spin-1 systems and QvdP in the deep quantum limit. 
The effect of synchronization blockade in the spin-1 system was also observed experimentally \cite{PhysRevLett.125.013601}.
Later, the idea of synchronization blockade was extended to larger spin systems \cite{tan2022half}, where the condition for synchronization blockade was qualitatively found to depend on spin length (integer vs. half-integer).

A system-agnostic, symmetry-based framework for understanding quantum synchronization blockade was introduced in \cite{PhysRevA.108.022216}. In that work, a no-go theorem was established using group-theoretic arguments: a finite $N$-level system whose dynamics fully explore the $su(N)$ algebra, and therefore possess no nontrivial symmetry constraints, cannot exhibit synchronization blockade. By contrast, systems whose dynamics are restricted to a subgroup algebra, such as $su(M)$ with $M<N$, can support blockade. This can be understood directly from the generalized synchronization measure in Eq.~(\ref{eq:generalized_sync_measure}). When $M=N$, each coherence contribution $z_{jk}$ is associated with a distinct $\xi_{jk}$, preventing interference between different coherence channels. In contrast, for $M<N$, symmetry constraints allow multiple coherence pathways to contribute to the same $\xi_{jk}$. This degeneracy is not merely a mathematical coincidence but reflects the existence of physically indistinguishable transition pathways enforced by the symmetry. As a consequence, the corresponding coherence contributions can interfere destructively, canceling one another and suppressing the net synchronization signal, driving the system into the blockaded regime. In this picture, synchronization blockade emerges fundamentally as a symmetry-enforced destructive interference effect among coherence channels, closely analogous to the formation of dark states in quantum optics, where interference between transition amplitudes suppresses the response to resonant driving.

Recently, the coexistence of quantum synchronization and synchronization blockade phenomenon was investigated in coupled twin limit-cycle oscillators \cite{kehrer2025twin}.
Another interesting phenomenon of mediating the quantum synchronization via synchronization blockade was explored in \cite{PhysRevA.110.042203}.
Additionally, non-reciprocity has been proposed as another way to induce synchronization blockade in quantum systems \cite{9y9w-lw92}.

\subsection{Transient and other forms of synchronization}\label{Sec:transient synch}

Beyond the quantum scenario inspired by the classical mutual synchronization of coupled self-sustained oscillators \cite{pikovsky2001synchronization} as detailed in the previous subsections,  synchronization phenomena have also been reported in different quantum settings. In this section, we review the phenomena of transient (metastable) and steady (stable) synchronization first reported in coupled harmonic oscillators \cite{giorgi2012quantum} and qubits \cite{Giorgi2013Spontaneous} interacting with a common environment. As we review in this section, transient and steady synchronization emerges in a variety of quantum systems subject to open system dynamics (anticipated in Sec. \ref{Sect:Liouv}), as induced by coupling to an environment, a measurement process, or an engineered stochastic process and is characterized by a strong temporal correlation or similitude in the dynamics of coupled (dissimilar) units \cite{giorgi2019transient,bucaalgebraic2022}.

The main signature of transient synchronization is the emergence of a collective oscillation that dominates the long-time dynamics of the system. Despite being detuned, these component units oscillate at the same frequency with a fixed phase relation, without any fine-tuning of the initial conditions. 
This manifests in the dynamics of local expected values $\langle \hat{O}_j(t)\rangle$ and multitime correlations (see Sec.~\ref{Sect:meas}) and can witness persistent quantum correlations \cite{giorgi2012quantum,Giorgi2013Spontaneous,manzano2013synchronization}. 

\begin{figure}
\includegraphics[width=0.42\textwidth]{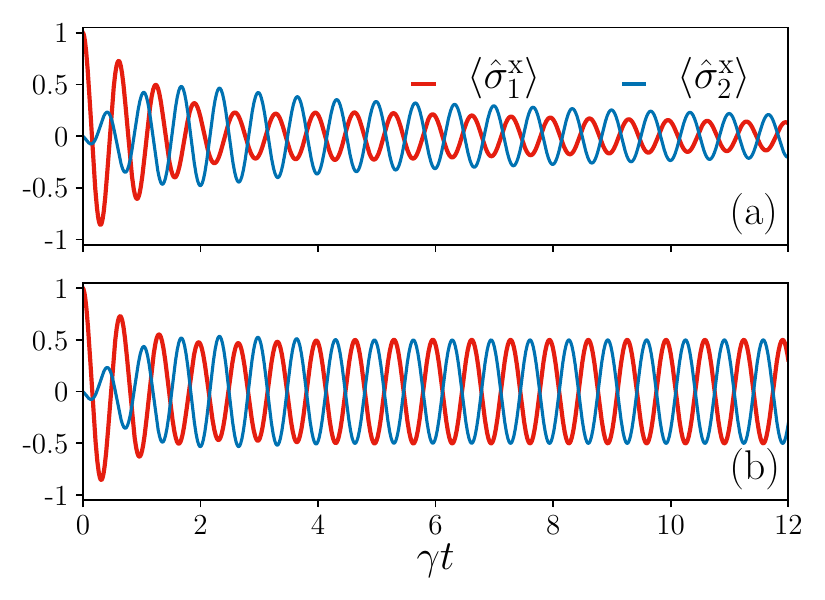}
    \caption{
    Transient (a) and steady (b) (anti-)synchronization in the dynamics of $\langle \hat{\sigma}_{1,2}^\mathrm{x}(t)\rangle$, emerging in a system of two coupled qubits with collective decay. The system is described by the Hamiltonian $\hat{H}=\frac{\omega_1}{2}\hat{\sigma}^\mathrm{z}_1+\frac{\omega_2}{2}\hat{\sigma}^\mathrm{z}_2+g(\hat{\sigma}^+_1\hat{\sigma}^-_2+\hat{\sigma}^+_2\hat{\sigma}^-_1)$ and jump operator $\hat{L}=\sqrt{\gamma}(\hat{\sigma}^-_1+\hat{\sigma}^-_2$). Using the definition $\omega_1=\omega_0+\frac{\delta}{2}$, $\omega_2=\omega_0-\frac{\delta}{2}$, the parameters have been fixed to $\omega_0=10\gamma$, $g=0.5\gamma$ and $\delta=0.8\gamma$ in (a), while $\delta=0$ in (b). This system was studied in detail in \cite{bellomo2017quantum}.}
    \label{fig:transient_sync}
\end{figure}

While transient synchronization can also be observed in the classical regime, it was first studied in quantum systems \cite{giorgi2012quantum,Giorgi2013Spontaneous} and displays a metastable character: It emerges after a short initial transient time, it manifests for a long window of time, and it eventually decays out when the system reaches its non-oscillatory stationary state, see Fig.~\ref{fig:transient_sync}(a).  This timescale separation originates from nonunitary processes that filter out most collective excitations in the system, except for the long-lived ones that are responsible for the synchronous response. Under some circumstances, this phenomenon can last indefinetly, beyond a metastable regime, leading to non-decaying oscillation 
as in Fig.~\ref{fig:transient_sync}(b). This is referred to as {\it steady} or stable synchronization and can be associated to the presence of symmetries and noiseless subsystems \cite{manzano2013avoiding,giorgi2019transient,cabot2018unveiling,bucaalgebraic2022}.

A formal description of transient and steady synchronization can be given in the framework of open quantum systems and is dictated by the spectrum of the Liouvillian for systems described by Lindblad master equations \cite{giorgi2019transient,bucaalgebraic2022}.  
In particular, in the simplest scenario —where synchronization is driven by a pair of slowly damped complex-conjugate eigenmodes— the real parts of their eigenvalues must satisfy
\begin{equation}
[\lambda^\text{Re}_{\bar{k}}|\ll|\lambda^\text{Re}_{k}|, \,\text{ for }\, \bar{k}\neq k,
\end{equation}
where $\bar{k}$ labels the synchronizing modes, while $k$ denotes all other eigenmodes with different frequencies. This pronounced spectral gap characterizes metastability and manifests itself in the system’s dynamics by producing a transient regime in which oscillations at a single dominant frequency prevail,
 as in Fig.~\ref{fig:transient_sync}(a), with synchronization frequency $[\lambda^\text{Im}_{\bar{k}}]$ generally different from other frequencies present. A more detailed discussion of the enabling mechanism of transient synchronization as a timescale separation induced by certain forms of dissipation can be found in a previous specialized review \cite{giorgi2019transient}.  
 
Transient synchronization induced by collective dissipation of two detuned system units interacting with a common bath was first reported for quantum harmonic oscillators \cite{giorgi2012quantum} and two-level systems \cite{Giorgi2013Spontaneous}, and extended to complex harmonic networks in Refs. \cite{manzano2013synchronization,cabot2018unveiling}. 
Beyond collective dissipation, transient synchronization can also be induced by local dissipation at inhomogeneous local decay rates, as first reported between a pair of qubits \cite{giorgi2016probing} (with applications in probing specific system properties as discussed in Sec.~\ref{sec:applications_sync}). The inhomogeneity of staggered local losses can also enable transient mutual synchronization in atomic lattices where, at difference from common scenarios, it is favored by an increase of spins detuning \cite{cabotquantum2019} (see also Sec.~\ref{sec:sync_and_TCs}). These forms of synchronization have been further reported in bio-inspired vibronic systems \cite{siwiak2019transient}, fermionic lattices \cite{tindallQuantum2020}, superconducting qubits \cite{cattaneo2021bath}, and chiral networks
\cite{lorenzo2022quantum}. Moreover, the relation with super/subradiance \cite{bellomo2017quantum}, or with exceptional points  \cite{cabot2021synchronization}, and non-Markovianity \cite{karpat2021synchronization} has also been established. 
 
In the special case $[\lambda^\text{Re}_{\bar{k}}]=0$, the synchronizing modes become undamped, the oscillatory dynamics persist indefinitely, leading to steady (stable) synchronization (Fig.~\ref{fig:transient_sync}(b)).  In this context,  \cite{bucaalgebraic2022} has built a general algebraic theory for synchronization, based on the spectral properties of the Liouvillian, and in which strong dynamical symmetries \cite{bucaNonstationaryCoherentQuantum2019} play a central role (see also Sec.~\ref{sec:sync_and_TCs}). Within this framework,  algebraic conditions for the emergence of (meta)stable synchronization in Lindblad systems were obtained \cite{bucaalgebraic2022}.  Such steady synchronization was first reported in \cite{Giorgi2013Spontaneous} for spins, and in \cite{manzano2013synchronization,manzano2013avoiding} for quantum harmonic oscillators, in which a connection with decoherence subspaces was unveiled. 
These first works showed that the synchronized response is generally accompanied by long-lived quantum correlations. Later, \cite{tindallQuantum2020} showed that steady synchronization can emerge due to the presence of strong dynamical symmetries. 
When these symmetries are broken perturbatively, transient synchronization is displayed instead \cite{tindallQuantum2020,bucaalgebraic2022}. 
The emergence of stable synchronization in different arrays has been reported when a suitable subset of qubits is subjected to amplitude damping, \cite{ccakmak2026synchronization} or under collisions at random times \cite{PhysRevA.106.022209}.

Synchronization has also been explored in chiral networks interacting in cascade with the environment \cite{lorenzo2022quantum}, where harmonic oscillators are coupled pairwise via unidirectional, traveling-wave channels rather than symmetric, bidirectional links \cite{Lodahl2017}. This leads to a non-reciprocal effect between paired oscillators, with each link assigning a distinct driver/target role to the two coupled nodes. As a result, the dynamics exhibit hybrid features of both mutual and driven synchronization, a behavior absent in conventional, non-directional networks, where the onset and pattern of synchronization depend strongly on network architecture \cite{lorenzo2022quantum}.
Non-reciprocal interactions in the presence of many-body effects are discussed in Sec.~\ref{sec:nonreciprocal}.
 
Beyond master equations, a minimal model of synchronization can be realized between two oscillators coupled to a tight-binding chain bath \cite{benedetti2016minimal}. Synchronization emerges between these nodes, which are at the boundary of the chain, during a time that lasts infinitely if the chain is infinite. The phenomenon is enabled by transient energy depletion and crosstalk, even between oscillators that are not directly coupled to each other. 
Transient synchronization has also been reported in collision models \cite{karpat2019quantum}, where non-Markovian effects have also been addressed \cite{karpat2021synchronization} and found to hinder the phenomenon.
Beyond coupling to an environment, transient and steady synchronization have also been induced by adding stochastic terms in a Hamiltonian spin system  \cite{schmolkenoise2022}, which, when averaging over stochastic realizations, effectively implements Lindbladian dynamics (see also Sec. \ref{sec:sync_and_TCs}). This has led to the experimental observation of steady synchronization in an array of superconducting qubits \cite{tao2025noise}. 
Stemming from the effective damping introduced by monitoring quantum systems, measurement can also induce synchronization \cite{schmolkeMeasurement-InducedQuantum2024}. 
Indeed, the variety of scenarios illustrates the ubiquity of this form of synchronization in open quantum systems.

\subsection{Trajectory approach to quantum synchronization}\label{subsec:fewbody_traj}

The possibility of quantum self-sustained oscillations and synchronization has also been explored within the framework of quantum trajectories, which arise, for instance, from the continuous monitoring of quantum systems (see Sec.~\ref{sec.IntroQS.trajectory}). At this level of description, the ability to perform measurements and acquire real-time information about the system dynamics enables the implementation of feedback strategies to enhance synchronization, as well as the analysis of the effects of thermal and quantum fluctuations.

\subsubsection{Entrainment along trajectories}

A natural setup to explore entrainment in single trajectories is that of a  QvdP oscillator subject to a harmonic drive (see Eq.~\eqref{eq:QvdP} and Eq.~\eqref{eq:qvdP_dr}). This possibility was considered in \cite{kato2021enhancement}, which focuses on continuous homodyne measurements (see Sec.~\ref{sec.IntroQS.trajectory}) of single-excitation decay processes implemented by the operator $\hat{L}=\hat{a}$ (recently reconsidered in \cite{nadolnyQuantum2026} with heterodyne trajectories). Continuous monitoring can enhance phase coherence relative to the case without measurement, as observed, e.g., in the concentration of the Wigner distribution (see Sec.~\ref{sec:phase_locking_measures}). The information gathered dynamically through the continuous measurements allows for the application of feedback operations on the oscillator, which can even further enhance the quality of the synchronization to an external signal \cite{kato2021enhancement}. A subsequent study \cite{shenEnhancingQuantum2023} considers similar effects and further demonstrates that squeezing can boost entrainment, especially near mean-field regimes. 

Phase locking to an external drive was recently explored in superconducting electrical circuits based on Josephson junctions \cite{hohe2025quantum}. These platforms naturally incorporate shot noise, that is quantum fluctuations, via the discrete nature of Cooper-pair tunneling in superconductors. The setup considered in \cite{hohe2025quantum} consists of a dc-biased Josephson junction that emits nonclassical light in a microwave cavity, effectively behaving as a nonlinear quantum oscillator capable of self-sustained oscillations. In the presence of an ac-drive, the phase of the oscillator stabilizes to nearly constant values, interrupted solely by rare $2\pi$ slips.

\subsubsection{Mutual synchronization along trajectories}

Several works have focused on synchronization between two quantum systems along quantum trajectories. Mutual synchronization of two QvdP oscillators (see~Eq.~\eqref{eq:QvdP}), dissipatively coupled through a mixed jump operator, has been systematically investigated in \cite{eshaqi-sani2020synchronization}.  In the limit of an infinite number of excitations, an effective classical behavior emerges, captured by mean-field equations of motion \cite{eshaqi-sani2020synchronization,leeEntaglementTongueAndQuantum2014}. 
Genuinely quantum behavior instead emerges in the low excitation limit \cite{lee2013quantum,leeEntaglementTongueAndQuantum2014}, which has been studied through quantum master equations in \cite{leeEntaglementTongueAndQuantum2014} and at the level of homodyne and heterodyne trajectories in \cite{eshaqi-sani2020synchronization,katoInstantaneousPhaseSynchronization2021,nadolnyQuantum2026}. 

\begin{figure}
\includegraphics[width=0.48\textwidth]{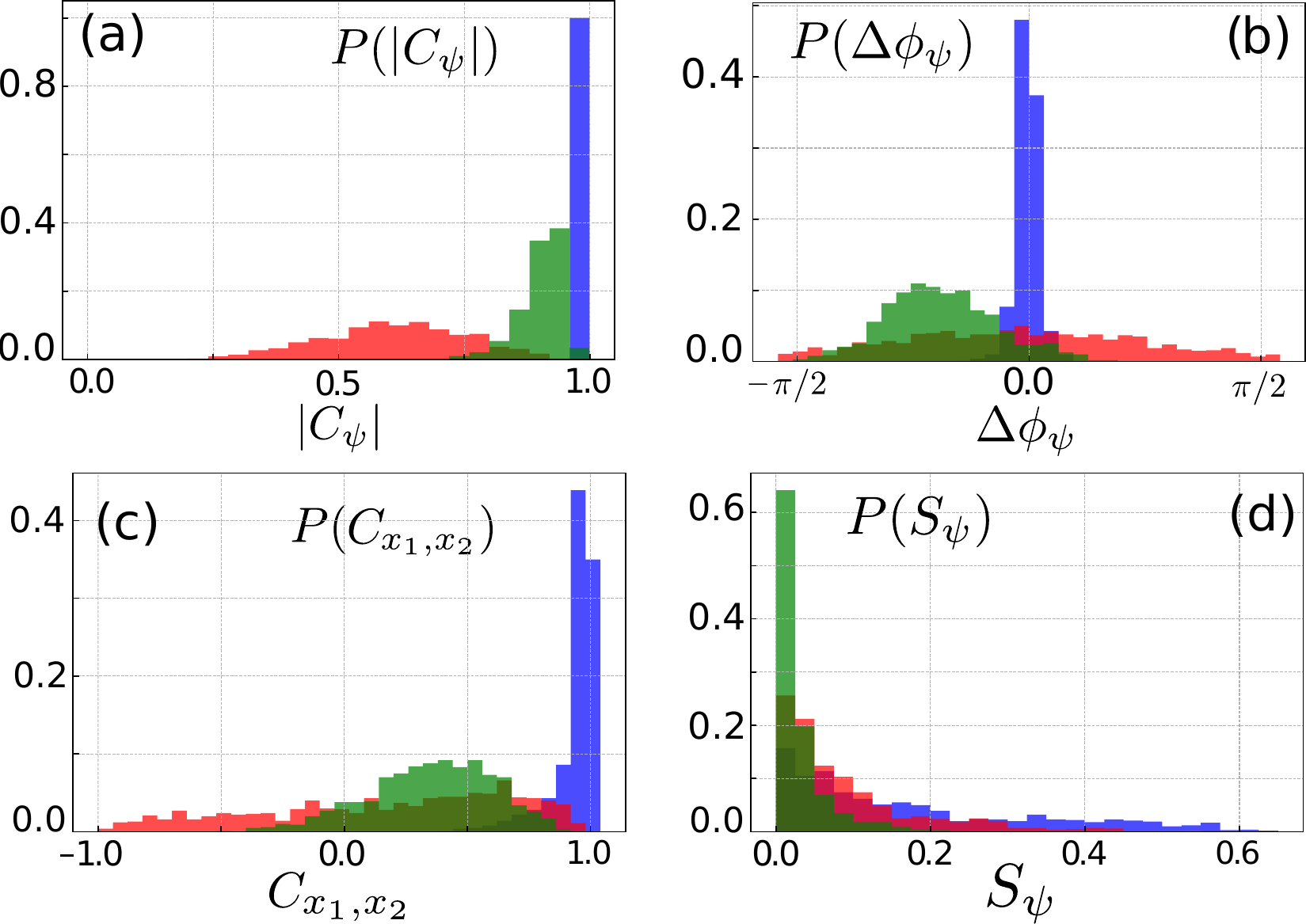}
    \caption{Probability density obtained from quantum trajectories of modulus (panel (a)) and phase (panel (b)) of the complex correlator $C_\psi$, of the Pearson correlation coefficient $C_{x_1,x_2}$ (panel (c)) and of the entanglement entropy $S_\psi$ (panel (d)), for different parameter regimes: $V=100\gamma_1$, $\Delta \omega=\gamma_1$ (blue bars), $V=5\gamma_1$, $\Delta \omega=\gamma_1$ (red bars), and  $V=20\gamma_1$, $\Delta \omega=20\gamma_1$ (green bars). 
    Figure adapted from \cite{eshaqi-sani2020synchronization}.}
    \label{fig:few_body_trajectories}
\end{figure}

In order to discuss how quantum synchronization manifests in trajectories, we briefly discuss the case of two vdP oscillators with independent linear gain at rate $\gamma_1$ and nonlinear loss at rate $\gamma_2$ (see Eq.~\eqref{eq:QvdP}). 
These oscillators are characterized by natural frequencies $\omega_1$ and $\omega_2$ with a detuning of $\Delta\omega = \omega_1 - \omega_2$ between them. 
The two oscillators are coupled through a dissipative channel $\hat{L}=\sqrt{V}(a_1 -e^{i\theta}a_2)$, where $a_i$ are the corresponding bosonic modes. 
Using quantum trajectories, one can sample the probability density of suitable measures of synchronization. 
Examples are shown in Fig.~\ref{fig:few_body_trajectories}(a-c) for three different parameter regimes corresponding to near perfect synchronization (blue bars), poor synchronization within the classical synchronization regime (red bars), and poor synchronization outside the classical synchronization regime (green bars). 
Specifically, the red bars are obtained for parameters that lie inside the corresponding classical Arnold tongue, which for this model can be obtained in the $\Delta \omega-V$ plane (similar to Fig.~\ref{fig:Introduction-self_sustained}(e)), and the green bars are related to poor synchronization outside such a classical Arnold tongue.  
For near-perfect synchronization, the distributions of the modulus ($|C_\psi|$) and phase ($\Delta\phi_{\psi}$) of the normalized correlation $C_\psi(t)$ (see Eq.~\eqref{eq:meas_quant_correl}), and of the Pearson coefficient $C_{x_1,x_2}$ (see Eq.~\eqref{eq:meas_temp_correl}) are sharply peaked around their mean.  
In the other two regimes, the distributions are significantly flatter, signaling poor synchronization even when mean indicators may suggest partial phase locking. 
Fig.~\ref{fig:few_body_trajectories}(d) shows the probability distribution of bipartite entanglement $S_\psi$ in the quantum trajectories. 
Strongly synchronized trajectories are characterized by distributions with fat tails extending toward large entanglement values. 
In contrast, trajectories displaying poor phase locking are associated with lower correlations.

Currents emitted by two QvdP oscillators independently coupled to two environments can, in principle, be experimentally measured. In~\cite{nadolnyQuantum2026}, the authors observe that this information is useful to estimate the phase relation between the oscillators as well as to estimate the spectral content of the current two-time correlation functions (see Sec.~\ref{sec:emission_measures}).  In quantum trajectories of coupled QvdP oscillators, mutual synchronization can also be induced by conditional photon detection when the emission channels are described by mixed jump operators \cite{katoInstantaneousPhaseSynchronization2021}. Even if the average open dynamics describes completely uncoupled oscillators, in quantum trajectories, the phase of the correlator can assume exactly the value $0$ or $\pi$ when photons are detected.

Mutual synchronization in quantum trajectories occurs in a variety of different setups. It was observed in a qubit strongly coupled to a driven-dissipative oscillator \cite{zhirov2008synchronization}:   beyond a critical coupling strength, the qubit locks to the oscillator's frequency and the system displays bistability, with markedly different long-lived metastable synchronized states. A similar phenomenology was observed in a setup consisting of two mechanically-coupled optomechanical systems \cite{weiss2016noise}. These platforms allow for the efficient exploration of the effects of quantum and thermal noise, and may reach experimental regimes where quantum noise dominates. 
Josephson junction devices have also been exploited to investigate mutual synchronization \cite{hohe2025quantum}. It observes that in these systems synchronization emerges through mutual frequency pulling: each oscillator provides a reference frequency for the other, causing their emission frequencies to converge. 

Another work addresses synchronization in optomechanical systems and considers an approach through Bohmian trajectories \cite{liAnalyzingQuantumSynchronization2022}. The quantum dynamics reduces to a classical trajectory ensemble governed by a classical Hamiltonian supplemented by a quantum potential.
Here, non-Gaussian initial states were investigated, displaying trajectories exhibiting alternating periods of enhanced and suppressed synchronization, as compared to the behavior of more classical coherent states, leading to significantly larger fluctuations of the Pearson's correlation factor.

\subsection{Experiments} 
Quantum synchronization in few-body systems has been experimentally demonstrated in a wide array of platforms. 
The earlier works \cite{PhysRevResearch.2.023026, PhysRevLett.125.013601} focused on realizing limit cycles in spin-1 systems (see Sec.~\ref{sec: finite-level}), entraining them to external classical fields and demonstrating quantum synchronization blockade. In \cite{PhysRevResearch.2.023026}, a digital simulation of quantum synchronization of a spin-1 system was realized on an NISQ computer consisting of five qubits in a star-shaped topology. A spin-1 subspace was constructed with two of these five qubits. 
The other three qubits acted as ancilla qubits, which helped realize the dissipation necessary for realizing a limit cycle. 
Particularly, dissipation was engineered using control unitaries applied on the qubits forming the spin-1 system and the ancilla qubits, followed by measurements and reset of the ancillas. 
Further, the drive Hamiltonian on the spin-1 subspace was implemented by controlled two-qubit gates. 
The synchronization measure, estimated from the experimentally obtained density matrix, was used to verify typical features of quantum synchronization. In \cite{PhysRevLett.125.013601}, the authors demonstrated synchronization in a dilute ensemble of laser-cooled spin-1 $^{87}$Rb atoms.
Here, the spin-1 system was formed by the hyperfine ground state manifold of the $^{87}$Rb, on which dissipation was engineered by coupling the states from this ground state manifold to an excited state. 
The coherence between the hyperfine levels was retrieved from the optical fields \cite{PhysRevLett.86.783}, which then served as a synchronization measure exhibiting Arnold tongue behavior. 

Phase synchronization in a four-level system consisting of a pair of interacting nuclear spins at room temperature was observed in \cite{PhysRevA.105.062206}. 
In particular, the authors studied a sample of sodium fluorophosphate molecules dissolved in D$_2$O to experimentally realize a nondegenerate 4-level system from the spin-1/2 nuclei of $^{19}$F and $^{31}$P in the sample. 
In the absence of external influences, the system is in a thermal state, and the Husimi distribution lacks phase localization.
 However, when subjected to an external drive resonant with the transition of the $^{31}$P nucleus, the system exhibits localization in the phase-space and an Arnold tongue upon detuning.
An interferometric method to measure the phase-space Husimi distribution of a quantum state was developed in this work, thereby bypassing the need for expensive quantum state tomographic measurements.

Recently, quantum synchronization was also demonstrated in trapped-ion systems.  
In \cite{sciadv.ady5649}, the axial motion of the Ca$^+$ ion in a Paul trap was engineered to implement a QvdP (see Sec.~\ref{sec:QvdP}) and its entrainment to an external drive was studied. 
The Wigner function of the motion degree of freedom was obtained by mapping it to the spin population of the internal degree of freedom of the ion, which was further measured. 
When an external drive was applied, the QvdP became entrained, as captured by the mean resultant length of a circular distribution, which 
exhibited an Arnold tongue. 
Further, by applying an additional squeezing drive, they showed that synchronization could be enhanced by small squeezing in a direction perpendicular to the external drive, as well as by linear dissipation.
In another experiment \cite{liu2025observation}, mutual synchronization between two QvdPs formed by the motional modes of a two-ion crystal composed of $^{40}$Ca$^+$ and $^{44}$Ca$^{+}$ ions trapped in a Paul trap was studied. 
Nonlinear dissipation and collective dissipative coupling between the modes were engineered using ancilla qubits formed by the internal degrees of freedom of the ions.  
The relative phase between the oscillators showed locking, which was only accessible through the joint readout of both oscillators. 
Furthermore, trapped $^{171}$Yb$^{+}$ ions were used to demonstrate synchronization in a single qubit system along quantum trajectories and an Arnold tongue was also observed upon tuning the drive parameters \cite{PhysRevResearch.5.033209}.

\section{Synchronization in Quantum Many-Body Systems}\label{sec:Many_Body_QSynch}

We have so far focused on the theoretical frameworks used to characterize quantum synchronization and their realizations in few-body coupled quantum oscillators and spin systems. 
These systems constitute the quantum counterpart of Huygens' seminal synchronization paradigm, generalized to the quantum regime. 
However, synchronization acquires qualitatively new features in many-body settings. 
As is well established in classical nonlinear dynamics, the collective synchronization of a large ensemble of interacting oscillators can emerge through a dynamical phase transition between incoherent and phase-locked states.
This transition to a phase-locked state is provided by the {\it Kuramoto model}~\cite{acebron2005,gupta2018} formulated for large ensembles of coupled phase oscillators as discussed in Sec.~\ref{sec:classical_entrain_mutual}. In the following section, we will review how this paradigmatic model has provided deep insight into many-body synchronization at the quantum level and inspired models that exhibit similar features (for a review on many-body open systems see~\cite{SciPostPhysLectNotes.99}). 
However, many-body synchronization in quantum systems extends beyond the Kuramoto model to systems that are topologically protected and those that exhibit metastable oscillations. 
The timescale of these metastable oscillations can also scale with system size, leading to continuous time crystals that break the continuous time-translation symmetry of the system. 
These interesting phenomena will also be discussed below.

\subsection{The Kuramoto model in quantum systems}
The Kuramoto model has recently been explored in the quantum setting. In some cases, it has aided the understanding of quantum dynamics and in others it has led to the development of generalized quantum models. These generalized quantum models generally reduce to the Kuramoto model in the classical limit. In \cite{witthaut2017classical}, the authors analyzed a model of an interacting Bose-Hubbard-like quantum many-body system that exhibits a delocalization-to-strong-correlation transition. 
It was found that the behavior of the entanglement was directly linked to the onset of synchronization, as in the Kuramoto model. 
The Kuramoto model was also used as an external drive for a quantum system~\cite{Bastidas_2025}, and synchronized phases were found in cavity-QED setups~\cite{valencia-tortoracrafting2023}.

Studies have also focused on building both semiclassical \cite{hermoso2014synchronization} and fully quantum \cite{delmonte2023quantum} extensions 
of the Kuramoto model. 
These studies consistently show that quantum noise raises the critical coupling strength required for synchronization relative to the classical case.
A way to study the Kuramoto model for a quantum rotor is to follow the Caldeira-Leggett approach~\cite{caldeirainfluence1981,leggett1987dynamics} and start from a system-environment model that reduces to the Kuramoto model in the classical limit. The semiclassical limit was considered in \cite{hermoso2014synchronization} where a quantum Langevin equation was derived. The solution of the associated dynamics leads to a shift to a larger value of the coupling. As expected, the presence of quantum fluctuations further suppresses synchronized motion. This is even more evident as one enters the quantum regime, as studied in \cite{delmonte2023quantum}, where, in the overdamped regime, the quantum model exhibits a phase transition from an incoherent motion phase to a synchronized one. The phase transition here occurs at any temperature including the zero temperature. 
Quantum versions of the Kuramoto model were further explored in different contexts. In~\cite{wangsachdev2020} it was employed to analyze the superconducting phase fluctuations in a spin-1/2 version of the Sachdev-Ye-Kitaev model. In~\cite{lohequantum2010} a network of quantum oscillators was studied, in which quantum states are distributed among nodes connected by means of unitary transformations, resulting in the onset of synchronization and a non-abelian generalization of the Kuramoto model.

Kuramoto-like models have further emerged as versatile frameworks in quantum many-body physics, providing insights into quantum network synchronization \cite{lohequantum2010,deville2019synchronization,antonelli2017model}, entanglement generation \cite{witthaut2017classical}, conditional Ramsey spectroscopy \cite{xu2015conditional}, and superconductivity \cite{velasco2021unconventional,wangsachdev2020}. 
It has been useful in understanding synchronization in coupled optomechanical systems \cite{cabot2017dynamical,ghang-geng2019quantum,sun2024quantum,weiss2016noise,ludwig2013quantum,holmes2012synchronization,heinrich2011collective}. Kuramoto models have also inspired a body of work in light-matter interacting systems such as quantum dipolar gases \cite{zhu2015synchronization} and spin-ensembles \cite{xu2014synchronization}, where two entities each with a macroscopic number of oscillators, synchronize with each other. This phenomenon, called \textit{macroscopic synchronization}, is discussed below.

\subsection{Macroscopic synchronization} 
Macroscopic synchronization is an emergent property of complex quantum systems and is characterized by the collective synchronization of networks of quantum oscillators. 
A key feature that distinguishes macroscopic synchronization is the preservation of quantum coherence even as the number of oscillators increases, thereby exhibiting phenomena such as synchronization blockade absent in their classical counterparts \cite{zhu2015synchronization, nadolny_macroscopic_2023}. In this context, an important open question is whether there exist general properties of dissipative processes that can robustly generate macroscopic quantum coherence. 
A constructive method to generate such synchronized quantum states is provided in \cite{PhysRevA.104.012410}. 
Here, a quantum channel is defined to be a \textit{quantum state synchronizer} if and only if, for any arbitrary input state, it gives a quantum state whose reductions to single-party states are all identical. 
The asymptotic scaling of multipartite mutual information, which can be a measure of quantum synchronization (see Sec.~\ref{Sect:meas}), in permutationally invariant systems can be captured using the standard mean-field description as proposed in \cite{zwfx-m4nq}.

Permutationally invariant systems have been widely discussed in the context of macroscopic synchronization despite the fact that the set of all synchronized quantum states is larger than the set of permutationally invariant states \cite{xu2014synchronization, zhu2015synchronization, nadolny_macroscopic_2023, peng2025macroscopic}.  
Investigations of macroscopic synchronization, however, are often hindered by the exponential growth of the Hilbert space with system size, necessitating methods that reduce computational complexity, such as adiabatic elimination, mean-field theory, or cumulant expansions.

One of the earliest efforts to understand macroscopic synchronization in physical systems is the study of two driven ensembles of two-level atoms \cite{xu2014synchronization}. 
These ensembles were considered to be detuned both from one another and from the lossy single-mode cavity to which each ensemble is collectively coupled. 
In the bad cavity limit, the cavity can be adiabatically eliminated, allowing the two ensembles to become dissipatively coupled. 
When the effective coupling exceeds the detuning between the oscillator ensembles, the system undergoes a synchronization transition at a critical pump strength. 
The linewidth of the output field exhibits a sharp enhancement at this transition and is ultranarrow in the synchronized phase, indicating rigid phase locking between the two ensembles. 
We note that this synchronization dynamics has also been explored experimentally in \cite{weiner2017phase}.

Macroscopic synchronization can also be achieved in dipolar systems, where superradiant dipoles can be engineered to have a synchronized steady state in the presence of a sufficiently large incoherent repumping rate \cite{zhu2015synchronization}. 
The repumping counteracts the fast decay in superradiant systems by maintaining population inversion, thereby stabilizing a highly synchronized state. This mechanism also appears in other collective spin platforms, where collective decay arises from a lossy cavity mode \cite{xu2015conditional} or from the normal modes of a trapped‑ion crystal \cite{shankar2017steady-state}.
In a spatially inhomogeneous array of incoherently driven strongly coupled dipoles, the incoherent collective photon decay rate of the dipoles was shown to be the most decisive factor in synchronization \cite{zhu2015synchronization}. 
This is similar to \cite{peng2025macroscopic} where the authors examined synchronization between a bosonic mode and a molecular collective mode in an optomechanical cavity, and showed that synchronization increases with increasing cavity decay rate. 
Moreover, when cooperative photon decay follows a power law dependence on the distance between any two dipoles, the nature of synchronization changes qualitatively \cite{zhu2015synchronization}. 
In particular, the system transitions from having global synchronization to having spatially localized synchronized clusters as the decay exponent becomes comparable to the spatial dimension of the dipole array.

Macroscopic manifestations of quantum synchronization blockade have also been observed in ensembles of coupled three-level atoms where each atom is incoherently driven to $\ket{1}$ with rates $\gamma_+$ and $\gamma_-$ from levels $\ket{0}$ and $\ket{2}$, respectively \cite{nadolny_macroscopic_2023}. 
The asymmetry in the energy differences between the transitions $\ket{1}\leftrightarrow\ket{2}$ and $\ket{2}\leftrightarrow\ket{3}$ is quantified by the parameter $K$. 
In the absence of any asymmetry ($K = 0$), the blockade occurs when the coupling strength exceeds a threshold and $\gamma_+/\gamma_- = 1$, much like a single three-level atom coupled to an external drive \cite{PhysRevLett.121.053601, PhysRevA.99.043804} (see also Sec.~\ref{sec: finite-level}). 
However, unlike the latter case, blockade is not lifted for every non-zero value of the asymmetry parameter, $K$. 
Specifically, the system synchronizes for $K<0$, but blockade persists for $K>0$ due to phase frustration arising from a single atom attempting to anti-align with the mean-field of the ensemble. 
The authors further considered the synchronization between two ensembles of three-level atoms, each detuned from the other by an amount $\delta$. 
Here, detuning plays an additional role in determining the regimes of quantum synchronization blockade. 
Synchronization is observed only if $|K| \sim |\delta|$, that is, in regimes where the dominant transitions are resonant. 
Each oscillator now reacts to the mean field of the entire group, leading to another case of phase frustration that imposes an additional blockade for $K > \delta$.

\subsection{Synchronization beyond the Kuramoto model} \label{sec:beyond_kuromoto}

Synchronization in many-body quantum systems goes beyond the applications and generalizations of the Kuramoto paradigm. Indeed, beyond limit-cycles, we have already presented different metastable and stable quantum synchronization in Sec.~\ref{Sec:transient synch}, where networks of oscillators and qubits synchronize under the action of dissipation into the environment, measurement
process, or for on average in stochastic ensembles.
More broadly, a formal approach to characterize synchronization was put forth in \cite{bucaalgebraic2022}. 
According to this approach, a many-body system with $N$ subsystems is said to be synchronized if for local observables $\hat{O}$ at two subsystems $i$ and $j$, 
\begin{equation}
    \langle  \hat{O}_i(t)\rangle = \langle  \hat{O}_j(t)\rangle
    \;\;\; \mbox{with}\;\;\;\;
    \frac{d}{dt} \langle  \hat{O}_i(t)\rangle \ne 0
    \label{condition-synchro}
\end{equation}
for some or all pairs $i,j$. 
The algebraic conditions for the existence of oscillatory solutions, corresponding to imaginary eigenvalues of the Lindbladian, have also been characterized \cite{bucaalgebraic2022}. This says that the Lindbladian has an eigenstate $\hat \sigma$ with purely imaginary eigenvalue, $\mathcal{L}[\hat \sigma] = i \lambda \hat \sigma$ with $\lambda \in \mathbb{R}$, if there is a dynamical symmetry defined by the operator $\hat A$, satisfying the conditions \cite{bucaalgebraic2022}:
\begin{eqnarray}
 [\hat L_i, \hat A] \hat \rho_{\rm ss}&=&0, \nonumber \\
 \left( i[\hat H,\hat A] + \sum_i [\hat L_i^\dagger, \hat A] \hat L_i\right) \hat \rho_{\rm ss} & = &-i\lambda \hat A \hat \rho_{\rm ss},
\end{eqnarray}
where $\hat \rho_{\rm ss}$ is the steady state of the Lindbladian ($\mathcal{L}[\hat \rho_{\rm ss}] = 0$) and $\hat \sigma = \hat A \hat \rho_{\rm ss}$.
The criterion encompasses the well-known case of decoherence-free subspaces, where there are subspaces immune to dissipation (formally, dark states $|D_j\rangle$ with $\hat L_i |D_j \rangle = 0$) over which the Hamiltonian term can induce a coherent dynamics. 
More importantly, however, it also describes intricate scenarios where non-stationary states are not decoupled from the environment ($\hat L_i \hat \sigma \hat L_j^\dagger - \{\hat L_i^\dagger \hat L_i,\hat \sigma\}/2 \neq 0$), meaning that the persistent dynamics now arise from the competition between coherent Hamiltonian dynamics and dissipation. 
Lastly, given a full-rank steady state, the criterion takes the form of a ``strong dynamical symmetry'' \cite{bucaNonstationaryCoherentQuantum2019}, revealing a potential route that generalizes many-body scars \cite{serbynQuantumManybodyScars2021} to dissipative systems. 
Specifically, in the absence of dissipation, the criterion resembles a spectrum-generating algebra \cite{Bohm:1988hq,serbynQuantumManybodyScars2021}, with $\hat A$ acting as a ladder operator that generates a tower of equidistant eigenstates $\hat \sigma_{nm} = \hat A^n \hat \rho_{\rm ss} (\hat A^{\dagger})^m$ with $\lambda_{nm} \propto (n-m)$, a trait that is characteristic of many-body scars and ergodicity-breaking in closed systems.  

This formalism distinguishes between transient (metastable) synchronization, in which observables synchronize for an intermediate time frame $t \in [\tau, T]$ ($T \gg \tau$), and steady-state (stable) synchronization, where they synchronize for all $t \geq \tau$ where $\tau$ is some transient time period. More generally, the underlying mechanisms behind these two forms of synchronization are not exclusive to many‑body systems and can arise in a variety of settings (see Sec. \ref{Sec:transient synch}).
Table \eqref{table.synchronization} summarises the different forms and classification of synchronization, with a further discrimination based on the extent of synchronization among the subsystems. 
Specifically, synchronization in many-body systems can lead to a fraction of the system synchronizing due to environmental properties or interactions among its constituents. Depending on the system under consideration, the synchronized fraction of the system can be finite or scale with the system size. Alternatively, the entire system could synchronize at finite system sizes, or it could arise as a consequence of spontaneous symmetry breaking in the thermodynamic limit.

\begin{table*}
    \centering
    \begin{tabular}{|c|c|c|c|}\hline
         \multicolumn{4}{|c|}{ 
         \makecell{ \\ \textbf{Synchronization:} \quad  $\langle \hat O_i (t)\rangle = \langle \hat O_j (t)\rangle $,\quad  $\frac{d}{dt} \langle \hat O_i \rangle \neq 0$ \vspace{0.3cm} } 
         } \\ \hline
         \textbf{Stability}  &  \textbf{Extension} &  \textbf{Robust} & \textbf{Complete} \\ \hline 
         \makecell[l]{$\cdot$ Stable: $\forall t \geq \tau$   \\   \hline $\cdot$ Metastable: $\forall t \in [\tau,T], T \gg \tau$}
  & 
  \makecell[l]{$\cdot$ Finite: $i,j = 1,...,k$ \\ \hspace{1.05cm}  $k < N$ \\ \hline $\cdot$ Full ensemble: $\forall i,j$}  
  &  
 \makecell[l]{ $ \hat O_i(t) = \hat P_{i,j} \hat O_j(t) \hat P_{i,j}$ \\ $\hspace{0.5cm}  \frac{d}{dt}\hat O_i \neq 0$ } 
  &  
  \makecell[l]{ Robust $\forall \hat O$} 
  \\ \hline 
    \end{tabular}
    \caption{Different forms of synchronization discussed in \cite{bucaalgebraic2022} for identical coupled subsystems.
    Two subsystems are synchronized once they share an observable $\hat O_{i,j}$ whose expectation value satisfies the condition given by Eq.\eqref{condition-synchro}. (Stability) Synchronization can be either stable, arising after a characteristic time $\tau$ and lasting indefinitely in time, or metastable, persisting only until a finite time $T$ (see also Sec.~\ref{Sec:transient synch}). 
    (Extension) The number of synchronized pairs in the system can be either finite or encompass the full ensemble of subsystems. (Robustness) If Eq.\eqref{condition-synchro} extends to the operator level, the synchronization is robust, with $\hat P_{i,j}$ the permutation operator exchanging subsystems $i$ and $j$. (Completeness) Moreover, if this condition holds for all observables, the synchronization is called complete. 
    }
    \label{table.synchronization}
\end{table*}

An example for when a fraction of a finite system gets synchronized was proposed in~\cite{schmolkenoise2022} and experimentally detected in~\cite{tao2025noise}. They~\cite{schmolkenoise2022} considered the dynamics of an $XY$ spin chain in a transverse field and found that, in the presence of white noise, stable synchronization occurs between the magnetization of the two ends of an arbitrary-length quantum $XY$ model with a transverse field. 
Here, a closed system is subject to noise, or measurements, described by a Hermitian operator. When the dynamics admits decoherence-free subspaces associated with different eigenvalues of the noise operator, the system probabilistically selects one of these subspaces and features noiseless asymptotic oscillations. These oscillations can determine synchronization (or anti-synchronization) of different degrees of freedom of the system, which can be quantified via the Pearson correlation coefficient \cite{tao2025noise} (see also Sec. \ref{Sect:meas}). 
Here, the time scale at which synchronization sets in can be controlled by tuning the noise amplitude, and the optimal speed is closely linked to the Lieb–Robinson bound for the propagation of information through the chain. These theoretical predictions were later experimentally confirmed in~\cite{tao2025noise} using a chain of superconducting transmon qubits. In this example, synchronization is intimately linked to entanglement between the two endpoint spins, thus confirming the genuinely quantum nature of the effect. 

Synchronized motion between boundary observables can also rely on topological protection. Topological quantum synchronization of fractionalized spins in an AKLT chain, subject to an engineered dissipation, has been investigated in \cite{wachtlertopological2024}. The local spin-1 degrees of freedom at the two ends of the chain are anti-synchronized, and their topological nature is reflected in their robustness to perturbations that break the inversion or permutation symmetry. Inherent topological protection in quantum synchronization was extensively explored~\cite{wachtlertopological2023} in a variety of networks of van der Pol oscillators, reinforcing the observation that synchronization at the edges that is protected over a wide range of disorder is a generic feature of these systems.

A synchronized motion in a many-body setting can also be realized by properly tailoring the dissipation's type and spatial dependence. Aiming also at an experimental realization with atoms in a 1D  optical lattice, \cite{cabotquantum2019} analyzed the dynamics of a dimer lattice governed by an $XX$ model in a transverse field and subjected to a staggered, local dissipation. Here, the inhomogeneity of the losses, with a well-defined wave-vector, is the key to achieving synchronized motion (which appears at intermediate times since the Hamiltonian of the system commutes with the total number of excitations; therefore, due to losses, the steady state is the one with all the ions in the ground state). Optimizing the parameters, the intermediate regime can be extended sensibly. 
Synchronization has also been studied in networks of quantum systems, with applications ranging from the distribution and sharing of quantum states across a network \cite{li2017network} to the emergence of global or partial synchronization through local tuning of a single node. It has also been shown that two initially separable and uncoupled nodes can be synchronized and entangled by embedding them in a common network \cite{manzano2013synchronization}.

Beyond stable and metastable synchronization (see Table~\ref{table.synchronization}), there can be situations in which the synchronized transient may become longer with system size, eventually developing into a limit cycle \cite{iemini_boundary_2018,  mattesEntangledTimecrystalPhase2023, duttaQuantum2025, li2026generalinterpretation}. This limit cycle can be a consequence of a true spontaneous symmetry breaking in the thermodynamic limit. A particularly striking example of this phenomenon is the emergence of time crystals, phases of matter arising from the breaking of time-translation symmetry in open quantum systems \cite{iemini_boundary_2018}. In the following sections, we will discuss time crystals and their relationship to synchronization in detail.

\subsection{Synchronization and time crystals}\label{sec:sync_and_TCs}

Spontaneous symmetry breaking has long been a foundational principle in physics, revealing how order and structure can emerge from underlying symmetries of a system \cite{Goldenfeld_1992,Sachdev_2000}.  It occurs across a wide range of energy scales, from the complex behavior of condensed-matter systems to high-energy physics. Although traditionally associated with spatial arrangements, such as crystallization from liquid phases or magnets aligning their spins, symmetry breaking can also extend into the realm of time itself. 
The concept of breaking time-translation invariance has sparked both curiosity and controversy 
\cite{Wilczek2012,PhysRevLett.109.163001,PhysRevLett.110.118901,PhysRevLett.111.070402,PhysRevLett.111.029301, Nozieres_2013,Volovik2013,PhysRevA.91.033617} since its first proposal in \cite{Wilczek2012}. 
Unlike conventional crystals, which repeat periodically in space, time crystals (TCs) exhibit persistent, spontaneous oscillations in time, defying conventional notions of equilibrium and thermalization. 
Soon after the proposal of time crystals, a no-go theorem~\cite{Watanabe2015} ruled out their existence for closed systems (with short-range interactions) in thermal equilibrium. 
Consequently, the research shifted focus towards systems out of equilibrium. 
A significant advancement was first made with the discovery of Floquet time crystals in systems that exhibit many-body localization \cite{PhysRevLett.116.250401, PhysRevLett.117.090402}. 
In these systems, certain observables were shown to oscillate at higher multiples of the driving period, thereby breaking the discrete time-translation symmetry imposed by the external drive; see \cite{Zaletel2023,Sacha_review} for comprehensive reviews of this topic.

On the contrary, continuous time-translation symmetry can be spontaneously broken in many-body open systems~\cite{iemini_boundary_2018,bucaNonstationaryCoherentQuantum2019}. The defining hallmark of a continuous time crystal is the spontaneous transition of an interacting many-body system into robust, self-sustained oscillations in response to a time-independent drive. Specifically, it emerges in a time-independent system, given that
\begin{enumerate}[(a)]
    \item there exists a macroscopic observable that spontaneously breaks the time-translation symmetry and exhibits persistent periodic oscillations,
    \item the oscillations withstand small perturbations to the system’s dynamics,
    \item the oscillations stem from the collective interactions among the system’s constituents, persisting indefinitely in time only in the thermodynamic limit.
\end{enumerate} 
 The collective oscillations in the system have been linked to the emergence of purely imaginary eigenvalues in the eigenspectrum of the Lindbladian. This constitutes coherence synchronization in many-body open quantum systems \cite{PhysRevA.105.L020401}. This also fits in with definition of synchronization given in Eq.~\eqref{condition-synchro}, where the synchronized transients in finite systems become steady-state synchronization in the thermodynamic limit.

\begin{figure}
\includegraphics[width=0.49\textwidth]{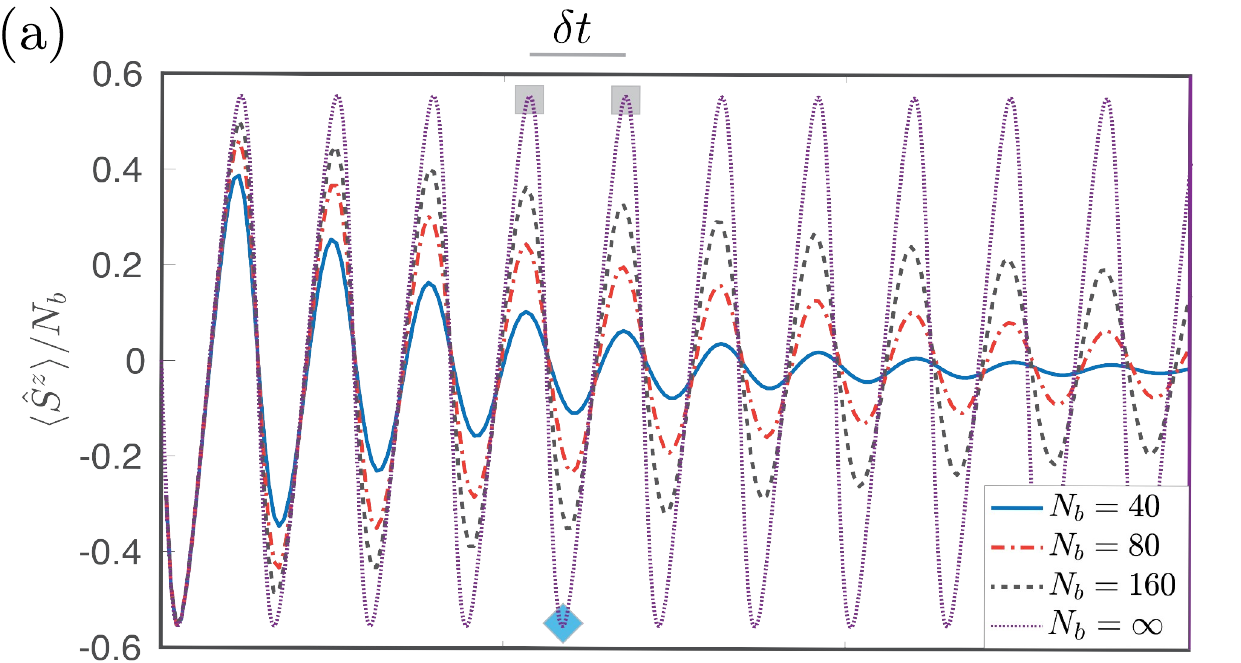}
\includegraphics[width=0.235\textwidth]{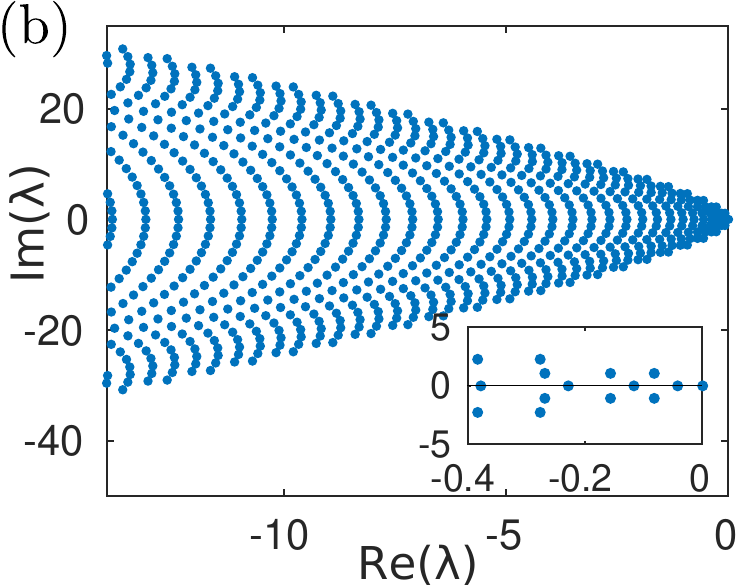}
\includegraphics[width=0.235\textwidth]{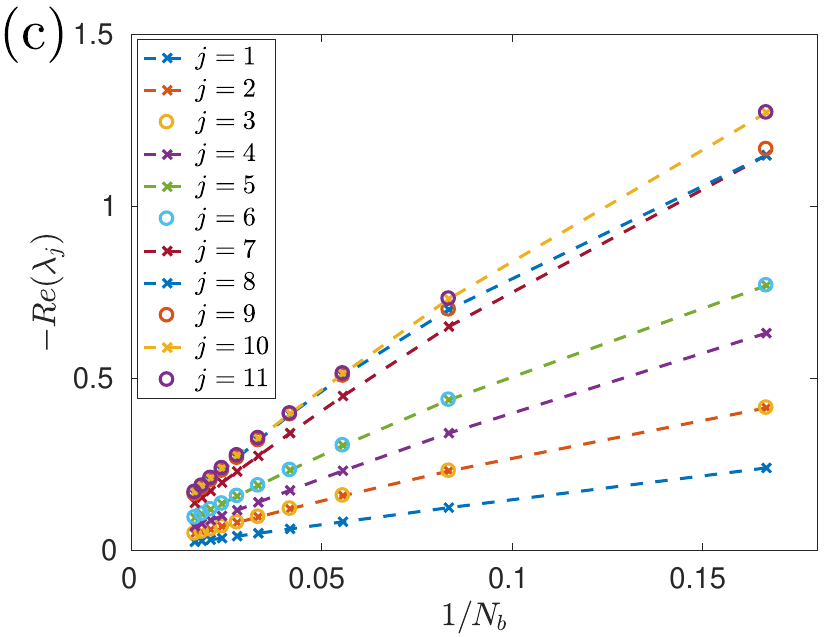}
    \caption{
    \textbf{(a)} Dynamics of a macroscopic observable (here, total spin magnetization) in a CTC, for different system sizes.  The observable exhibits oscillations whose lifetime increases with system size. In the asymptotic limit ($N \rightarrow \infty$), spontaneous symmetry breaking occurs leading to persistent self-sustained oscillations.
    \textbf{(b)} The Lindblad spectrum for a finite system size, with the existence of eigenvalues close to $\lambda = 0$ with a nonnull imaginary term.
    \textbf{(c)}  In a finite-size scaling, these eigenvalues have a real part  vanishing in the thermodynamic limit.
    Figures (a-c) correspond to the spin-model of Eq.~\eqref{eq.master.equation.BTC} with $\omega_o/\kappa = 1.5$, and are reprinted from \cite{iemini_boundary_2018}.
    }
    \label{fig:BTC_schematic}
\end{figure}

An illustration of the typical macroscopic dynamics in a CTC is shown in Fig.~\ref{fig:BTC_schematic}(a).
These properties ensure that the self-sustained oscillations do not arise from specific fine-tuned parameters or few-body processes but rather reflect a genuine many-body phase. Moreover, as expected from spontaneous symmetry breaking phenomenology, these are stable phases only in the thermodynamic limit. For instance, a single spin can precess indefinitely in time under the influence of a static magnetic field; however, it lacks any many-body character of a real phase of matter. 
Following the early proposals, CTCs have been predicted in a large variety of systems, and more recently experimentally tested, as we discuss in Sec.\ref{sec.Models} and Sec.\ref{sec.Experiments}.

The prototypical model for CTCs is called boundary time crystals~\cite{iemini_boundary_2018}, emphasizing that they emerge as a boundary critical phenomenon which can be viewed as arising from a larger closed system-environment evolution. 
While the combined system (boundary) and environment must obey the no-go theorem~\cite{Watanabe2015}, the boundary itself constitutes a vanishingly small (but still macroscopic) fraction of the whole. Thus, the boundary alone can evade the no-go theorem and form a time crystal, while leaving the rest (environment) time-translationally
invariant (see Fig.~\ref{fig:BTC_schematic}).
Interestingly, CTCs are more closely aligned with the original proposal in time-independent systems \cite{Wilczek2012}, while operating in a dissipative framework.

In order to explore concrete examples of CTCs, one faces the challenge of analyzing the dynamics of many-body open quantum systems.
The existing literature, so far, has mostly focused on the simpler, yet nontrivial, case of Markovian systems where the evolution is governed by a Lindblad master equation. 
The spectral properties of the Lindbladian superoperator play a key role in identifying CTCs. Specifically, a CTC emerges if the Lindbladian has gapless excitations with nonvanishing imaginary values \cite{iemini_boundary_2018,duttaQuantum2025}. 
Precisely, when (i) the real part of the Lindbladian spectrum exhibits a vanishing gap, leading to a non-equilibrium steady-state subspace (only) in the thermodynamic limit, and (ii) this subspace contains eigenvalues with nonzero imaginary parts, ensuring sustained periodic dynamics. Figure \ref{fig:BTC_schematic}(b,c) illustrates this case. 
The presence of gapless excitations can lead to a non-commutativity between the macroscopic and long-time limits, i.e. $\lim_{N \rightarrow \infty} \lim_{t \rightarrow \infty} \hat \rho(t) \neq \lim_{t \rightarrow \infty} \lim_{N \rightarrow \infty} \hat \rho(t)$, serving as a powerful sufficient criterion to identify such excitations in general Lindbladian dynamics. This concept was investigated in a class of spin-boson systems with conserved angular momentum \cite{souzaSufficientConditionGapless2023}.
Here, the extensive steady state correlations generated at finite system sizes \cite{carolloExactSolutionBoundary2022,lourencoGenuineMultipartiteCorrelations2022} are incompatible with the exact mean-field predictions in the thermodynamic limit, revealing the non-commutativity of the two limits and the presence of gapless modes. This breakdown underscores the critical role of correlations in sustaining these phases.

It is worthwhile to note here that boundary TCs are conceptually distinct from the other interesting approach to CTCs that relies on the concept of dynamical symmetries, discussed in Sec.~\ref{sec:beyond_kuromoto}, which provides an algebraic criterion for the existence of purely imaginary eigenvalues \cite{bucaalgebraic2022}. These eigenvalues guarantee non-stationary dynamics within their respective subspaces, which under suitable conditions can also result in synchronized dynamics (see Secs.~\ref{Sec:transient synch} and \ref{sec:beyond_kuromoto}). Crucially, the criterion is system-size independent, meaning it applies even to few-body systems exhibiting non-decaying oscillations. 
Existing studies have focused on systems where the symmetry is already present at finite systems \cite{bucaNonstationaryCoherentQuantum2019,bookerNonstationarityDissipativeTime2020,tindallQuantum2020,bucaalgebraic2022,campbell_stability_2025}, or it emerges from a non-interacting single-particle excitation which decouples from the environment (i.e. becomes decoherence-free) for large enough system sizes \cite{alaeian_exact_2022}. Explicitly constructing these emergent dynamical symmetries for robust many-body CTCs, however, remains an interesting open challenge.

  \subsection{Models}
\label{sec.Models}
 There are various models that exhibit CTCs. We discuss them, highlighting their distinct dynamical features.

 \subsubsection{Spin models}
 \label{sec.CTCs.spinmodels}

 Among the various models exhibiting CTCs, spin-only Lindbladians have been the most explored. A paradigmatic example is a dissipative spin system with all-to-all correlations mediated by the environment \cite{iemini_boundary_2018}, commonly used to describe cooperative emission phenomena \cite{puri_exact_1979,drummond_observables_1980,carmichaelAnalyticalNumericalResults1980,lawande_non_resonant_1981}. The system consists of $N$ spin-$1/2$ particles with Lindbladian dynamics,
\begin{equation}
     \frac{d}{dt} \hat \rho = i\omega_o \left[ \hat \rho,\hat S_x\right]
     + \frac{2\kappa}{N}\left(  \hat S_- \hat \rho \hat S_+ - \frac{1}{2}\{\hat S_+\hat S_-,\hat \rho\}
     \right)
     \label{eq.master.equation.BTC}
\end{equation}
where $\hat S_\alpha = \frac{1}{2}\sum_{j=1}^N \hat \sigma_j^\alpha$ are collective spin operators, $\hat \sigma_j^\alpha$ the Pauli matrix for the $j$th spin along $\alpha=x,y,z$, and $\hat S_{\pm} = \hat S_x \pm \hat S_y$ are collective raising/lowering operators. The total angular momentum $S$ is conserved, and we assume its maximum value $S = N/2$. Dissipation is encoded in a single jump operator describing collective emissions, establishing an infinite-range dissipative coupling mediated by the environment. Analytical steady-state solutions reveal two distinct phases \cite{puri_exact_1979,drummond_observables_1980,carmichaelAnalyticalNumericalResults1980,lawande_non_resonant_1981,hannukainen_dissipation-driven_2018}: (i) for strong dissipation ($\omega_0/\kappa < 1$) the system relaxes to a stationary state with finite magnetization $\langle \hat  S_z \rangle_{ss} < 0$; (ii) for weak dissipation ($\omega_0 /\kappa > 1$) this alignment vanishes. 
Remarkably, in the latter regime spins self-organize into sustained oscillations that persist in the thermodynamic limit breaking time-translation symmetry and forming a boundary TC \cite{iemini_boundary_2018}, as seen in Fig.~\ref{fig:BTC_schematic}(a). This behavior is also captured by a semiclassical mean-field description of the collective spin magnetization $\hat{m}^\alpha_N=\hat{S}^\alpha/N$. Owing to the all-to-all dissipative coupling, this approach is exact as $N\rightarrow \infty$ \cite{PhysRevA.85.013817,PhysRevLett.126.230601,souzaSufficientConditionGapless2023}, and factorizes expectation values $\langle \hat m^\alpha_N \hat m^\beta_N \rangle \approx \langle \hat m^\alpha_N \rangle \langle \hat m^\beta_N \rangle$ yielding a set of nonlinear dynamical equations. In the boundary TC phase, fixed points appear in pairs and, within a Jacobian analysis, reveal purely imaginary eigenvalues \cite{piccittoSymmetriesConservedQuantities2021,prazeresBoundaryTimeCrystals2021,nakanishi_continuous_2024}, indicating neutral stability and leading to multiple closed periodic orbits determined by the initial state. 

From a spectral perspective, the Lindbladian differs markedly between phases: it is gapped for $\omega_o/\kappa < 1$, and gapless for $\omega_o/\kappa > 1$   (see Fig.~\ref{fig:BTC_schematic}(b,c)) reflecting persistent oscillations in the thermodynamic limit \cite{iemini_boundary_2018}. Gapless excitations feature bands of imaginary eigenvalues spaced by a fundamental frequency ($\Gamma = \sqrt{\omega_o^2 - \kappa^2}$ in the thermodynamic limit \cite{carolloExactSolutionBoundary2022}) that matches the magnetization oscillation frequency. This frequency is incommensurate with the driving $\omega_o$, showing oscillations arise from the intrinsic competition between coherent dynamics and dissipation.

The boundary TC exhibits remarkable resilience. Under Hamiltonian perturbations such as $\hat H = \sum_p \omega_{x,p} (\hat S^x)^p + \omega_{z,p} (\hat S^z)^q$, the crystal persists provided dissipation explicitly breaks the Hamiltonian symmetry \cite{piccittoSymmetriesConservedQuantities2021,wang_dissipative_2021,prazeresBoundaryTimeCrystals2021}. For even $p$ the Hamiltonian has a $\mathbb{Z}_2$ rotational symmetry around the x-axis broken by the environment, preserving the boundary TC; for odd $p$ the time-crystalline order is destroyed. The boundary TC maintains stability in the presence of finite-temperature environments and non-Markovian effects \cite{piccittoSymmetriesConservedQuantities2021,Carollo_2024,dasStabilizing2026}. With dissipative power-law decaying interactions, the boundary TC persists for sufficiently long-ranged interactions \cite{passarelliDissipativeTimeCrystals2022}. Inhomogeneities in the initial state break permutational invariance, leading to dynamics across different symmetry sectors; the boundary TC survives and can even yield richer phenomena such as chimera and clustered synchronization \cite{PhysRevA.109.032204,solanki2024exotic}.

Extensions of the boundary TC spin model have revealed rich behaviors.  Replacing two-level spins with higher-dimensional $d$-level systems, or coupled spin ensembles, can lead to the coexistence of multiple limit cycles, a route to chaotic dynamics \cite{prazeresBoundaryTimeCrystals2021,b6gq-z5nf,postavovaClassical2026,shenDissipative2026},  non-resonant stabilization \cite{wangNonResonant2026} and highlight underlying parity-time symmetries \cite{nakanishiDissipativeTimeCrystals2023,nakanishiLindbladian2025a}. Coupling a boundary TC to additional static spin ensembles, where the boundary TC acts as a nucleation center, can initiate (seed) synchronized dynamics in the surrounding spins \cite{hajdusekSeedingCrystallizationTime2022}. 
 Furthermore, measurement-induced time crystals have been proposed, leveraging the quantum Zeno effect in a central-spin model \cite{krishnaMeasurementInducedContinuousTime2023}: frequent measurements of the central spin can generate an effective Lindbladian evolution on the peripheral spins, inducing boundary TC-like behavior without direct driving. Recently, multiple time-translation symmetry breaking has been proposed in coupled spin systems, where one subsystem first breaks continuous time-translation symmetry, followed by discrete symmetry breaking in another subsystem, giving rise to a hierarchical time crystal phase \cite{schumannHierarchical2026}.

\subsubsection{Bose-Hubbard dimer}

Another class of time-crystal systems involves bosons, notably the Bose-Hubbard dimer model describing two nonlinear photonic cavities coupled via photon hopping under driving and dissipation. The Hamiltonian is
\begin{equation}
 \hat H = \sum_{i=1}^2 \Delta \hat a^\dagger_i \hat a_i +  U \hat a^{\dagger^2}_i \hat a_i^2 + F_i (\hat a^\dagger_i + \hat a_i) - J (\hat a^\dagger_1 \hat a_2 + \hat a^\dagger_2 \hat a_1),
\end{equation}
with $\Delta$ detuning, $U$ on-site interaction, $F_i$ driving amplitude, and $J$ intermode coupling. The thermodynamic limit here is defined by rescaling parameters with $N$ (e.g., $U \rightarrow \tilde{U}/N$, $F_i \rightarrow \sqrt{N}\tilde{F}$), where $\langle \hat a^\dagger \hat a \rangle \propto N$ as $N\to\infty$ \cite{PhysRevA.95.012128}.

Various Lindblad dissipators yield distinct time-crystalline orders. 
Under homogeneous driving, nonlocal bonding dissipation $\mathcal{D}[\hat a_B]$, with $\hat a_B = (\hat a_1 + \hat a_2)/\sqrt{2}$, decouples the normal modes of the system, inducing limit-cycle evolution in one of the mode while damping the other \cite{lledoDissipativeTimeCrystal2020,lledoDrivenBoseHubbardDimer2019}.
 Time crystal behavior can also emerge in the presence of inhomogeneous driving ($F_1 \neq F_2$) under local dissipation ($\kappa \sum_i \mathcal{D}[\hat a_i]$) \cite{seiboldDissipativeTimeCrystal2020a}, or in absence of driving through incoherent hopping ($\kappa_{12(21)}  \mathcal{D}[\hat a^\dagger_{1(2)} \hat a_{2(1)}]$) \cite{solankiGeneration2025}, exhibiting characteristic bistability with limit-cycle attractors.
Quasiperiodic attractors can also arise in dimers with nonlinear tunneling, incoherent gain, and nonlinear dissipation, as shown in \cite{nowoczyn_universal_2025}, where their quantum‑to‑classical crossover is analyzed.  
A recent work has also indicated that time crystals can emerge even in a simplified framework involving a single-mode cavity \cite{liTime2024}. This model lacks intracavity interactions ($U$), but incorporates both linear gain and damping processes ($\kappa \mathcal{D}[\hat a] + g \mathcal{D}[\hat a^\dagger]$) alongside nonlinear damping effects ($\eta \mathcal{D}[\hat a^2]$). In a semiclassical treatment, the model reduces to a driven van der Pol oscillator. 
Intriguingly, for small driving strength, the oscillating limit-cycle frequency is set by the detuning $\Delta$ \cite{liTime2024}, while for larger ones both the oscillating frequency and the bifurcation type depend on the interplay between several system parameters \cite{weiss2017underdamped,navarrete-benlloch2017general}. Similarly, \cite{cabot2024nonequilibrium} analyzed the emergence of limit-cycle and subharmonic oscillations in the squeezed QvdP oscillator and discussed them in terms of different types of time-crystalline order.

\subsubsection{Spin-boson models}

Different forms of spin-boson systems were explored in the context of CTCs. These models describe $N$ spins coupled collectively to a dissipative bosonic mode. In the rotating frame, the Hamiltonian becomes,
\begin{equation}
\label{eq.spin.boson.models}
\hat H  =
\hat H_B + \hat H_S +
\sum_{k} \frac{1}{\sqrt{N}} (\lambda_k \hat a \hat S_k + \text{h.c.}) + \frac{U_\alpha}{N} (\hat a^\dagger \hat a) \hat S'
\end{equation}
where $\hat a^\dagger$ and $\hat a$ are bosonic ladder operators operators ($[\hat a, \hat a^\dagger] = 1$).
The first term describes the boson-only contribution, $\hat H_B = \omega_b \hat a^{\dagger} \hat a$, with $\omega_b$ denoting the effective bosonic detuning. The spin-only term, $\hat H_S$, and operators $\hat S_k$ and $\hat S'$ can take different forms, either structureless \cite{dograDissipationinduced2019,bucaDissipationInducedNonstationarity2019,ferioliNonequilibriumSuperradiantPhase2023,mattesEntangledTimecrystalPhase2023,zhang_dissipation-induced_2025}, described by collective spin operators $\hat S^{\alpha=x,y,z}$ (similar to the models discussed in Sec.~\ref{sec.CTCs.spinmodels}), or spatially structured, as realized, for example, in Bose-Einstein condensates trapped in optical lattices \cite{kesslerEmergentLimitCycles2019,piazza_self-ordered_2015,kongkhambut_observation_2022,kongkhambut_observation_2024,skulte_realizing_2024}.
Interactions occur via coherent boson exchange ($\lambda_k$) and dispersive light shifts ($U_\alpha$), with photon losses at rate $\kappa$ described by $\kappa \mathcal{D}[\hat a]$.

A key feature of these models is that the bosonic field mediates long-range interactions among the spins (in fact, for some regimes adiabatic elimination recovers boundary TC dynamics with collective spin dissipation \cite{mattesEntangledTimecrystalPhase2023}). When the bosonic mode becomes macroscopically populated, the mediated interactions can drive the spins into self-organized, many-body ordered states. Crucially, here dissipation not only introduces decoherence, but also actively shape the dynamics triggering structural instabilities in the otherwise steady ordered spin configurations. In some regimes, dissipation even prevents settling into any static ordered phase, inducing continuous self-sustained transitions among multiple unstable spin arrangements and leading to many-body limit cycles and CTCs \cite{dograDissipationinduced2019,kongkhambut_observation_2022}.

\subsubsection{Lattice models}

Lattice dissipative systems with localized constituents under coherent coupling and local dissipation have been largely studied. 
They are generally described by,
\begin{equation}
 \hat H = \sum_{\alpha,\beta, \langle ij \rangle } J_{ij} \hat O_i^\alpha \hat O_j^\beta + \sum_{\alpha, j}h_j \hat O_j^\alpha
\end{equation}
with $\hat O_i^\alpha$ local 
operators acting on the $i$'th site of the $d$-dimensional lattice, $\langle ij \rangle$ runs over close (e.g. nearest-neighboor) sites, $J_{ij}$ and $h_i$ denote interaction and field terms, respectively.
Dissipation is introduced via local photon loss/pumping or spin relaxation/excitation/dephasing.

Analysing the dynamics of these systems is typically complex,  requiring a combination of semiclassical and quantum treatments. A common strategy involves identifying mean-field limit cycles and testing robustness against quantum fluctuations using methods like truncated Wigner approximation, cluster mean-field, and cumulant expansions \cite{SciPostPhysLectNotes.99}. These approaches reveal limit-cycle phases in Bose-Hubbard models with cross-Kerr nonlinearities \cite{jinPhoton2013,scarlatellaDynamical2021}, optomechanical arrays \cite{ludwig2013quantum}, isotropic \cite{yangEmergent2025} and anisotropic \cite{chanLimit-CyclePhase2015,owenQuantum2018} Heisenberg spin lattices, Rydberg atoms \cite{leeAntiferromagnetic2011,qianPhase2012}, and quantum contact-process spin models \cite{xiangSelforganized2024}. In these phases, all local operators synchronize into persistent collective oscillations.

Key questions remain, particularly regarding stability in lower dimensions. Some studies suggest quantum fluctuations destabilize limit cycles in spin-half systems beyond a critical dimension \cite{chanLimit-CyclePhase2015,owenQuantum2018}, while others indicate finite connectivity can preserve synchronization if balanced by stronger incoherent driving in bosonic systems \cite{scarlatellaDynamical2021}.  Resolving these questions is crucial for a deeper understanding of (Continuous Time Crystals) CTCs in quantum lattices.

Moreover, recently there has also been growing interest in searching for CTCs without an underlying semiclassical picture. Investigations into spin-$1$ lattices suggest that this is possible \cite{russo_quantum_2025,wang_boundary_2025}. Using a combination of numerical methods (cluster mean-field, cumulant expansion and MPS) persistent oscillations absent in mean-field dynamics were observed in both 2D lattices with short-range interactions and 1D chains with long-range interactions.  Interestingly, these dynamics are also not present in equivalent spin-half systems, mirroring recent experimental observations of CTCs in Rydberg atomic gases  \cite{wu_dissipative_2024,wadenpfuhl_emergence_2023}. These findings raise fundamental questions about the necessary conditions for such classes of strictly quantum CTCs.

\subsubsection{Non-reciprocal models}
\label{sec:nonreciprocal}

Non-reciprocal interactions break the conventional action-reaction symmetry, meaning that the influence of one subsystem on another is not equally reciprocated. 
An interesting consequence of this is the emergence of limit cycles. The lack of reciprocity can create a persistent imbalance, preventing the system from settling into a stable equilibrium and causing it to oscillate in a repeating pattern. In Sec.~\ref{Sec:transient synch} we introduced synchronization in chiral networks of harmonic oscillators \cite{lorenzo2022quantum}, and we address here many-body non-reciprocal phenomena.
While non-reciprocal interactions can induce limit cycles even in finite systems, in some cases, the stability of these oscillations depends crucially on many-body correlations, becoming robust only in the thermodynamic limit — aligning the phenomenon with the formation of CTCs \cite{fruchartNonreciprocal2021,avniNonreciprocal2025,nadolny_nonreciprocal_2025}. 
This mechanism differs fundamentally from cases such as the Kuramoto model, which require nonlinearity; indeed, such a nonlinearity may even suppress synchronized motion in non-reciprocal systems \cite{raskatla_continuous_2024}.

To better illustrate the underlying dynamical mechanism, one can revisit the classical Kuramoto model of coupled oscillators (Eq.~\eqref{eq:_classical_kuramoto}) with a modification, which is to consider a bipartition of the system into two groups, $A$ and $B$ that interact asymmetrically, with coupling strengths $J_{AB} \neq J_{BA}$ \cite{fruchartNonreciprocal2021}. In particular, when $J_{AB} \sim - J_{BA}$, oscillators in group $A$ attempt to align with those in $B$, while those in $B$ attempt to anti-align with $A$. These frustrated goals induce a ``chase-and-run" behavior: group $A$ chases $B$, which simultaneously tries to escape. In a noiseless system, this motion eventually reaches an equilibrium alignment or anti-alignment (unless at exact fine-tuning $J_{AB} = -J_{BA}$). However, in the presence of frequency disorder ($\omega_i \neq \omega_j$) or white noise in Eq.~\eqref{eq:_classical_kuramoto}, the system is repeatedly kicked out of these metastable configurations, initiating new chase cycles. The fluctuations of this motion decrease as $1/\sqrt{N}$ \cite{fruchartNonreciprocal2021}, making the dynamics increasingly rigid due to many-body correlations, thus establishing in the thermodynamic limit a stable (noise-activated) limit cycle CTC.

Various models have been explored in this context, mostly in classical systems such as spin lattices \cite{avniNonreciprocal2025,avniDynamical2025,hanai_nonreciprocal_2024}, arrays of optically driven nanowires \cite{liu_photonic_2023,raskatla_continuous_2024,liuBreaking2024}, levitated or self-propelled particles \cite{morrell_nonreciprocal_2025,eversActive2023}, as well as in quantum spin ensembles with all-to-all interactions \cite{nadolny_nonreciprocal_2025}.

\subsection{Correlations in continuous time crystals}
\label{sec.CTCs.correlations}
As discussed in Sec.~\ref{sec.CTCs.spinmodels}, an effective strategy to identify the emergence of CTCs is to investigate the dynamics of average magnetization operators or suitably rescaled bosonic ones. 
In collective models, the evolution of these observables is accurately captured by a semiclassical mean-field theory (see also discussion in Sec~\ref{Subsec:meanfield} and Sec.~\ref{sec.CTCs.spinmodels}). 
For initial states with low correlation content (e.g.~product states or thermal states of gapped Hamiltonians), the average magnetizations converge, at any fixed time $t$ and in the large $N$ limit, to multiples of the identity \cite{PhysRevLett.126.230601}. 
This fact, together with the collective character of the dynamics, significantly simplifies their treatment and allows one to probe the CTC phase transition through the observation of persistent oscillations in macroscopic properties. 
However, these observables effectively behave as deterministic variables and therefore fail to capture fluctuations or correlations. 

One way to explore correlations, their impact, and their dynamical evolution in CTCs is to consider initial states featuring strong correlations, e.g.~superpositions of product states, as done in \cite{mukherjeeSymmetries2024}. In these cases, average operators do not generally converge to deterministic quantities and exhibit finite covariances in the thermodynamic limit, i.e. $\chi_{\alpha\beta}=\lim_{N\to\infty}(\langle \hat{m}^\alpha_N \hat{m}^\beta_N \rangle - \langle \hat{m}^\alpha_N  \rangle \langle \hat{m}^\beta_N \rangle)\neq 0$.  As a consequence, the factorization of expectation values, $\langle \hat{m}_N^\alpha \hat{m}_N^\beta \rangle \approx \langle \hat{m}_N^\alpha \rangle \langle \hat{m}_N^\beta \rangle$ (see discussion in Sec.~\ref{sec.CTCs.spinmodels}), completely misses the role played by initial correlations \cite{mukherjeeSymmetries2024}. 
A way to improve upon mean-field theory is to employ a cumulant expansion, which results in a set of hierarchical equations of motion \cite{kuboGeneralizedCumulant1962,kiraCluster-expansion2008,meiserProspectsfor2009,henschelCavityQEDwith2010,kirtonSuperradiantAndLasing2018,zensCriticalPhenomena2019} (see however Refs.~\cite{fowler-wrightDetermining2023,carolloNonGaussian2023} where it is shown that, in certain collective models, the scaling of the interaction term with the system size may impact on the validity of both mean-field theories and cumulant expansions).
Remarkably, retaining correlations only up to second-order cumulants, by setting the third-order cumulants to zero, already captures the dynamics of the CTC for initial states given by a superposition of two spin coherent states \cite{mukherjeeSymmetries2024}. 

Although successful at incorporating strong initial correlations, cumulant expansions remain focused on the evolution of sample-mean operators $\hat{m}_\alpha^N$ and their moments. These observables form a commutative algebra in the thermodynamic limit, since $[ \hat{m}_\alpha^N,\hat{m}^N_\beta]\sim 1/N$. As such, they can at most account for classical correlations. 
An important question thus concerns how to describe genuine quantum correlations, such as entanglement, in CTCs. In what follows, we review two complementary approaches to the investigation of correlations in CTCs, mainly focusing on the boundary TC model of Eq.~\eqref{eq.master.equation.BTC}.  The first approach is based on investigating correlations in the stationary state of the dynamics as the system size $N$ increases. The second one exploits instead bosonization procedures that can describe the evolution of quantum fluctuations around the semiclassical behavior. 

\begin{figure}
\includegraphics[width=0.48\textwidth]{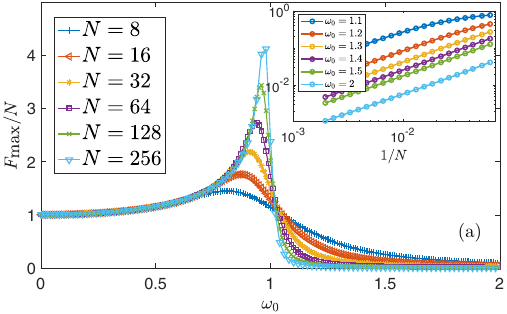}
    \caption{Quantum Fisher information in the stationary state of the boundary TC model. The maximum of the quantum Fisher information, $F_{\rm max}$, optimized over measurements of the collective spin operators and rescaled by $N$. This quantity shows multipartite entanglement solely in the stationary phase, diverging as one approaches the phase transition point. In the CTC phase, $F_{\rm max}$ becomes sub-extensive. Figure adapted from \cite{lourencoGenuineMultipartiteCorrelations2022}.}
    \label{fig:QFI_stationary}
\end{figure}

\subsubsection{Stationary correlations in finite systems}
Let us focus on the boundary TC model of Eq.~\eqref{eq.master.equation.BTC}. For any finite system size, the quantum state converges to a stationary one in the long-time limit ($t\to\infty$). This also happens for parameter regimes where a CTC phase is expected in the thermodynamic limit, and is due to the fact that the gap of the Lindblad dynamical generator remains open for any finite $N$ (see Fig.~\ref{fig:BTC_schematic}). This highlights that the limits $t, N\to\infty$ do not commute within the CTC phase \cite{souzaSufficientConditionGapless2023}. 

The stationary state $\hat{\rho}_{\rm ss}$ for the boundary TC model can be found exactly. It can be expressed as \cite{puri_exact_1979,carmichaelAnalyticalNumericalResults1980,hannukainen_dissipation-driven_2018,lourencoGenuineMultipartiteCorrelations2022}
\begin{equation}
\hat{\rho}_{\rm ss}=\frac{1}{\mathcal{N}}\hat{\eta} \hat{\eta}^\dagger\, , \quad \mbox{ with }\quad \hat{\eta} =\sum_{j=0}^N \left(\frac{\hat{S}_-}{-i\omega_o N/2}\right)^j\, ,
\label{eq.rho.stat}
\end{equation}
and $\mathcal{N}$ being a normalization constant. Analogous solutions can also be constructed for related models with or without permutation symmetries \cite{robertsexactsolution2023}. From the stationary state, it is then possible to compute several correlation measures. 

A comprehensive characterization of genuine multipartite correlations for the boundary TC model was performed in \cite{lourencoGenuineMultipartiteCorrelations2022}. The authors explored the behavior of genuine $k$-partite correlations \cite{girolamiQuantifyingGenuineMultipartite2017} which provide the contribution to the total correlations encoded in states of at most $k$ particles. 
This quantity can be computed as the additional information accessible to an observer when their access is extended from states of at most $k - 1$ subsystems to states of at most $k$ subsystems.
It was found that the genuine multipartite correlations, for at most $k=10$ subsystems, are subextensive for parameter regimes $\omega_0<\kappa$ while they become extensive in regimes where a CTC phase emerges in the thermodynamic limit $(\omega_0\ge \kappa$). 
This suggests that the CTC phase is characterized by strong correlations. 

However, the genuine multipartite correlation cannot distinguish between quantum and classical correlations. In order to explore the nature of correlations for the boundary TC, the authors of \cite{lourencoGenuineMultipartiteCorrelations2022} also considered the behavior of the quantum Fisher information $F_{\rm max}$ (see Fig.~\ref{fig:QFI_stationary}), maximized over all possible measurements of collective spin operators. Such a quantity shows a completely different picture: regimes where a CTC phase is expected are characterized by subextensive quantum Fisher information, which cannot detect any signature of quantum correlations. On the other hand, parameter regimes associated with the stationary phase are characterized by $F_{\rm max}>N$, which signals the presence of entangled states. Moreover, close to the phase transition point, $F_{\rm max}$ diverges with $N$, indicating a superextensive character of $F_{\rm max}$ and multipartite entanglement. 

This suggests that the extensive correlations in the CTC phase measured by  genuine multipartite correlations are of a statistical nature. An intuition for this fact can be obtained by thinking of single realizations of the CTC dynamics (see Sec.~\ref{sec:manybodytraj}). In the CTC phase, each realization of the dynamics exhibits oscillatory behavior. For finite systems, however, these oscillations are never perfectly in phase across different realizations. As a result, averaging over them produces a quantum state that effectively becomes a statistical mixture of oscillations with different phases. This averaging leads to a state that is indeed stationary (i.e.,~not time-dependent anymore) and characterized by strong statistical correlations arising from averaging over different oscillatory patterns.

\subsubsection{Bosonization and quantum fluctuations}
\label{sec.CTCs.correlations.bosonization}
Another way to explore classical and quantum correlations in many-body systems is to study quantum fluctuations. The fundamental idea is to perform a self-consistent expansion of the collective operators around their semiclassical behavior \cite{kesslerDissipativePhase2012,bucaDissipationInducedNonstationarity2019,benattiQuantumFluctuationsIn2017,benattiQuantumSpinChain2018,leroseBridgingEntanglementDynamics2020}. This approach entails taking the large $N$ limit first and then following the dynamics of the system over time, and is thus complementary to one presented earlier. Such a procedure, when considering states with low correlation content, typically gives rise to a bosonic algebra.

One way to see how bosonic degrees of freedom can emerge from spin ensembles is by moving from the average magnetization operators $\hat{m}_\mu^N$ to the fluctuations $\hat{F}_\mu^N=(\hat{S}_\mu -\langle \hat{S}_\mu\rangle)/\sqrt{N}$. 
These operators account for deviations of the collective operators $\hat{S}_\mu$ from their average behavior and are reminiscent of fluctuation variables appearing in central limit theorems. 
Their variance serves as a susceptibility for the mean-field order parameters. Furthermore, 
\begin{figure}
\includegraphics[width=0.37\textwidth]{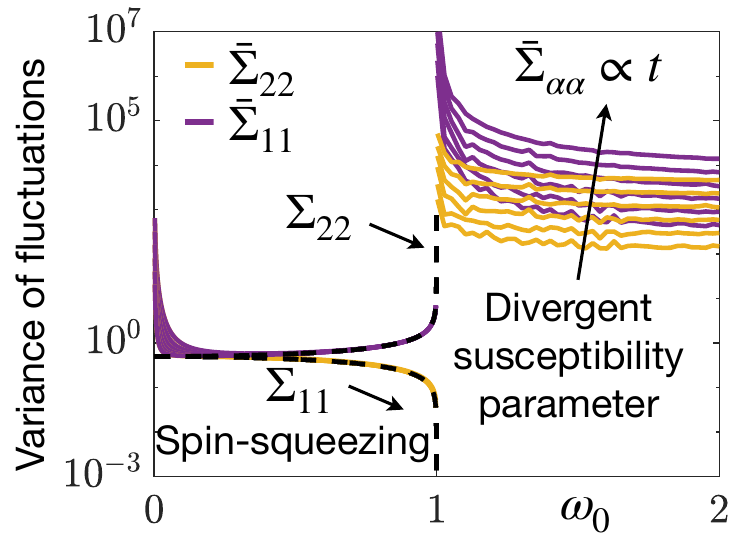}
    \caption{Time-averaged variance of the quantum fluctuation operators. The plot shows the time-averaged variances, $\bar{\Sigma}_{\alpha\alpha}$, of the normal modes of the quantum fluctuation operators for the system in the thermodynamic limit. In the stationary phase, we observe that one mode displays (spin) squeezing approaching zero near the phase transition point. In the CTC phase, instead, the variances remain time-dependent and grow indefinitely with time, as shown by the behavior of the time-averaged $\bar{\Sigma}_{\alpha\alpha}$. Figure adapted from \cite{carolloExactSolutionBoundary2022}.}
    \label{fig:QF_correlations}
\end{figure}
the commutator between two fluctuation operators does  not vanish in the thermodynamic limit, but is proportional to the average magnetizations  
\begin{equation}
[\hat{F}_\mu^N,\hat{F}_\nu^N]=i\sum_{\xi} \varepsilon_{\mu\nu\xi} \hat{m}_\xi^N\, ,
\end{equation}
where $\varepsilon_{\mu\nu\xi}$ is the fully anti-symmetric tensor. As such, when considering initial states with weak correlations, the above commutator converges to a multiple of the identity. This highlights the bosonic character of quantum fluctuations. 

For collective models, such as the ones reviewed in the previous section, the quantum dynamics preserves the Gaussian character of initial states of fluctuation operators \cite{benattiQuantumFluctuationsIn2017,benattiQuantumSpinChain2018}. One can therefore describe correlations and the properties of quantum fluctuations by exploiting measures for Gaussian continuous-variable systems based on the covariance matrix of quantum fluctuations, $\Sigma$. This approach has been used in~\cite{carolloExactSolutionBoundary2022} to characterize spin-squeezing by looking at squeezing parameters of the bosonic quantum fluctuations. 

\begin{figure*}[!ht]
\includegraphics[width=\textwidth]{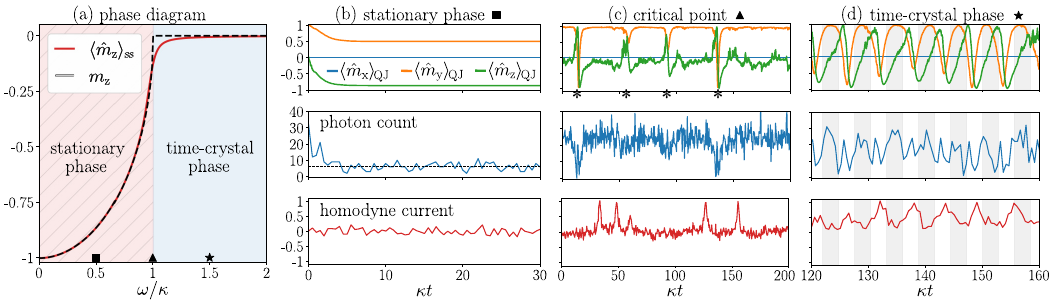}
    \caption{Quantum trajectories of continuous time crystals. (a) The plot shows the stationary phase diagram for the boundary TC model of Eq.~\eqref{eq.master.equation.BTC} in the semi-classical approximation and for the case of $N=100$. (b-d) Representative quantum trajectories showing the expectation value of the collective spin operators and of the time record of photon emission and of the homodyne current. 
    Figure taken from \cite{cabotQuantumTrajectoriesDissipative2023}.}
    \label{fig:many_body_trajectories}
\end{figure*}

As shown in Fig.~\ref{fig:QF_correlations}, the smallest variance of the quantum fluctuations, denoted here by $\Sigma_{11}$ and obtained as the smallest symplectic eigenvalue of their bosonic covariance matrix, vanishes upon approaching the critical point from the stationary phase. This behavior indicates divergent spin squeezing and multipartite entanglement and is consistent with the results for the quantum Fisher information shown in Fig.~\ref{fig:QFI_stationary}. The largest variance, $\Sigma_{22}$, instead diverges approaching the critical point, resembling the divergence of the susceptibility parameter close to second-order phase transitions \cite{carolloExactSolutionBoundary2022}. In the CTC phase, instead, the system displays oscillations and the variance of quantum fluctuations grows indefinitely with time, as it happens to finite-$N$ simulations of the genuine multipartite correlations \cite{lourencoGenuineMultipartiteCorrelations2022}. This also entails the absence of quantum correlations in the CTC phase and the growth of the variance suggests that statistical correlations grow due to dephasing of the oscillatory dynamics (see also discussion in \cite{chanLimit-CyclePhase2015}).  The latter aspect is also supported by the indefinite growth with time of the system entropy as captured by quantum fluctuations \cite{Carollo_2024}.  

These results on correlations in CTCs suggest that these phases are dominated by classical statistical correlations. However, in certain models such as driven-dissipative spin-boson models and Bose-Hubbard dimers, it has been shown that  CTCs can also host collective quantum correlations, including entanglement and quantum discord \cite{mattesEntangledTimecrystalPhase2023,solankiGeneration2025}. In the following section, we will discuss the quantum and classical correlations in single realizations of the dynamics.

\subsection{Trajectories approach to continuous time crystals} \label{sec:manybodytraj}

Quantum trajectories in CTCs were systematically investigated in \cite{cabotQuantumTrajectoriesDissipative2023}, focusing on the boundary TC model. 
In the stationary phase  (see~Fig.~\ref{fig:many_body_trajectories}(a)), trajectories rapidly approach a time-invariant state that is nearly pure. Correspondingly, time-records of the photon count and of the homodyne current (see Fig.~\ref{fig:many_body_trajectories}(b)) show a fast relaxation around a steady behavior, with essentially white-noise fluctuations. At criticality, the system displays markedly different trajectory dynamics (see~Fig.~\ref{fig:many_body_trajectories}(c)). Trajectories alternate between extended time intervals characterized by small fluctuations and abrupt events involving large fluctuations (highlighted with a `$*$' in Fig.~\ref{fig:many_body_trajectories}(c)). The mean time between these sudden fluctuations follows a power-law scaling, captured by an effective Langevin equation for a phase variable $\varphi$ (such that $m_y=\sin \varphi$ and $m_z=\cos \varphi$, where $m_{\mu}$ are the expectation values of the magnetization operators). 

In the CTC phase (see Fig.~\ref{fig:many_body_trajectories}(d)),  quantum trajectories exhibit persistent nondecaying oscillations. Noise and finite-size effects lead to relative dephasing between different trajectories, so that average observables display exponentially damped oscillations.  Oscillations also directly appear in photon-count and homodyne-detection records. An analysis of homodyne-current fluctuations reveals the coexistence of dynamical phases in the CTC regime, characterized by both positive and negative expected values of the current. This coexistence reflects the fact that, in finite systems, the CTC phase corresponds to a mixture of distinct oscillatory patterns, suggesting a possible indirect experimental signature of the transition.

In \cite{linkRevealingTheNature}, a boundary TC model with an additional collective dephasing process was analyzed.
It explores the phase transition of the model using quantum trajectories defined with a complex Wiener noise. The authors observe that the system transition can be understood in terms of changes in the topology of different trajectories. Neglecting the presence of noise and considering the system in the thermodynamic limit, they observe that in the stationary phase, trajectories approach a stable fixed point whereas in the CTC phase they approach a limit cycle. In finite systems (see also the previous discussion on coexistence), they observe that a single trajectory can visit different limit cycles. In this case, a single trajectory spans the whole phase space, thus giving rise to a stationary state on average. 
In \cite{nadolny_nonreciprocal_2025}, two collective spin ensembles are considered, which are coupled via nonreciprocal interactions. In certain parameter regimes, the system enters a CTC phase characterized by self-sustained oscillations. 
At the level of quantum trajectories, the CTC exhibits mutual synchronization between the spin ensembles. This can be observed from the relative phase of the spin-ladder operators but could also be, in principle, inferred experimentally from the system time-averaged emission signal.

Quantum trajectories of many-body systems can display {\it measurement-induced entanglement phase transitions}. Tuning the relative strength of the coherent dynamics and the dissipation may lead to qualitatively different entanglement content in the stochastically evolving quantum state. 
An example is the emergence of transitions between area-law entangled phases at large measurement rates and volume-law entangled phases at weak measurement rates. These transitions have also been explored in quantum trajectories of CTCs \cite{Passarelli_2024, delmonteMeasurementInducedPhase}. It was shown that the stationary half-system entanglement entropy, evaluated along single quantum trajectories, exhibits a non-analytic behavior at the critical point associated with the CTC transition. 
In particular, the derivative of the entanglement entropy diverges logarithmically at the critical point, while the entanglement entropy itself scales sub-logarithmically in the CTC phase and remains area law in the stationary phase \cite{Passarelli_2024}. 
These findings were supported by further analytical results presented in \cite{liEmergentDeterministicEntanglement}, where an analytic treatment of quantum trajectories (analogous to the bosonization discussed in Sec.~\ref{sec.CTCs.correlations.bosonization}) demonstrated that, in the thermodynamic limit, correlations become effectively deterministic.

\subsection{Experiments}
\label{sec.Experiments}

Experimental observations of continuous time crystals have been reported across diverse physical platforms. In this section, we briefly review these studies.

\begin{figure*}[!htp]
\includegraphics[width=0.46\textwidth]{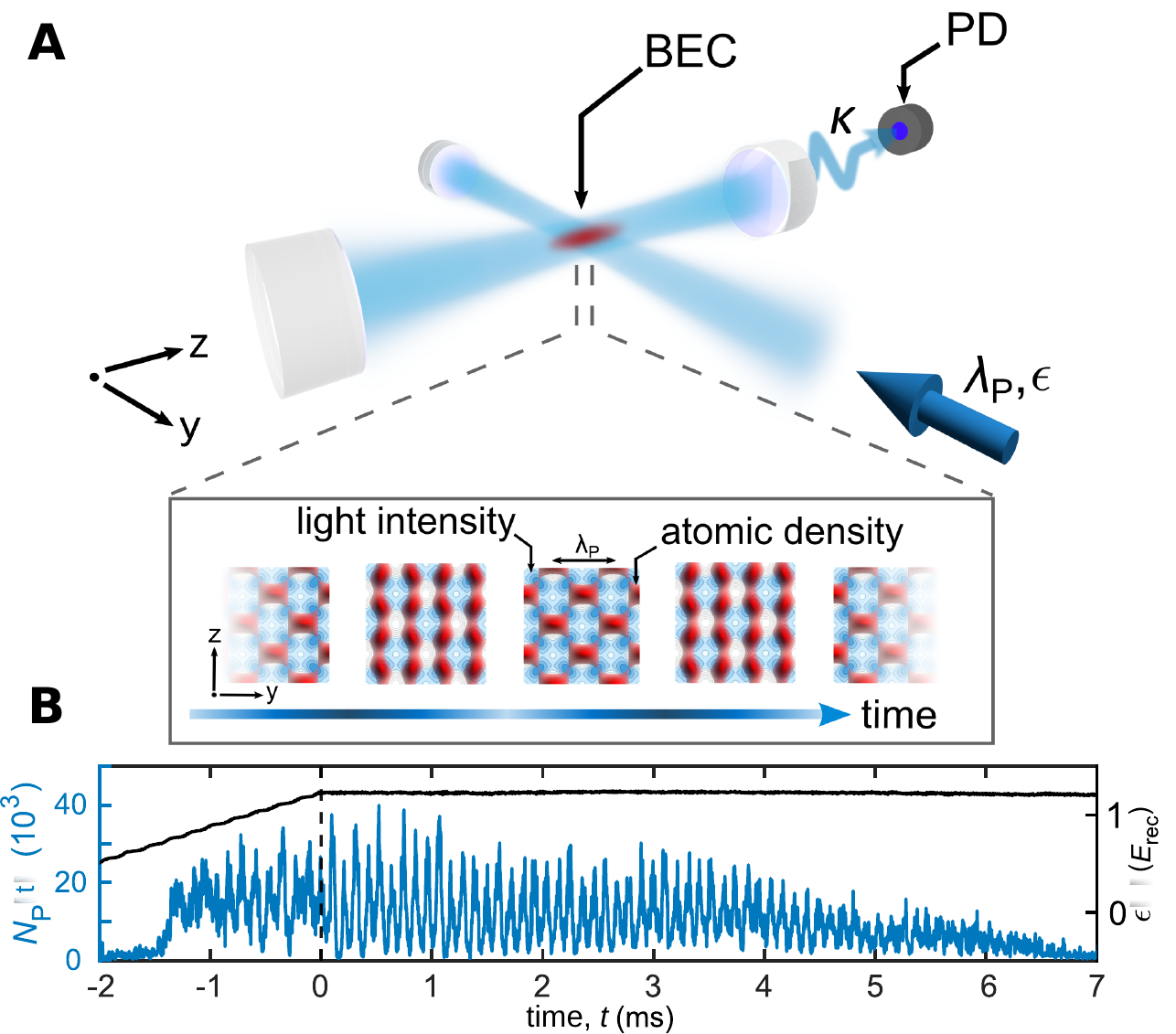}\includegraphics[width = 0.45\textwidth]{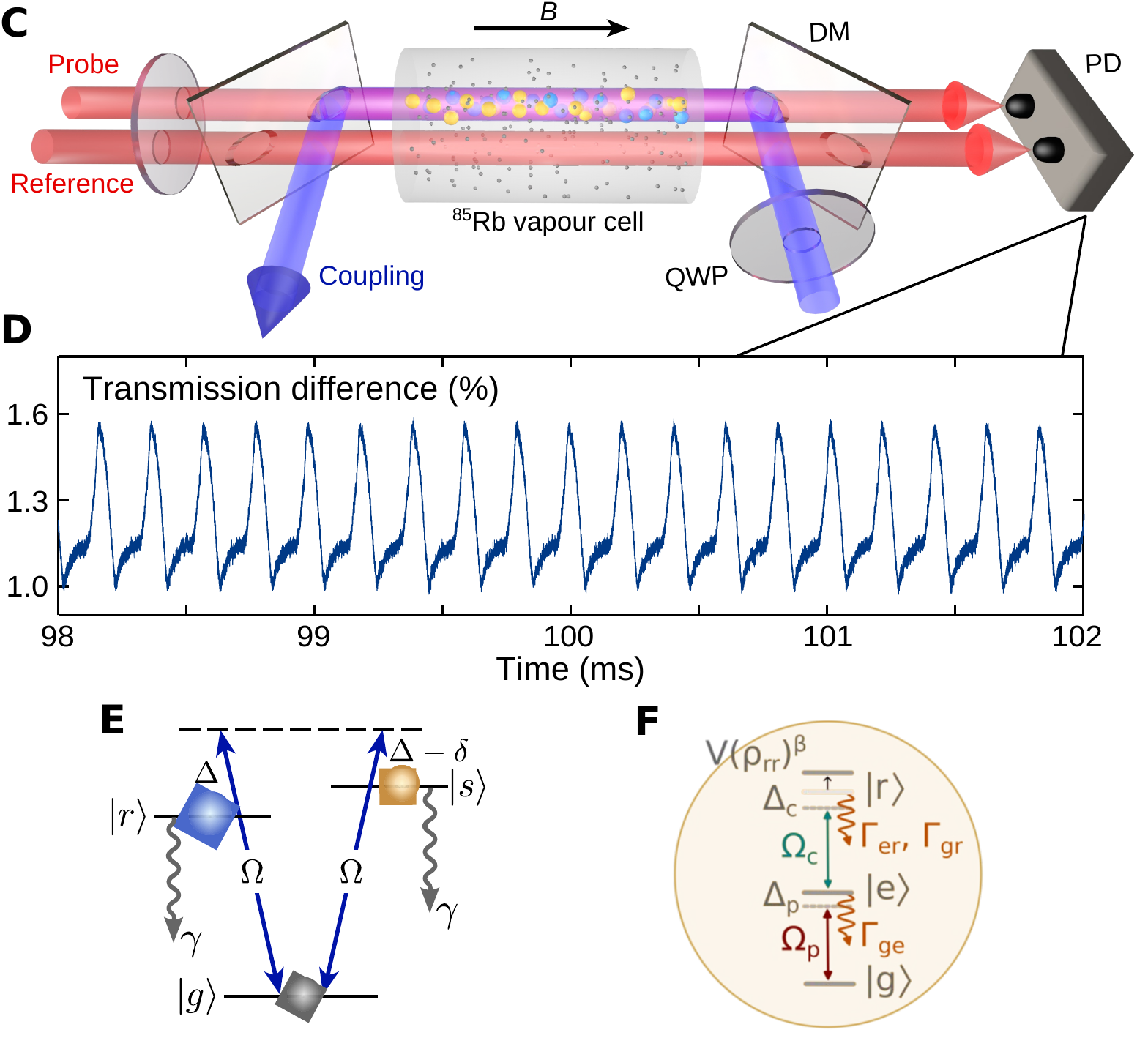}
    \caption{
     \textbf{(a)}  A schematic representation of an atom-cavity system, pumped transversely by an optical lattice.
(Inset) A numerical simulation of the photon field (blue) and the BEC atomic density (red), showing CTC dynamics with persistent oscillations in their spatial patterns.
 \textbf{(b)} A single experimental realization capturing this dynamics indirectly by measuring the intracavity photon number $N_P(t)$ (blue), which exhibits robust and long-lasting oscillations of several milliseconds while the pump strength (black) is held constant.
\textbf{(c)} Schematic diagram of the CTC in a Rydberg gas setup: a probe and reference beam propagate in parallel through a room-temperature $^{85}$Rb-vapour cell, driving the atoms into Rydberg states. The density of atoms in the Rydberg states can be inferred from the polarisation of the propagating beams,
observed via the transmission difference in the photon detector (PD). \textbf{(d)}  Single-run transmission difference for Rydberg states, showcasing the persistent CTC oscillations throughout the entire $125$ ms experiment time window.
\textbf{(e-f)} Different level schemes involving the two Rydberg states $|r\rangle $ and $|s\rangle$, as (e) V-type \cite{wu_dissipative_2024} or (f) $\mathsf{\Xi}$-type \cite{wadenpfuhl_emergence_2023}.
 Figures (a-b) are reprinted from \cite{kongkhambut_observation_2022}, while (c-e) are reprinted from \cite{wu_dissipative_2024} and (f) is reprinted from \cite{wadenpfuhl_emergence_2023}
   }
    \label{fig:CTC_experiments}
\end{figure*}

\emph{Atom-cavity.--- } A key realization of CTCs involves a Bose-Einstein condensate (BEC) trapped in an optical cavity and subjected to an external transverse pump laser field \cite{kongkhambut_observation_2022} (see Fig.~\ref{fig:CTC_experiments}(a,b)). The system is described by the Hamiltonian of Eq.~\eqref{eq.spin.boson.models} with,

\begin{eqnarray}
\hat H_S &=& \sum_{i=1}^N \hat p_i^2/2m + V_{\rm ext} \cos^2(k \hat y_i), \\
\hat S_k &=& \sum_{i=1}^N \cos(k \hat y_i) \cos(k \hat z_i), \quad \hat S' = \sum_{i=1}^N \cos^2(k \hat z_i) ,
\end{eqnarray}
with the cavity (pump) field oriented along the z(y)-axis, and $\hat y_i$ and $\hat z_i$ conjugate position operators. The first term accounts for atomic motion and the pump-induced standing wave $k$, while the second and third terms describe cavity-mediated interactions (creating a dynamical square-lattice potential dependent on atomic positions) and light shift, respectively.

At low pump intensity, the BEC remains homogeneous with no photons in the cavity. As the pump strength increases, the BEC transitions to a self-organized superradiant checkerboard density wave, with a macroscopic population in the cavity mediating infinite-range atomic interactions. For intermediate pump strengths, neither the checkerboard nor the homogeneous phase is stable. Instead, due to comparable atomic and cavity timescales, the system undergoes persistent oscillations between these configurations, forming a CTC phase.

In a different atom-cavity experiment,
a two-component BEC is coherently coupled to the cavity through two different spatial atomic configurations \cite{dograDissipationinduced2019}. The Hamiltonian interaction is described by:
\begin{eqnarray}
\hat H_{SB} = \frac{1}{\sqrt{N}} \left[ \lambda_D \hat x (\hat S^x_{+1} + \hat S^x_{-1}) + \lambda_S \hat p (\hat S^x_{+1} - \hat S^x_{-1}) \right]
\end{eqnarray}
where $\hat x = (\hat a+ \hat a^\dagger)$ and $\hat p = i(\hat a -\hat a^\dagger)$ represent the orthogonal cavity quadratures, while $\hat S^x_{m_F=\pm 1}$ are the angular momentum operators corresponding to the $m_F=\pm 1 $ Zeeman states. The first term in the Hamiltonian, with coupling rate $\lambda_D$, links one of the quadratures to the density mode (DM) of the atomic ensemble, favoring a spatial checkerboard density modulation in the Bose gas. The second term, with rate $\lambda_S$, couples the other quadrature to the spin mode (SM), promoting a spin density modulation instead. Depending on which cavity quadrature is excited, either the density or spin modulation dominates. However, cavity dissipation introduces an effective coupling between the DM and SM through the cavity's phase response.
 Remarkably, when this dissipative coupling becomes sufficiently strong, it triggers a structural instability in both spatial modes.  Instead of settling into a static configuration, the system continuously cycles through the different spatial patterns, forming a limit-cycle CTC phase \cite{dograDissipationinduced2019,bucaDissipationInducedNonstationarity2019}.  Notably, the same system can also exhibit discrete time-translation symmetry breaking when subjected to an additional periodic drive in certain dynamical regimes \cite{kongkhambut_observation_2024}. 

Beyond BEC platforms, the experimental realization of strong collective coupling between atomic ensembles driven on a metastable transition and a high‑finesse optical cavity mode has enabled the observation of the dissipative superradiant phase transition \cite{song2025dissipation}. 
This platform opens a route for exploring nonequilibrium phases in collective spin \cite{carmichaelAnalyticalNumericalResults1980, iemini_boundary_2018} and spin–boson systems \cite{mattesEntangledTimecrystalPhase2023}.
We remark that, despite a cavity being typically employed to enhance collective light-matter interactions in extended spin ensembles, a recent experiment has achieved similar effects in free space \cite{ferioliNonequilibriumSuperradiantPhase2023}.
In the experiment, a pencil-shaped cloud of cold atoms in free space (thus interacting with a continuum of free-space electromagnetic modes) was optically driven along its main axis, leading to a nonequilibrium superradiant phase transition. Remarkably, the atomic dynamics were accurately described by the boundary TC model (Eq.~\eqref{eq.master.equation.BTC}), revealing the emergence of the CTC in the system, and
opening new possibilities for studying such collective phenomena beyond conventional cavity-based setups. 
Later theory-experiment analyses showed that the boundary TC model can capture only some of the experiment's observations. In fact, more complicated models are required to fully explain it, which must, in some way, include spatial degrees of freedom in the description \cite{goncalves_driven-dissipative_2025, PRXQuantum.5.040335}.

 \emph{Rydberg gases.---}  These systems consist of a gas of three-level atoms at room temperature, where ground-state atoms ($|g\rangle$) are driven to two Rydberg states ($|r\rangle$  and $|s\rangle$) under different configurations, such as V-type \cite{wu_dissipative_2024} or $\mathsf{\Xi}$ (cascade)-type \cite{wadenpfuhl_emergence_2023} schemes, with dissipation causing decay towards the ground state, see Fig~.\ref{fig:CTC_experiments}(c)-(e). The single-atom Lindbladian term is modeled as,
\begin{equation}
\hat H_i =  \sum_{\alpha \neq \beta}\Omega_{\alpha \beta} \hat \sigma_i^{\alpha\beta}+ \Delta_r \hat n_i^r + \Delta_s \hat n_i^s, \,\,\,  \hat L_{i,\alpha \beta} = \sqrt{\Gamma_{\alpha \beta}} \hat  \sigma_i^{\alpha \beta}
\end{equation}
where $\hat \sigma_i^{\alpha \beta}=|\alpha\rangle \langle \beta|_i$ is the atomic transition operator for the $i$'th atom, $\hat n_i^\alpha = |\alpha \rangle\langle \alpha|_i$ are density operators 
and the indices account for the specific configuration  V (or $\mathsf{\Xi}$)-type. The atoms interact via a power-law coupling, $\hat H_{i,j} = V_{ij} \hat n_i^r \hat n_j^r $ 
with
$V_{ij}  = C_6/|\vec{r}_i - \vec{r}_j|^6$, the van der Waals interaction between Rydberg atoms located at $\vec{r}_i$ and $\vec{r}_j$. In the V-type configuration, each Rydberg state establishes its own bistable stationary phase. Interactions, however, can couple the Rydberg states. Due to the coexistence of the bistable phases under certain parameter regimes, their dynamical competition can result in a limit cycle phase. In this non-stationary phase, the interactions can facilitate the excitation of one Rydberg state at the expense of the other, and vice-versa, leading to persistent oscillations between them and the formation of a CTC \cite{wu_dissipative_2024}. Indications of a CTC were also observed in the $\mathsf{\Xi}$-type configuration \cite{wadenpfuhl_emergence_2023}.
In both cases, a minimal three-level atom is essential, and it has been shown using a meanfield approach that an effective two-level atom description does not support limit cycles \cite{wadenpfuhl_emergence_2023}. Nevertheless, a transient synchronized dynamics can still be observed in a two-level Rydberg setup due to inhomogeneities in the gas, which lead to the formation of spatial clusters and collective oscillations on the order of milliseconds \cite{ding_ergodicity_2024}. Recent observations have also revealed more complex dynamics in Rydberg platforms, including period-doubling bifurcations in CTCs and routes to chaotic behavior \cite{jiaoObservation2025,liu_bifurcation_2025}.

\emph{Solid-state platforms.--- }
 A prominent realization of a CTC involves an electron-nuclear spin system in a semiconductor, continuously driven by circularly polarized laser light \cite{greilich_robust_2024}. The electron and nuclear spins in the system interact through hyperfine coupling. Here, the electron spins polarize the nuclear spins to generate an Overhauser field, which in turn influences the electron spin dynamics. 
 This process creates a nonlinear feedback loop. 
 Further, the optical pumping continuously injects spin-polarized electrons into the system, compensating for the energy losses and sustaining these nonlinear dynamics. This drives the system into stable, periodic auto-oscillations with coherence times exceeding hours, forming a CTC. Another observation was also done in a driven-dissipative exciton-polariton system within a semiconductor microcavity \cite{carraro-haddad_solid-state_2024}. Here, it is the coupling between excitons and photons, sustained by continuous optical pumping, that generates a CTC whose oscillations lock to the phonon frequency.

 A classical counterpart to CTCs has been observed in a 2D array of plasmonic metamolecules supported on nanowires and illuminated by light \cite{liu_photonic_2023}. The incident radiation mediates many-body interactions among the metamolecules, modeled as coupled classical oscillators. Above a critical driving power, the system undergoes a spontaneous synchronization transition, forming a classical CTC. Interestingly, this synchronization is comprehensively explained as a nonreciprocal phase transition \cite{raskatla_continuous_2024}, where nonreciprocal forces arise from radiation pressure due to asymmetric scattering (e.g., from metamolecules of varying size or geometry).  Finally, there are recent indications that CTCs may also arise in non-reciprocal acoustically levitated particles \cite{morrell_nonreciprocal_2025} or erbium-doped crystals \cite{chen_realization_2023}.

\section{Thermodynamics of Quantum Synchronization}\label{sec:thermodynamics}

Quantum thermodynamics has emerged as an active field of research that studies non-equilibrium systems in terms of the thermodynamically relevant variables \cite{vinjanampathy2016quantum, binder2018thermodynamics, deffner2019quantum,  roadmap2026roadmap}.
A natural question in quantum thermodynamics is the energetic cost of sustaining non-equilibrium states resulting from competitions between external drives and dissipation.
In the context of synchronization, resources such as the minimal work required to sustain synchronization and thermodynamic constraints on the rate of synchronization have been investigated.
It was shown that for a collection of harmonic oscillators that are dissipatively coupled to each other via a non-Hermitian Hamiltonian and initialized in a thermal equilibrium state, the minimal cost of synchronization can be quantified in terms of the synchronization measures \cite{PhysRevLett.133.020401}.
In the limit of a large number of oscillators, the cost scales extensively, and the asymptotic minimal work required for the synchronization of quantum systems is greater than that needed in their corresponding classical counterparts.
Synchronization has also been used to engineer violations of the thermodynamic uncertainty relations (TURs) \cite{razzoli2024synchronization}, which
are constraints on fluctuations of currents of nonequilibrium steady states \cite{seifert2012stochastic}. These relations can provide bounds on dissipation necessary to achieve a reduction in energy current fluctuations.  
The authors consider two harmonic oscillators that are strongly coupled to a common thermal bath, which synchronizes them, and weakly coupled to another thermal reservoir that is modulated by an external drive.
This system has been shown to exhibit local violation of TUR in the regime when they are highly synchronized.

One of the practical applications of quantum thermodynamics is the development of quantum thermal machines. 
Three-level systems dissipatively coupled to hot and cold baths generate limit-cycle steady states that are diagonal in the absence of external coherent drive or couplings.
Such three-level systems when coupled to an external drive, such as in a three-level Scovil-Schulz-DuBois maser operation \cite{PhysRevLett.2.262}, exhibits quantum synchronization.
Here, the system phase-locks to the drive, generating coherence in the energy eigenbasis \cite{PhysRevE.101.020201, PhysRevA.105.L020401}.
The finite power output in thermal machine operation can be attributed to the energy coherence \cite{PhysRevE.68.016101} generated by synchronization.
It has been shown that the steady-state power of these machines can be upper-bounded, with the upper bound being proportional to the synchronization measure defined using the $l_{1}$ norm of coherence, which is saturated when the system is resonantly driven \cite{PhysRevE.101.020201}.
The efficiency of the thermal machine only depends on the ratio of transition frequencies of the system and is independent of its dynamics.
Whereas the power generated by the thermal machine depends on the ratio of temperatures at which the machine is operated, and the sign of the steady-state power depends on whether the thermal machine is operating in the engine or refrigerator regime.

Another interesting direction is the study of coherence generation when several systems are coherently coupled to each other.
Coupling two three-level systems using Hamiltonian terms that do not commute with the bare Hamiltonians can generate coherence and synchronize them \cite{PhysRevA.105.L020401}, allowing them to operate as thermal machines.
When the detuning in the system vanishes, coupling different systems generates coherence and synchronization. 
However, it does not generate power \cite{PhysRevA.105.L020401}, hence uncoupling engine performance from synchronous behavior.
Moreover, when the detuning is non-vanishing, it has been shown that synchronization can not be achieved via energy-conserving couplings.
Hence, synchronization in a non-degenerate system can occur only with an energy non-conserving interaction and will always involve non-vanishing power generation.
In scenarios with external drives and coherent couplings between constituents, competition and cooperation between entrainment and mutual synchronization can occur. This phenomenon has been investigated in multidimensional generalizations of the Scovil-Schulz-DuBois masers, where the thermal machine acts as an engine (refrigerator) in the regime of competition (cooperation) between entrainment and mutual synchronization \cite{PhysRevLett.131.030401}.
Note that applications of many-body synchronization in thermodynamics have also been explored in various settings, as we review in the next section, along with other applications.

\section{Applications}\label{sec:applications_sync}

As noted briefly in the introduction, classical synchronization is fundamental to a wide range of technologies, including communication networks, clock distribution systems in computing, power grid stability, laser and oscillator arrays, and precision measurement instruments \cite{pikovsky2001synchronization,strogatz2015nonlinear,JAALAM20161471}. 
In these contexts, locking the phase and frequency is essential for ensuring robustness and scalability.
Beyond engineering applications, synchronization serves as a key framework for comprehending coordinated behavior across biological, chemical, and mechanical systems \cite{repp2013sensorimotor,strogatz2015nonlinear,ARENAS200893}. This includes processes such as neural activity and the regulation of circadian rhythms. The extensive relevance of synchronization underscores its importance as a foundational concept for managing collective dynamics. 
This understanding is particularly motivating for applications in quantum platforms, where similar synchronization phenomena can be harnessed for applications, such as sensing, control, and information processing, that surpass classical limits.

Quantum synchronization phenomena have been investigated for applications, mainly, in metrology and sensing across a range of scenarios, from a few-body systems to mesoscopic atomic ensembles. 
The ultimate precision of such a sensor is bounded by the Quantum Fisher Information (QFI), denoted by $F^{q}_\theta(t)$, where the minimal estimation uncertainty associated with a parameter $\theta$ scales as $\Delta \theta(t) \propto 1/\sqrt{F_\theta(t)}$ \cite{Braunstein1994,Liu2019}. 
This establishes the QFI as a measure of the sensor's maximum performance. While classical Fisher information is restricted to a linear scaling in resources (shot-noise limit), $F^{c}_\theta(t) \propto Nt$, quantum sensors can leverage coherence and entanglement to achieve the so-called Heisenberg limit, $F_\theta(t) \propto N^2 t^2$ \cite{montenegroReview2025a}.  
In this context, quantum synchronization provides a dynamical mechanism through which correlated evolution and phase locking in few-body and many-body systems can enhance collective sensitivity and enable access to these fundamental precision bounds. Beyond quantum metrology, quantum synchronization has also been proposed for applications in quantum information processing and broader quantum technologies. In particular, it enables frequency multiplexing, where multiple signals are transmitted simultaneously through a single physical channel by encoding information in distinct synchronized Liouvillian eigenmodes of coupled quantum systems via suitable preparation of the initial state. More recently, quantum synchronization in many-body quantum systems has been proposed as a mechanism for storing qubit states in collective eigenmodes, thereby providing protection against decoherence and offering a potential route toward robust quantum memories. Quantum synchronization has also found applications in quantum thermodynamics, particularly in improving the performance of quantum heat engines, also detailed in the previous section. It has also been suggested as a resource for satellite-based positioning systems, where synchronized time signals distributed across satellite networks could enhance timing precision, improve positioning accuracy, and increase communication reliability in space-based technologies. In addition, quantum synchronization has naturally found relevance in the context of quantum clocks, where it has been explored as a mechanism for stabilizing and coordinating quantum timekeeping devices. These developments collectively point to synchronization as a versatile resource across quantum technologies. A detailed discussion of these applications is presented in the subsequent sections, separating those applications that involve few-body dynamics from those involving many-body correlations.

\subsection{Applications involving few-body systems}

Sensing applications with QvdP have been analyzed at both its critical point separating the fixed point regime from the limit-cycle regime~ \cite{dutta2019critical,li2025exact}, and within the entrained regime \cite{vaidya2025quantum}. 
In \cite{dutta2019critical}, susceptibility to changes in driving strength at criticality was investigated and found that for weak drives this susceptibility diverges as a power law, while for larger driving strengths negative susceptibilities are possible.
In \cite{li2025exact}, the QFI for estimating the nonlinear damping strength was found to display Heisenberg scaling in the number of excitations and measurement time at the critical point. 
In \cite{vaidya2025quantum}, it was shown that the QFI for estimating driving strength was also enhanced in the entrained regime of the driven QvdP. 
Moreover, for large excitation numbers, this QFI can be efficiently exploited with simple quadrature measurements \cite{vaidya2025quantum}. 

The emergence of mutual synchronization between two coupled qubits has also found application in sensing. In particular, \cite{giorgi2016probing} showed that a sudden transition between in- and anti-phase synchronization allows the characterization of the spectral density of an environment. In this protocol, one accesses only the dynamics of a probe qubit, which becomes synchronized with a dissipative qubit coupled to the environment that is to be characterized. 

Beyond metrology, mutual synchronization emerging in a chain of qubits and at the level of homodyne quantum trajectories has been proposed as a way to implement a quantum form of frequency multiplexing \cite{schmolkeMeasurement-InducedQuantum2024}. In this system, depending on the initial condition, the chain of qubits synchronizes at different frequencies with different probabilities, which could be used to implement different communication channels using a single quantum system. 

In addition, quantum synchronization has potential uses in communication networks \cite{nande2023quantum} and space technologies, such as enhancing satellite-based positioning systems \cite{nande2023quantumsatellite,nande2024satellite}. 
In particular, satellite-based positioning and timing systems conventionally rely on atomic clocks, whose frequency stability can be degraded by external perturbations such as electromagnetic interference, strong magnetic fields, or even gravitational disturbances, leading to unwanted shifts in the clock signal. To overcome these limitations, recent works \cite{nande2023quantumsatellite,nande2024satellite} have explored synchronization-based alternatives using finite quantum systems. Specifically, a driven Tavis–Cummings model with a small number of qubits has been proposed, in which individual qubits are distributed across satellites and interconnected via optical intersatellite links. These links mediate quantum synchronization through light–matter interactions. By analyzing the spectral density of the emitted radiation, one finds that energy is concentrated at a common frequency across all qubits, yielding highly accurate, synchronized signals that can be exploited for robust timing, positioning, and communication in space-based quantum technologies, thereby ensuring the real-world application of quantum synchronization.

\subsection{Applications involving many-body systems}

\begin{figure*}
\includegraphics[width=\textwidth]{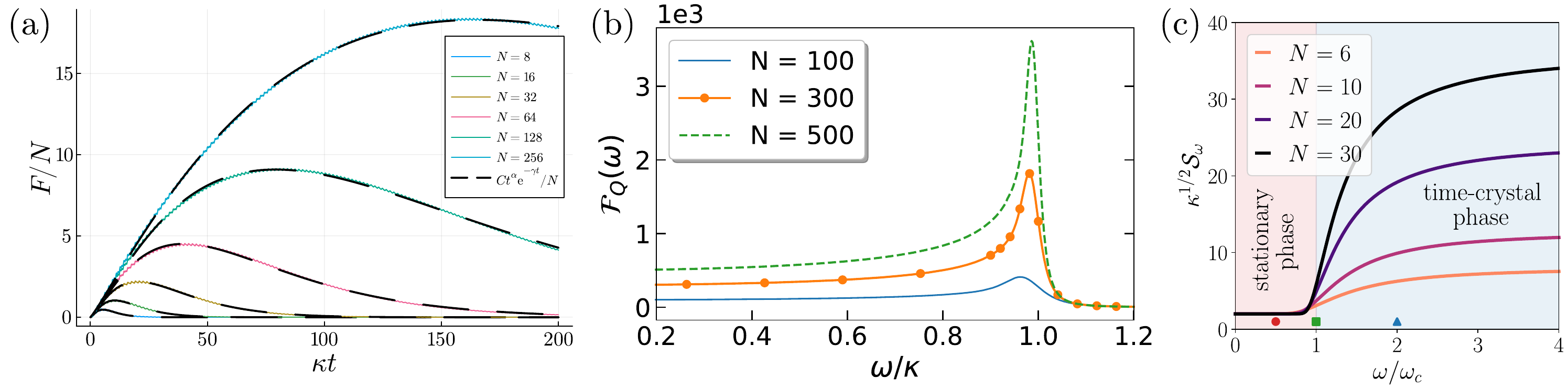}
    \caption{Metrology with CTCs (results shown considering the spin model of Eq.~\eqref{eq.master.equation.BTC} as a sensor).
    \textbf{(a)} Dynamics of the rescaled QFI for sensing the amplitude of an AC field in resonance with the CTC, at $\omega_o = 4\kappa$. The dashed lines correspond to a fit of the Ansatz $Ct^\alpha e^{-\gamma t}$, highlighting its moderate enhancement over the classical limit for exponentially long times ($\gamma\sim 1/N$). \textbf{(b)} In critical sensing, in order to estimate a static magnetic field ($\omega_o \hat S_x$),  the QFI of the steady state $F_Q(\omega_o)$ exhibits a peak close to the transition. The peak fits the form $F_Q^{\rm max} \approx  a N^b$, with $b = 1.345$. The $b > 1$ exponent evidences a quantum enhancement overcoming the SQL. \textbf{(c)} In continuous monitoring, to estimate the same static magnetic field, the sensitivity bound $S_\omega \equiv \lim_{t\rightarrow \infty} \sqrt{\mbox{QFI}/t}$  scales superlinearly with the number of spins in the CTC phase, revealing its quantum enhancement.
    Figures (a-c) are reprinted from \cite{gribbenBoundary2025}, \cite{montenegroQuantumMetrologyBoundary2023} and \cite{cabotContinuous2024}, respectively.
     }
    \label{fig:CTC_sensing}
\end{figure*}

Beyond their fundamental importance to non-equilibrium physics, many-body synchronization may also hold promise for practical applications. 
While research in this direction is still in its infancy, there has recently been significant interest in their potential for metrology protocols  \cite{montenegroQuantumMetrologyBoundary2023,pavlovQuantumMetrologyCritical2023a,cabotContinuous2024,midhaMetrology2025,gribbenBoundary2025,mattesDesigning2025,cabotQuantum2025,oconnorQuantumenhanced2025,arumugam_electric-field_2025,leeTimescalesSqueezingHeisenberg2025,li2025exact},
  and also first steps along quantum clocks \cite{viottiQuantum2025,singhQuantumThermodynamicsLimit2025}, nonequilibrium engines \cite{Carollo2020}, energy storage \cite{paulinoThermodynamics2024,yinFeedbackEnhanced2025}, quantum information processing \cite{esencan2026timecrystalspassivelyprotected} and perspectives on photonics and timetronics \cite{zheludev_time_2024}.

Mutual synchronization between two-level systems can lead to the emergence of a collective dipole with an enhanced coherence time. Examples include synchronization within an ensemble of two-level atoms \cite{xu2015conditional} or between two detuned ensembles \cite{xu2014synchronization} that emit in a common cavity mode, or synchronization of arrays of two-level systems with long-range dipole interactions \cite{zhu2015synchronization}. The collective enhancement of coherence time can then be tapped for metrology by Ramsey protocols, or even by continuously monitoring the light emitted through the common cavity mode \cite{xu2015conditional}.

CTCs can also be used as sensors to estimate an unknown parameter (such as a magnetic field or an abstract parameter governing the system dynamics), leveraging their unique dynamical properties. In a typical metrological protocol, the CTC sensor is exposed to an unknown field that influences it and is encoded in the system dynamics. After an interrogation time $t$, suitable measurements on the sensor can be performed to extract this information and infer the value of the parameter of interest \cite{Degen_review}.
Remarkably, CTC sensors have been shown to achieve a quantum-enhanced Fisher information in various scenarios (see Fig.~\ref{fig:CTC_sensing}):

(i) \textit{Critical Sensing} \cite{montenegroQuantumMetrologyBoundary2023,pavlovQuantumMetrologyCritical2023a,li2025exact}: exploiting the quantum phase transition from a CTC to a stationary phase, where the unknown static field is the parameter being estimated. Close to the transition, the NESS becomes highly susceptible to variations in the unknown field, leading to the quantum-enhanced Fisher information.

(ii) \textit{AC field sensing} \cite{gribbenBoundary2025}: by tuning the CTC frequency into resonance with an unknown AC field, the information from the field is coherently imprinted onto the sensor's dynamics. This results in a quantum-enhanced Fisher information with exponentially long interrogation times, determined by the lifetime of the CTC.

(iii) \textit{Continuous monitoring} \cite{cabotContinuous2024,midhaMetrology2025}: where information about the unknown field is extracted not only from the CTC's state but also from the radiation it emits into the environment. The quantum correlations between the system and its emissions provide an additional information channel, enabling quantum-enhanced sensing for the unknown field. Although the QFI is an abstract quantity defining the ultimate theoretical bound, practical measurement schemes have been proposed to achieve it. These include measuring the total magnetization of the NESS near a phase transition \cite{montenegroQuantumMetrologyBoundary2023}, or using photocounting schemes in continuously monitored settings \cite{mattesDesigning2025,cabotQuantum2025,oconnorQuantumenhanced2025}. These proposals help to bridge the gap between theory and experimental feasibility, underscoring the potential of CTCs as next-generation quantum sensors.

Steps towards applications of CTCs have also been taken in other areas, particularly in the realm of quantum thermodynamics. A first study investigated CTCs as autonomous quantum engines \cite{Carollo2020}. In that work, a cavity–atom system acts as the quantum working fluid, and is coupled to a movable mirror whose forced oscillations define the mechanical work output of the engine. Once the system enters the CTC phase, the engine can operate continuously without the need for external, time-dependent control protocols that are typically required for cyclic engines. 
A thorough characterization of the potential performance of CTCs in applications invariably relies, however, on their thermodynamic efficiency. 
A subsequent study \cite{PhysRevA.108.023516} developed a consistent thermodynamic framework for such engines, showing how persistent dissipated currents and collective many‑body effects govern their power and efficiency in the thermodynamic limit.
Indeed, even the aforementioned sensor application may be severely constrained by its intrinsic entropic cost \cite{gribbenBoundary2025}. An analysis of key thermodynamic figures of merit, such as heat current, work power, and entropy production, was performed for the spin model (Eq.~\eqref{eq.master.equation.BTC}) using a collision-model framework \cite{Carollo_2024}. This work was later extended to a pair of coupled CTCs, exploring their utility for energy storage \cite{paulinoThermodynamics2024} and demonstrating a high efficiency in storing energy for long periods of time.

Another compelling, and arguably natural application (nevertheless only recently explored) involves using CTCs as clocks. By defining a clock tick through the intrinsic dynamics of a CTC, such as a fixed number of emissions to the environment \cite{viottiQuantum2025} or the crossing of a magnetization threshold \cite{singhQuantumThermodynamicsLimit2025}, one can quantify the clock’s precision based on the mean interval between ticks and its variance. 
As these are autonomous machines \cite{Antonio_Marin_Guzman_2024}, this also prompts a quantitative analysis of their performance from a thermodynamic perspective, thus exploring fundamental limitations of timekeeping devices. 
Pushing these ideas to a spin model \cite{viottiQuantum2025}, it has been shown that a CTC can act as a robust clock with enhanced accuracy. Furthermore, the clock’s accuracy scales with the (square) of the thermodynamic entropic cost, with a quantum operation that violates classical thermodynamic uncertainty relations (TURs).

Recently, it has been demonstrated that a time-crystalline phase can be exploited for encoding quantum information in the form of a qubit state \cite{esencan2026timecrystalspassivelyprotected}. In particular, they show that the qubit state can be embedded in an emergent purely imaginary eigenmode of a driven-dissipative Bose–Hubbard dimer, thereby enabling protection of quantum information even in the presence of local dissipation. Moreover, the encoded information is found to be robust against global dephasing and phase perturbations, highlighting time-crystalline dynamics as a promising platform for autonomous quantum memory applications.

\section{Summary \& Outlook}\label{sec:summary_outlook}

Over the last few decades, mechanical systems have been engineered to operate in the quantum regime. Analogues and variations of well-studied nonlinear models have been realized on quantum simulation and computation platforms, including nanomechanical systems, optomechanical devices, trapped ions, and superconducting circuits. It is thus natural that synchronization has emerged as an important topic in quantum science. In this review, we provide a comprehensive survey of the theoretical foundations,  experimental developments, and emerging applications of quantum synchronization. Inspired by the ubiquitous applications of synchronization in classical systems, it is expected that quantum synchronization will play an increasing role in harnessing nonlinear phenomena to control the stability, coherence, and collective behavior of quantum technologies.

Motivated by technological developments that have put quantum synchronization at the center stage, considerable effort has been devoted to understanding the relationship between synchronization and quantum correlations in few and many-body quantum systems, as also highlighted by a recent overview \cite{schmolke2026synchronizationquantumregime}. Studies of quantum synchronization have revealed strong connections to classical nonlinear synchronization phenomena, as captured by the van der Pol model or the many-body Kuramoto model, while also raising important questions about the roles of quantum correlations and entanglement. In the many-body setting, these investigations lead to the notion of continuous time crystals, which provide a framework for persistent collective dynamics and spontaneous breaking of time-translational symmetry in the thermodynamic limit.

Despite the substantial progress reviewed here, several fundamental questions remain open. A unified classification framework capable of organizing the diverse mechanisms of quantum synchronization remains lacking. Particularly intriguing in this context are systems in which synchronization is absent at the mean-field level yet emerges in the full quantum dynamics. Investigating synchronization within the broader context of dissipation engineering, including its connections to decoherence-free subspaces, would also be highly insightful. Moreover, the thermodynamics of synchronization is currently a largely unexplored area whose development may provide new insights into the energetic costs, efficiencies, and fundamental limits of coordinated quantum behavior.

One of the most important challenges ahead is to identify concrete examples where synchronization directly enhances the stability and performance of quantum technologies. 
Quantum synchronization may become an important resource for robust operation in quantum computing, communication, machine learning and quantum sensing. 
For example, mutual synchronization between spatially separated quantum processors may enable distributed information processing without centralized control. Limit cycles naturally emerge in quantum Hopfield and Potts neural networks \cite{rotondo2018open,fiorelli2022phase}. 
Synchronization and entrainment may also provide useful tools in quantum reservoir computing and machine-learning-based forecasting. 
Furthermore, the use of synchronization for metrological applications has recently emerged as a particularly promising direction. Early studies across various platforms already reveal a diversity of mechanisms by which synchronization can enhance sensing performance, making this an emerging yet compelling field for further exploration \cite{dutta2019critical,li2025exact, vaidya2025quantum,montenegroQuantumMetrologyBoundary2023,pavlovQuantumMetrologyCritical2023a,cabotContinuous2024,midhaMetrology2025,gribbenBoundary2025,mattesDesigning2025,cabotQuantum2025,oconnorQuantumenhanced2025,arumugam_electric-field_2025,leeTimescalesSqueezingHeisenberg2025,li2025exact}.

Overall, the breadth of theoretical concepts, experimental realizations, and potential applications reviewed here highlights quantum synchronization as a rapidly developing interdisciplinary field. As quantum technologies continue to mature, synchronization is likely to become an increasingly valuable framework for understanding and engineering collective quantum dynamics, opening new opportunities at the intersection of nonlinear science, quantum information, many-body physics, and quantum engineering.

\section{Acknowledgements}

The authors acknowledge discussions with C. Bruder, E. Lutz, A. Mari, J. Keeling, J. Thingna, T.S. Mahesh, S. Ghosh, B. Bu\v{c}a, V. Mukherjee, R. Mattes, M. Schir\'o, P. Lucignano, A. Russomanno, G. Passarelli, V. Giovannetti, G. L. Giorgi, Leonardo da Silva Souza, Antônio C. Lourenço, Luis Fernando dos Prazeres, Eduardo I. Duzzioni. 
P.S. acknowledges support from the Alexander von Humboldt Foundation through a Humboldt research fellowship for postdoctoral researchers. 
A.C. acknowledges support from the Deutsche Forschungsgemeinschaft (DFG, German Research Foundation) through the Walter Benjamin programme, Grant No. 519847240 and from both
the Spanish Ministerio de Ciencia, Innovación y Universidades and Universitat de les Illes Balears through the Beatriz Galindo programme (BG24/00134). 
F.I. acknowledges the Brazilian funding agencies CAPES, CNPq (308637/2022-4), FAPERJ (E-26/210.236/2024, E-26/204.340/2025), and the Serrapilheira Institute (grant number Serra 2211-42166). 
M.K. acknowledges support by
Prime Minister’s Research Fellowship (PMRF) offered by the Ministry of Education, Govt. of India. 
Y.I. acknowledges the support of the Government of India’s DST-INSPIRE and IIT Bombay for the Asha Navani Fellowship.  
M.H. acknowledges support from the JST Moonshot R\&D program under Grant Numbers JPMJMS2061 and JPMJMS226C. 
I.L. acknowledges supported by the ERC grant OPEN-2QS (Grant No. 101164443). 
R.F. was supported by PNRR MUR project PE0000023- NQSTI and by the European Research Council (ERC)  (Project - RAVE, Grant agreement No. 101053159). 
R.Z acknowledges the Spanish State Research Agency, through the COQUSY project PID2022-140506NB-C21 and -C22 funded by MICIU/AEI/10.13039/50110001103 and by ERDF, EU. 
A.C., I.L. and R.Z. acknowledge funding from the QuantERA II programme (project CoQuaDis, DFG Grant No. 532763411) that has received funding from the EU H2020 research and innovation programme under GA No. 101017733. 
S.V. acknowledges funding under the Government of India’s National Quantum Mission
grants numbered DST/QTC/NQM/QC/2024/1 and
DST/FFT/NQM/QSM/2024/3.  S.V. also acknowledges
useful discussions at the International Centre for Theoretical Sciences (ICTS) during the programs with codes ICTS/qm1002025/01 and ICTS/qt2025/01. 
Views and opinions expressed are, however, those of the author(s) only and do not necessarily reflect those of the European Union or the European Research Council. Neither the European Union nor the granting authority can be held responsible for them.

\end{document}